%% file: stjean_etal_2019.tex
\documentclass[authoryear,sort&compress,3p,final]{elsarticle}

\newcommand{\review}[1]{#1}
\newcommand{\rereview}[1]{#1}
\newcommand{\sout}[1]{}

\input{tex/header.tex}
\journal{Neuroimage}

\begin{document}

\begin{frontmatter}

    \title{Reducing variability in along-tract analysis with diffusion profile realignment}

    \author[umc]{Samuel St-Jean\corref{corr}}\ead{samuel@isi.uu.nl}
    \author[cubric]{Maxime Chamberland}\ead{chamberlandm@cardiff.ac.uk}
    \author[umc]{Max A. Viergever}\ead{max@isi.uu.nl}
    \author[umc]{Alexander Leemans}\ead{A.Leemans@umcutrecht.nl}

    \cortext[corr]{Corresponding author}
    \address[umc]{Image Sciences Institute, University Medical Center Utrecht, Heidelberglaan 100, 3584 CX Utrecht, the Netherlands}
    \address[cubric]{Cardiff University Brain Research Imaging Centre (CUBRIC), School of Psychology, Cardiff University, Maindy Road, Cardiff, CF24 4HQ, United Kingdom}

\input{tex/abstract.tex}

\end{frontmatter}

\input{tex/intro.tex}

\input{tex/theory.tex}

\input{tex/method.tex}

\input{tex/results.tex}

\input{tex/discussion.tex}

\input{tex/conclusion.tex}

\FloatBarrier

\input{tex/appendix.tex}

\section*{Acknowledgments}

We would like to thank Victor Bodiut for useful comments and discussions.
Samuel St-Jean was supported by the Fonds de recherche du Québec - Nature et technologies (FRQNT).
Maxime Chamberland was supported by the Postdoctoral Fellowships Program from the Natural Sciences and Engineering Research Council of Canada (NSERC) (PDF-502385-2017)
This research is supported by VIDI Grant 639.072.411 from the Netherlands Organization for Scientific Research (NWO).
The funding agencies were not involved in the design, data collection nor interpretation of this study.
Declarations of interest: none.
Data were provided [in part] by the Human Connectome Project, WU-Minn Consortium
(Principal Investigators: David Van Essen and Kamil Ugurbil; 1U54MH091657) funded
by the 16 NIH Institutes and Centers that support the NIH Blueprint for Neuroscience Research;
and by the McDonnell Center for Systems Neuroscience at Washington University.

\bibliography{stjean_etal_2019}

\input{tex/supplementary.tex}

\end{document}

%% file: tex/header.tex
\usepackage[T1]{fontenc}
\usepackage[utf8]{inputenc}
\usepackage[english]{babel}
\usepackage{csquotes}

\usepackage{caption,setspace}
\usepackage{amssymb,amsmath}
\usepackage{xspace}
\newcommand*{\eg}{e.g.,\@\xspace}
\newcommand*{\ie}{i.e.,\@\xspace}

\usepackage{hyperref}
\hypersetup{
  colorlinks   = true, %
  urlcolor     = blue, %
  linkcolor    = red,  %
  citecolor    = red   %
}

\usepackage{flafter}
\usepackage{subcaption}
\usepackage{graphbox}
\usepackage[export]{adjustbox}
\usepackage{tikz}
\usepackage{mdframed}
\usepackage{xcolor}
\definecolor{lgray}{gray}{0.9}

\usepackage[section,below]{placeins}

\usepackage[algoruled]{algorithm2e}
\usepackage[nameinlink,capitalise]{cleveref}

\graphicspath{{images/}}
\DeclareGraphicsExtensions{.png,.jpg}

\newcommand{\bval}[1]{b = #1 s/mm$^2$}

\DeclareMathOperator*{\argmax}{arg\,max}

\makeatletter
\g@addto@macro\@floatboxreset{\centering}
\makeatother

\makeatletter
\providecommand*\setfloatlocations[2]{\@namedef{fps@#1}{#2}}
\makeatother
\setfloatlocations{figure}{ht}
\setfloatlocations{table}{ht}
\setfloatlocations{algorithm}{ht}

\newcommand*\annotatedFigureBoxCustom[8]{\node at (#4) [fill=#6,shape=circle,draw=#7,inner sep=1pt,text=#8] {\textbf{#3}};}
\newcommand*\annotatedFigureBox[3][1]{\annotatedFigureBoxCustom{none}{none}{#3}{#2 * #1}{white}{lgray}{black}{black}}
\newenvironment{annotatedFigure}[1]{\centering\begin{tikzpicture}\node[anchor=south west,inner sep=0] (image) at (0,0) { #1};\begin{scope}[x={(image.south east)},y={(image.north west)}]}{\end{scope}\end{tikzpicture}}
\newcommand*{\figuretitle}[1]{{\centering\large{{\fontfamily{fvs}\selectfont#1}}\par\medskip}}

%% file: tex/abstract.tex
\begin{abstract}

Diffusion weighted magnetic resonance imaging (dMRI) provides a non invasive virtual reconstruction of the brain's white matter structures through tractography.
Analyzing dMRI measures along the trajectory of white matter bundles can provide a more specific investigation than considering a region of interest or tract-averaged measurements.
However, performing group analyses with this along-tract strategy requires correspondence between points of tract pathways across subjects.
This is usually achieved by creating a new common space where the representative streamlines from every subject are resampled to the same number of points.
If the underlying anatomy of some subjects was altered due to, \eg disease or developmental changes, such information might be lost by resampling to a fixed number of points.
In this work, we propose to address the issue of possible misalignment, which might be present even after resampling,
by realigning the representative streamline of each subject in this 1D space with a new method, coined diffusion profile realignment (DPR).
Experiments on synthetic datasets show that DPR reduces the coefficient of variation for the mean diffusivity,
fractional anisotropy and apparent fiber density when compared to the unaligned case.
Using 100 in vivo datasets from the human connectome project, we simulated changes in
mean diffusivity, fractional anisotropy and apparent fiber density.
\rereview{Independent} Student's t-tests between these altered subjects and the original subjects indicate that regional changes are identified
after realignment with the DPR algorithm, while preserving differences previously detected in the unaligned case.
This new correction strategy contributes to revealing effects of interest which might be hidden by misalignment
and has the potential to improve the specificity in longitudinal population studies
beyond the traditional region of interest based analysis and along-tract analysis workflows.

\end{abstract}

\begin{keyword}
    Diffusion profile realignment\sep
    Along-tract analysis\sep
    Tractometry\sep
    Tractography\sep
    Diffusion MRI\sep
    White matter
\end{keyword}

%% file: tex/intro.tex
\section{Introduction}
\label{sec:introduction}

Diffusion weighted magnetic resonance imaging (dMRI) is a noninvasive technique which can be used to study micro-structure in living tissues
based on the displacement of water molecules.
Since neurological diseases (\eg multiple sclerosis (MS) \citep{Cercignani2018}, amyotrophic lateral sclerosis (ALS) \citep{Haakma2017a})
involve many processes that affect the density and properties of the underlying tissue,
the corresponding changes are reflected on scalar values extracted from dMRI \citep{Bodini2009}.
However, it remains challenging to accurately pinpoint the underlying cause as many of these changes (\eg axonal damage, demyelination)
may be reflected similarly by changes in measurements from dMRI \citep{Beaulieu2002}.
Such changes could even be due to acquisition artifacts or from the use of a different processing method during data analysis \citep{Jones2010},
making dMRI sensitive, but not necessarily specific, to the various mechanisms involved in those changes \citep{ODonnell2015}.
Accurate characterization of the underlying processes affecting scalar metrics computed from dMRI still remains an open question.

A successful application of dMRI is to reconstruct the structure of the underlying tissues, a process known as tractography (see, \eg \citet{Mori2002a,Jeurissen2017a} for a review).
Tractography enables a virtual reconstruction of the white matter bundles and pathways of the brain, which is central to preoperative neurosurgical planning \citep{Nimsky2016}
and at the heart of connectomics \citep{Sporns2005,Hagmann2007a}.
Over the last years, various analysis strategies have arisen to study scalar values computed from dMRI models.
Two popular schools of techniques consist in using anatomical regions of interests (ROIs), either by manual or automatic delineation \citep{Smith2006,Froeling2016a},
or using spatial information additionally brought by tractography to analyze scalar metrics along reconstructed bundles \citep{Cousineau2017a,Colby2012,Yeatman2012a,Corouge2006,Jones2005}.
Both approaches involve various user defined settings and have their respective criticisms and drawbacks;
ROI based analysis requires accurate groupwise registration \citep{Bach2014a},
whereas tractography based analysis needs to deal with false positives streamlines which can also look anatomically plausible \citep{Maier-Hein2017}.
One key point shared between these methods is that they both require some form of correspondence between the studied structure of interest for each subjects,
either by spatial registration to align the delineated ROIs \citep{Smith2006,Froeling2016a}
or along the streamlines by resampling to a common number of points \citep{Colby2012,Yeatman2012a}.
Tractography based approaches can analyze the voxels traversed by a specific white matter bundle in a
data driven way and reveal subtle local changes inside a bundle, while ROI based analysis
discards the 3D spatial information but reveal widespread changes in the bundle \citep{OHanlon2015}.
For tractography based analysis, metrics are either averaged by using all points forming a common bundle \citep{Wakana2007} or
collapsed as a representative pathway of the bundle \citep{Colby2012,Yeatman2012a,Cousineau2017a}
to study changes in scalar values along its length.
Once this per subject representative streamline has been defined,
it is used to index scalar values along the length of this pathway \citep{OHanlon2015,Szczepankiewicz2013}.
Recent applications include studying changes in diffusion metrics due to Alzheimer's disease (AD) \citep{Jin2017},
which helped to uncover changes in mean diffusivity (MD) along the fornix for example.
Studies in ALS patients also identified a diminution in fractional anisotropy (FA) along the corticospinal tract depending on the origin of the disease \citep{Blain2011}.
Information from other MRI weighting such as myelin water fraction maps derived from T2 relaxometry have also been included to study changes due to MS \citep{Dayan2015}.
As each subject respective morphology is different (\ie reconstructed bundles from different subjects vary in shape and size) just as in ROIs based analysis,
one needs to ensure correspondence between each segment of the studied bundle for all subjects.
This correspondence is usually achieved by creating a new common space where all of the subjects representative streamlines are resampled to a common number of points.
As noted by \citet{Colby2012}, resampling to the same number of points makes the implicit assumption that the end points (and every point in between)
are in correspondence across each subject.
\citet{Yeatman2012a} also mention that \enquote{it is important to recognize that the distal portions of the tract may not be in register across subjects},
even though the resampling step creates a new 1D space for point-by-point comparison.
\citet{ODonnell2009a} previously noticed the potential issue introduced by misalignment between subjects mentioning that
\enquote{improved cross-subject alignment is of interest [...] as the high-frequency variations seen in individual subjects [...] are smoothed in the group average}.
While many methods for registering dMRI volumes or streamlines were developed (see, \eg \citet{ODonnell2017a} for a review),
they do not directly address the issue of possible residual misalignment between the end points \emph{after} extracting the representative streamlines of each subject.
To ensure an adequate comparison between subjects, one must make sure that each streamline corresponds to the same underlying anatomical location.

\begin{figure}
    \includegraphics[width=\linewidth]{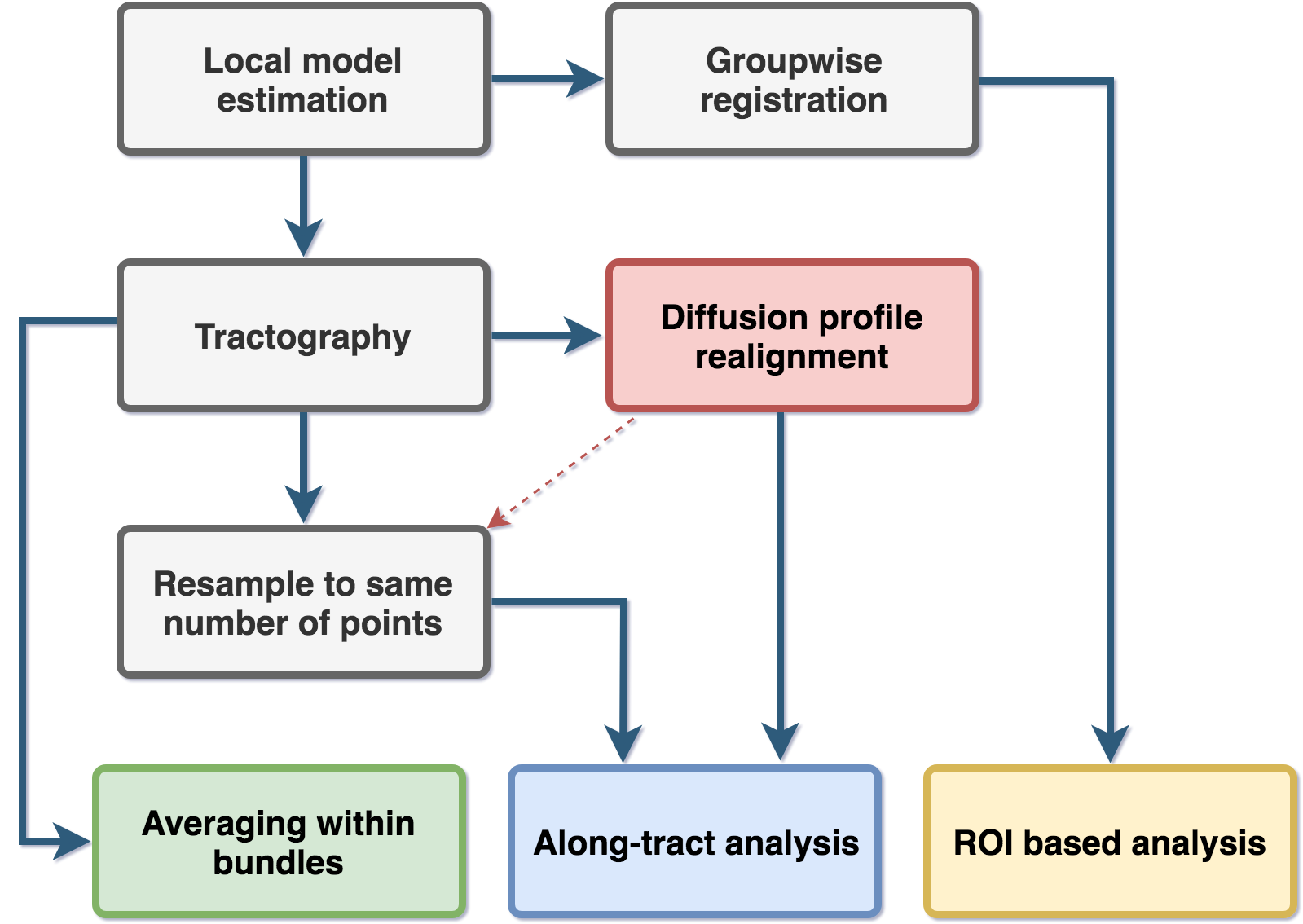}
    \caption{Flowchart of current approaches and the proposed methodology.
    The diffusion profile realignment inserts itself in existing along-tract analysis workflows (red box)
    by combining a different resampling strategy with a realignment step.
    It is also possible to resample each streamlines to a smaller number of points (red arrow) if desired.}
    \label{fig:flowchart}
\end{figure}

In the present work, which extends our preliminary work presented at the ISMRM \citep{St-Jean2016b}, we focus on the issue of possible misalignment between the final representative streamlines before conducting statistical analysis.
To prevent this issue, we propose to realign the representative streamline of each subject while ensuring that the distance between each point is preserved, by resampling to a larger number of points than initially present.
This strategy preserves the original 1D resolution of each subject, allowing a groupwise realignment based on maximizing the overall similarity
by using the Fourier transform.
After this realignment, points from individual streamlines which are identified as outliers can be discarded
as they would not overlap with the rest of the subjects.
The representative, and now realigned, streamlines can be resampled to a lower number of points
such as approximately one unit voxel size to facilitate group comparison and statistics \citep{Colby2012}.
\cref{fig:flowchart} shows an example of a typical workflow to analyze dMRI datasets and
how the proposed diffusion profile realignment (DPR) methodology can be used in preexisting pipelines.

%% file: tex/theory.tex
\section{Theory}
\label{sec:theory}

Each subject's representative streamline is a 3D object, but the scalar metrics extracted along the tract can be viewed as a discrete 1D signal that may be non-stationary.
In this work, we consider the 1D scalar metric profile to be a discrete signal equally sampled at each step of the tractography
which has a value of 0 outside the region delineated by the bundle it represents.
We now present a realignment technique for 1D signals \review{based on maximizing the cross-correlation function (CCF).}
\sout{and how the result can be refined to a spatial resolution below the step size used during tractography.}

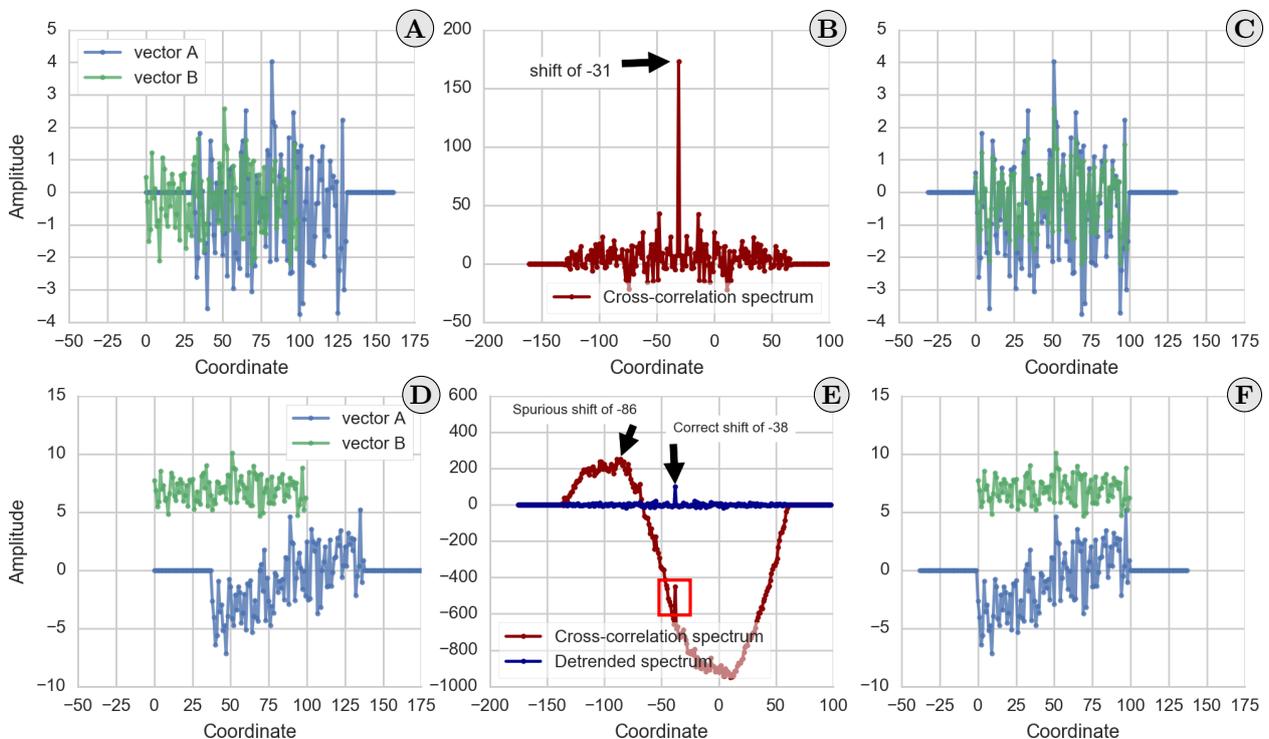
\begin{figure}
    \begin{annotatedFigure}{\includegraphics[width=1.0\linewidth]{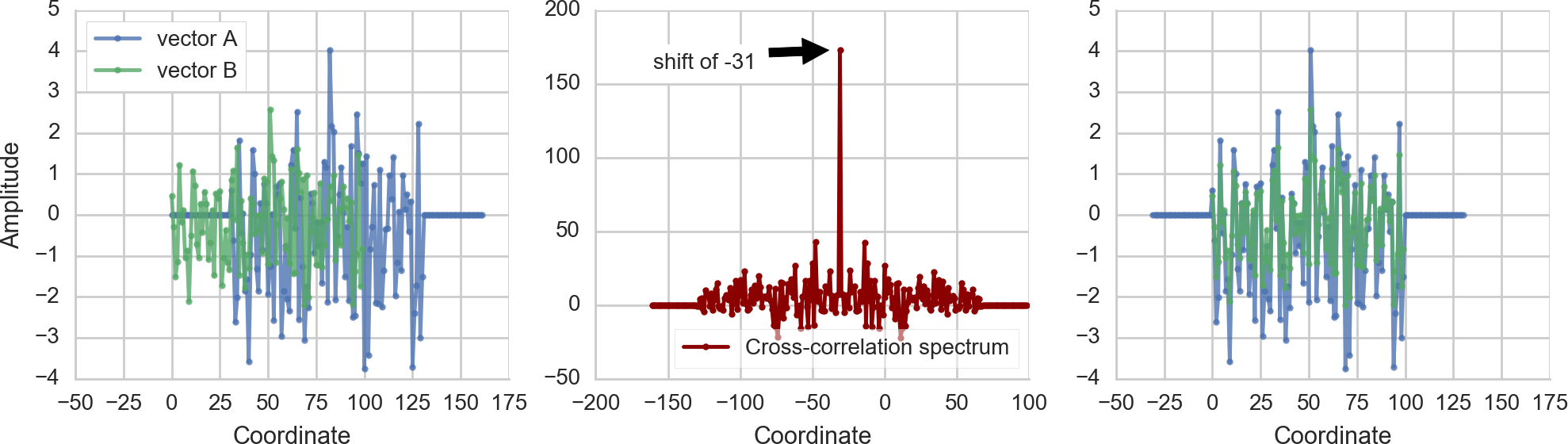}}
        \annotatedFigureBox{0.325,0.98}{A}
        \annotatedFigureBox{0.655,0.98}{B}
        \annotatedFigureBox{0.988,0.98}{C}
    \end{annotatedFigure}
    \begin{annotatedFigure}{\includegraphics[width=1.0\linewidth]{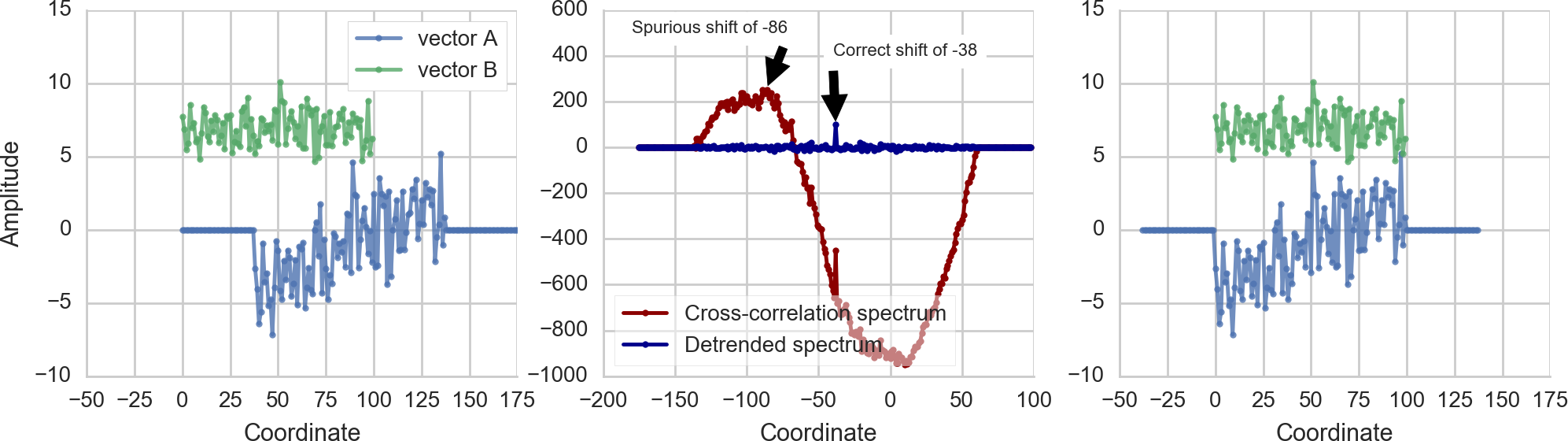}}
        \annotatedFigureBox{0.33,0.98}{D}
        \annotatedFigureBox{0.658,0.98}{E}
        \annotatedFigureBox{0.988,0.98}{F}
        \draw[color=red,very thick] (0.52, 0.35) rectangle (0.545, 0.45);
    \end{annotatedFigure}
    \caption{A synthetic example of the CCF between two randomly generated vectors.
    The top graphs showcase how the CCF spectrum can be used to find the displacement
    required to realign two different vectors by finding its maximum.
    \textbf{A)} Two vectors which are displaced with respect to each other, where vector B has a different amplitude from vector A.
    \textbf{B)} The cross-correlation spectrum, where the peak indicates the required shift to maximize the overlap between both vectors.
    \textbf{C)} The vectors after realignment, which is the exact displacement that had been applied.
    On the bottom graphs, removing the linear trend and normalizing the vectors satisfies the assumption of stationarity required by \cref{eq:crosscorr}
    and allows recovering the correct shift.
    \textbf{D)} Two unaligned vectors of different amplitude where vector B is also non stationary.
    \textbf{E)} The cross-correlation spectrum with detrending and normalization (in blue) and without these steps (in red).
    The detrended version recovers the correct shift, while the original CCF exhibits a variation in amplitude which hides the correct peak
    as a local (red box), but not global, maxima due to non stationarity.
    \textbf{F)} The vectors after realignment with the shift as computed by the detrended CCF.
    Both vectors are now realigned after shifting vector B with the shift computed in \textbf{E)}.
    }
    \label{fig:cc_example}
\end{figure}

We can define the CCF using the fast Fourier transform (FFT) \citep{Cooley1965} as

\begin{equation}
    \text{CCF}(x, y) = \mathcal{F}^{-1} (\mathcal{F} (x) \odot \mathcal{F}(y)^*),
    \label{eq:crosscorr}
\end{equation}
where $\mathcal{F}$(x) and $\mathcal{F}^{-1}$(x) is the Fourier transform of $x$ and its inverse,
$^*$ is the complex conjugation and $\odot$ the pointwise Hadamard product.
\review{The required shift to realign the vectors is given at the maximum coordinate of the CCF.}
\review{The CCF measures the similarity between two vectors $x$ and $y$ assuming}
that the data is 1) stationary, 2) equally spaced between all points and 3) normally distributed \citep{Platt1975,Denman1975a}.
Stationarity can be achieved by fitting and subtracting a low degree polynomial %
from each vector before computing the cross-correlation, see, \eg \citet{Box2008,Stoica2004} and references therein for more details.
Equal spacing between each points can be obtained by resampling the data.
The normality assumption seems less of an issue for large samples in practice \citep{Platt1975}.
If the two vectors $x$ and $y$ have a different amplitude, the cross-correlation can be normalized by
subtracting the mean and dividing by the standard deviation of each vector beforehand \citep{Lewis1995}.
The shift computed \review{at the maximum of the CCF} is an integer displacement which can be refined
\review{by finding the maximum of the parabola around this point}. %
\cref{fig:cc_example} shows an example of the cross-correlation for both the stationary and non stationary case on two vectors.
The first vector was randomly sampled from the standard normal distribution $\mathcal{N}(0 ,1)$.
The second vector was generated from the first vector by changing the offset and amplitude
and then zero padding it at both end to create an artificial displacement.

%% file: tex/method.tex
\section{Materials and methods}
\label{sec:method}

To evaluate the proposed realignment procedure, we 1) generated synthetic datasets comprised of crossing bundles
and 2) compared realignment on in vivo datasets with an altered version of their diffusion metrics.
We now detail the various steps needed to perform an along-tract analysis and
how the proposed realignment algorithm can be applied before performing a statistical analysis between subjects.

\subsection{Resampling strategies for comparison between subjects}
\label{sec:resampling}

Various resampling strategies have been discussed in previous along-tract frameworks,
with a common idea advocating resampling all representative streamlines to the same number of points.
In \citet{Cousineau2017a}, the authors used a fixed number of points by resampling all of the studied bundles to 20 points while \citet{Yeatman2012a} instead used 100 points.
\citet{Colby2012} opted for resampling each bundle based on their average group length, ensuring that approximately one point per voxel was present.
In this representation, each point of the streamlines is considered to correspond to the same anatomical
location across subjects and is therefore blind to the intrinsic variance in shape or length between subjects.
As each representative streamline most likely had a different length initially, the distance in millimeters between each sampled coordinate will be different for each subject.
If the underlying anatomy of some subjects was altered due to, \eg disease or developmental changes,
such information might be lost by resampling to a fixed number of points as a first step.
\review{This can be prevented by ensuring that} the new sampling resolution is at least equal or larger than the initial resolution used during tractography.

\review{As a bundle is comprised of many individual streamlines, they are usually collapsed to a single representative pathway to facilitate subsequent analysis.
This representative streamline is therefore an aggregation of many streamlines of various length
and can be obtained either by averaging \citep{Yeatman2012a,Colby2012} or by finding representative clusters \citep{Cousineau2017a}.
Other assignment strategies towards a single representative pathway have been discussed in \citep{ODonnell2009a,Corouge2006}.
To ensure correspondence during this aggregation step, individual streamlines are usually resampled to a common number of points for all subjects.
While this resampling is needed to obtain the representative streamline,
it may also reduces the sampling resolution from the original streamlines given by the step size used for tractography if not enough points are kept.
The representative streamline of each subject may also have a different orientation altogether and therefore might need to be flipped,
ensuring that they share a common coordinate system \citep{Colby2012}.}

In the present work, we instead advocate a novel two-step resampling strategy which builds upon the classical resampling strategy.
After extracting the representative streamlines $(S_1, \dotsc, S_n) \text{ for } i = 1, \ldots, n$ of each subject,
each representative streamline $S_i$ is defined by its number of points $N_i$ and the distance between its points $\delta_i$.
All streamlines are first resampled to $M_i = N_i \times \delta_i / \delta_{\min}$ points,
ensuring an equal distance between each point $\delta_{\min} = \min (\delta_1, \dotsc, \delta_n)$.
In the end, the streamlines still have a different number of points $M_i \ge \min (N_1, \dotsc, N_n)$
and points at the same coordinates across subjects do not implicitly assume to represent the same anatomical location.
However, the distance $\delta_i$ between each point $M_i$ is now constant across subjects.
While this idea may seem counterintuitive, the motivation behind this choice is due to \cref{eq:crosscorr},
which relies on the FFT, and as such, needs equally sampled vectors and benefits from a high sampling resolution.

\begin{figure}
    \begin{annotatedFigure}{\includegraphics[width=\linewidth]{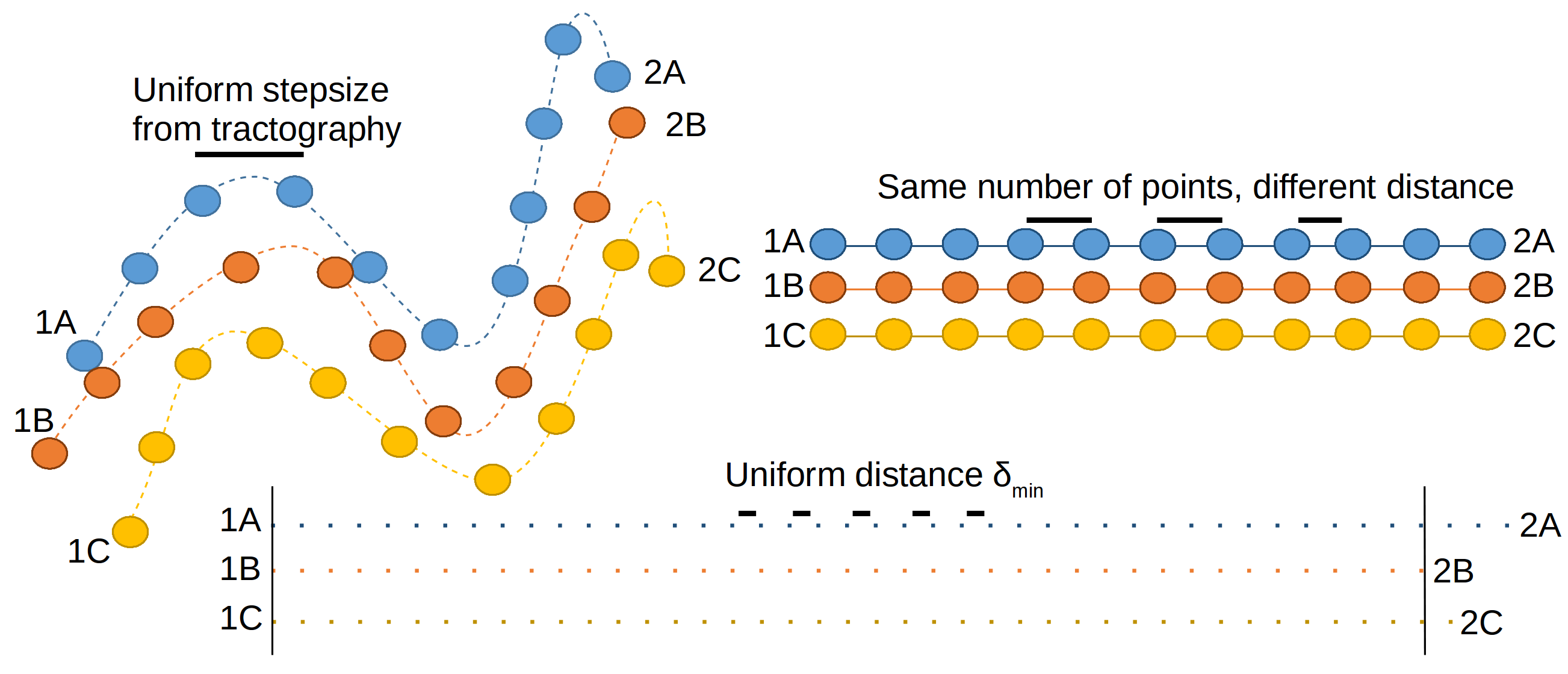}}
        \annotatedFigureBox{0.28,0.7}{A}
        \annotatedFigureBox{0.75,0.8}{B}
        \annotatedFigureBox{0.55,0.38}{C}
    \end{annotatedFigure}
    \caption{An example of the classical and proposed resampling strategies on three representative streamlines.
    In \textbf{A)}, three representative streamlines which have different shapes and lengths
    \review{with their start (1A, 1B and 1C) and end points (2A, 2B and 2C) at different spatial locations.}
    In \textbf{B)}, the classical strategy of resampling to the same number of points (circles) introduces a common space to easily compare them.
    However, the end points of the underlying anatomies are artificially aligned when compared to their original representation \review{and each point is at a different distance (black lines)}.
    In \textbf{C)}, the proposed resampling strategy ensures that the distance $\delta_{\min}$ \review{(black lines)} between every point is constant.
    Even though each streamline length is different \review{as indicated by the location of the end points}, they can now be realigned to identify the common anatomical positions between all subjects.
    }
    \label{fig:resampling_example}
\end{figure}

After the displacement has been applied, one can use the classical resampling strategies presented by other authors,
therefore making our approach fully compatible with already existing analysis techniques.
We opted to use the methodology of \citet{Colby2012} since more than one point per unit voxel size would not carry additional
information from the original data.
This also alleviates further complications arising from multiple comparisons \citep{Benjamini1995}
for the subsequent statistical analysis one seeks to apply afterwards.
\cref{fig:resampling_example} illustrates schematically the classical resampling versus our novel two-step resampling strategy.

\subsection{Proposed algorithm for diffusion profile realignment}
\label{sec:algorithm}

DPR works in three steps once the 1D profiles have been resampled to an equal spacing as presented in \cref{sec:resampling}.
We also ensure stationarity of the data by fitting and subtracting a polynomial of degree one (\ie a straight line) to each subject.
It is important to mention here that this step is only to satisfy the stationarity assumption of \cref{eq:crosscorr} and does not modify the extracted diffusion profiles afterwards.

Firstly, a matrix of displacement is computed between every pairs of subjects
and subsequently refined with parabola fitting as previously defined in \cref{sec:theory}.
A maximum possible displacement in mm is then chosen.
From the displacement matrix, the subject realigning the largest number of streamlines inside this maximum displacement is chosen automatically as the template subject.
As \cref{eq:crosscorr} is symmetric, realigning subject $A$ to $B$ or subject $B$ to $A$ will have the same outcome in practice.

\begin{figure}
    \begin{annotatedFigure}{\includegraphics[width=0.55\linewidth]{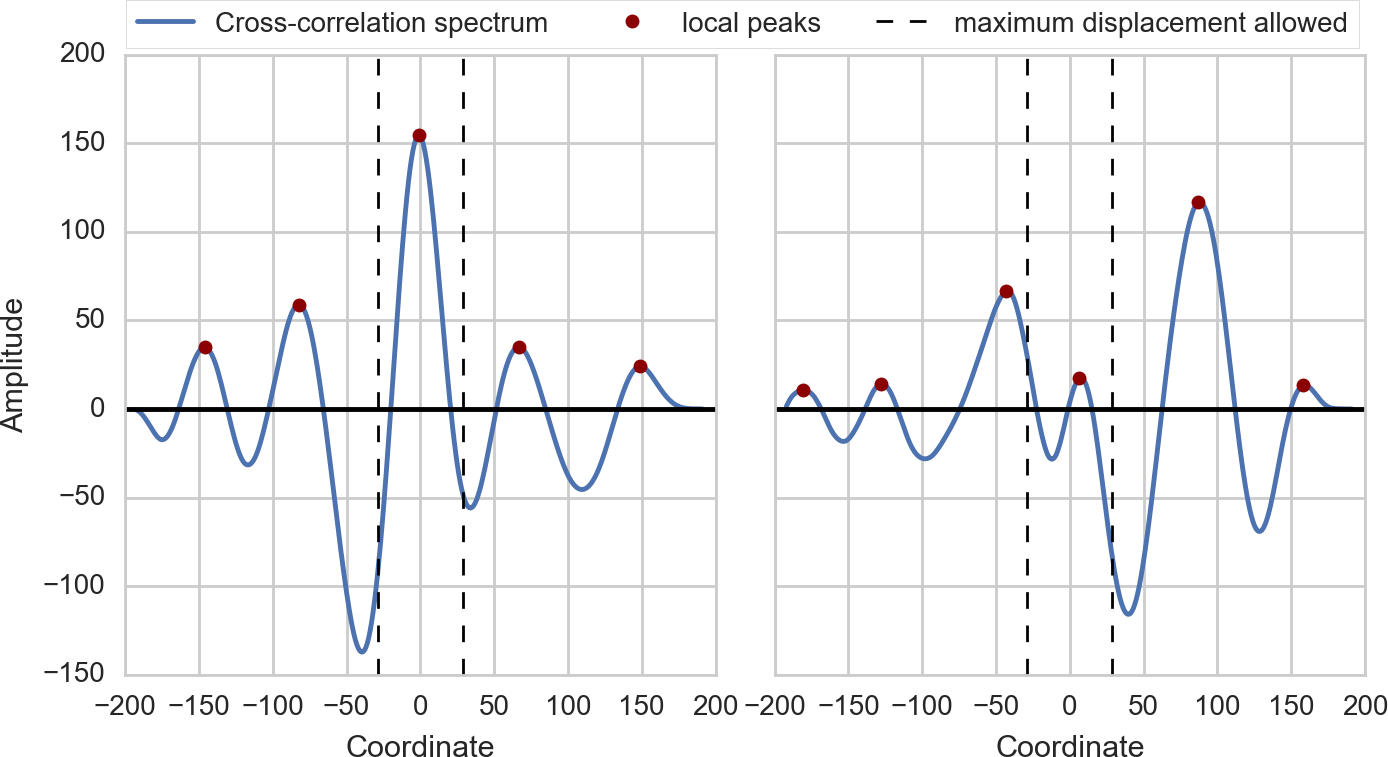}}
        \annotatedFigureBox{0.12,0.88}{A}
        \annotatedFigureBox{0.585,0.88}{B}
    \end{annotatedFigure}
    \hfill
    \begin{annotatedFigure}{\includegraphics[width=0.43\linewidth]{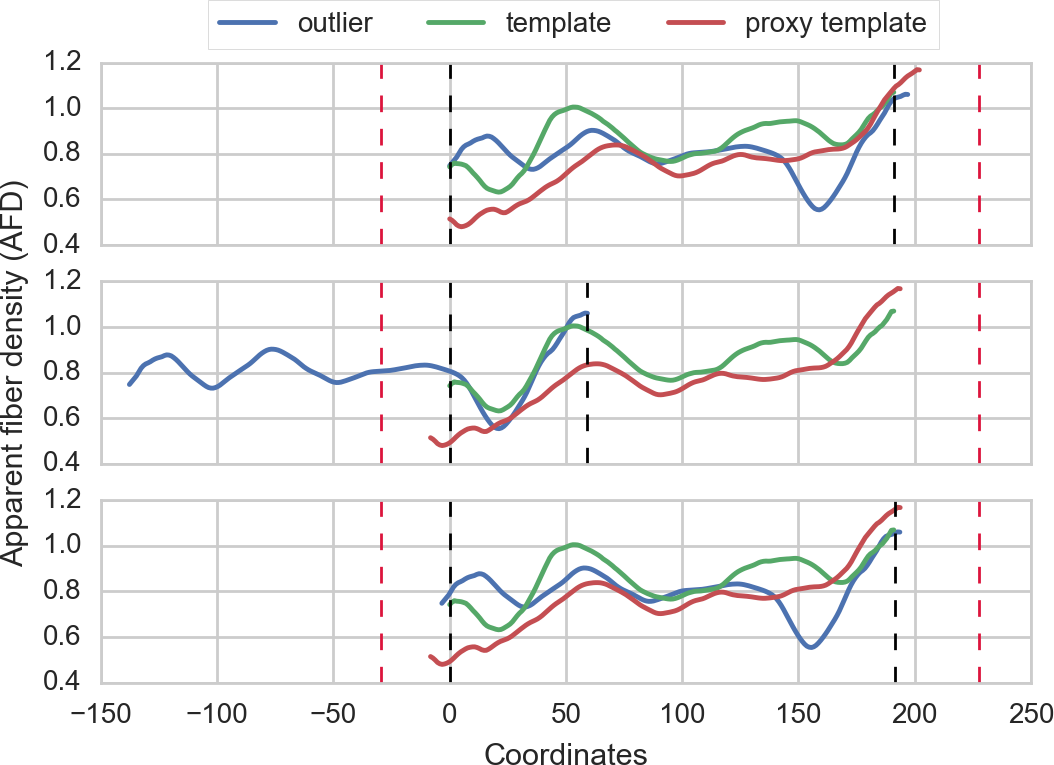}}
        \annotatedFigureBox{0.135,0.89}{C}
        \annotatedFigureBox{0.135,0.605}{D}
        \annotatedFigureBox{0.135,0.32}{E}
    \end{annotatedFigure}
    \caption{An example of a cross-correlation spectra (left) and finding a new template to realign outliers (right) using the HCP datasets.
    On the \textbf{left}, a threshold of 15\% of the total streamline length is selected as the
    maximum allowed displacement (dashed vertical lines).
    \textbf{A)} A streamline with the global maximum of the CCF inside the chosen threshold.
    The maximum indicates the shift needed to realign it to the template.
    \textbf{B)} A streamline with a local maximum, but not the global maximum, of the CCF inside the chosen threshold.
    In this case, the two streamlines would not be realigned together as only small shifts should be needed for realignment.
    On the \textbf{right}, an example of realigning an outlier subject (in blue) to the original template (in green)
    via the closest matching new template (in red) using the AFD metric.
    The black dashed bars indicate the region where all three streamlines fully overlap
    and the red dashed bars shows the maximum allowed displacement of 15\%.
    \textbf{C)} The three streamlines before realignment.
    \textbf{D)} Realigning the blue streamline with the template (in green) as given by the maximum of the CCF results in an outlier as in case \textbf{B)}.
    \textbf{E)} To circumvent the issue, a new template (in red) is found amongst the non-outlier subjects
    which minimizes the total displacement with the original template.
    The blue streamline is therefore not an outlier anymore as it now lies inside the displacement threshold as in case \textbf{A)}.
    }
    \label{fig:spectrum_of_retemplater}
\end{figure}
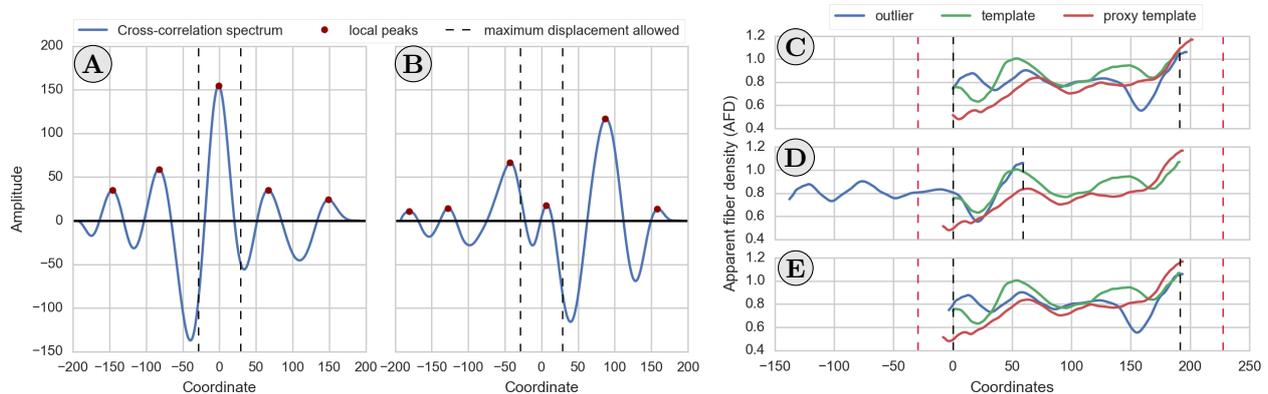

Secondly, all outliers with a displacement larger than the chosen threshold from the first step are realigned
with the help of a new per-streamline template.
For each outlier, a new template is selected amongst the remaining non-outlier subjects which
minimizes the total displacement between the original template
from the first step and the current outlier.
If the new minimum displacement is inside the chosen threshold, the subject which was previously an outlier
is now registered through this new template.
If no new template providing realignment inside the threshold can be found, then this subject
is declared as an outlier and is not realigned at all.
\cref{fig:spectrum_of_retemplater} shows the spectra of a normal subject and of an outlier for spectra computed with \cref{eq:crosscorr} from the HCP datasets.
Even if the optimum displacement lies outside the chosen threshold,
the outlier can still be realigned by finding a new template subject.

Finally, after realigning all the admissible streamlines to the template,
there will be a different number of overlapping subjects for each coordinate.
Just as ROIs were previously used to truncate the bundles' end points (recall \cref{fig:along_bundle_example}),
the resulting aligned streamlines should be truncated once again to reduce their uncertainty
since not all coordinates have the same number of overlapping streamlines anymore.
A pseudocode version of the proposed algorithm is outlined in \cref{sec:appendix}.
Our reference implementation is freely available as a standalone\footnotemark\ \citep[{[Code]}][]{st-jean2019c},
and will also be included in \textit{ExploreDTI} \citep{Leemans2009a}.
We also make available the synthetic datasets and metrics extracted along the representative streamlines of the HCP datasets
which are used in this manuscript \citep[{[Dataset]}][]{St-jean2018b}.
\footnotetext{\url{https://github.com/samuelstjean/dpr}}

\subsection{Datasets and acquisition parameters}
\label{sec:experiments}
\paragraph{Synthetic datasets}

A synthetic phantom consisting of 3 straight bundles crossing in the center at 60 degrees
with a voxel size of 2 mm was created with phantomas \citep{Caruyer2014}.
Each bundle has some partial voluming present on the outer edge to mimic the white matter / gray matter interface.
We simulated 64 diffusion weighted images (DWIs) using gradient directions uniformly distributed
on a half sphere and one \bval{0} image with a signal-to-noise ratio (SNR) of 10, 20 and 30 with uniformly distributed Rician noise and a noiseless reference volume.
Two distinct diffusion weightings of \bval{1000} and \bval{3000} were used, producing a total of 8 different synthetic datasets.
The SNR was defined as $\text{SNR} = S_0 / \sigma$, where $S_0$ is the non-diffusion weighted signal and $\sigma$ is the Gaussian noise standard deviation.

\paragraph{HCP datasets}

100 subjects (50 males, 50 females) from the in vivo Human Connectome Project (HCP) database \citep{VanEssen2012a} aged between 26 and 30 years old were selected.
All 18 \bval{0} volumes were kept along with the 90 volumes at \bval{3000} in order to maximize the angular resolution \citep{Tournier2013}.
The acquisition parameters were a voxel size of 1.25 mm isotropic,
a gradient strength of 100 mT/m, a multiband acceleration factor of 3 and TR / TE = 5520 ms / 89.5 ms.
We used the minimally preprocessed datasets which are already corrected for subject motion,
EPI distortions and eddy currents induced distortions \citep{VanEssen2012a}.

\subsection{Local model reconstruction and fiber tractography}

We used the constrained spherical deconvolution (CSD) algorithm \citep{Tournier2007} with a recursive calibration
of the response function \citep{Tax2014}
and spherical harmonics of order 8 to estimate the fiber orientation distribution functions (fODFs).
We also computed the diffusion tensors using the REKINDLE approach \citep{Tax2015} to exclude potential outliers from the data.
We subsequently computed the apparent fiber density (AFD) maps \citep{Raffelt2012a,DellAcqua2013} from the fODFs
and the FA and MD maps from the diffusion tensors \citep{Basser1996} in all experiments.
Whole-brain deterministic tractography was performed using the fODFs
with \textit{ExploreDTI} \citep{Leemans2009a} with a step size of 0.5 mm,
a fODFs threshold of 0.1 and an FA threshold of 0.2 for all datasets.
The angle threshold, seeding grid resolution and streamlines length threshold used during
tractography were different for the synthetic and HCP datasets as detailed below.

\paragraph{Tractography parameters for the synthetic datasets}

\begin{figure}
    \begin{mdframed}[backgroundcolor=black]
        \centering
        \begin{annotatedFigure}{\includegraphics[width=0.27\linewidth]{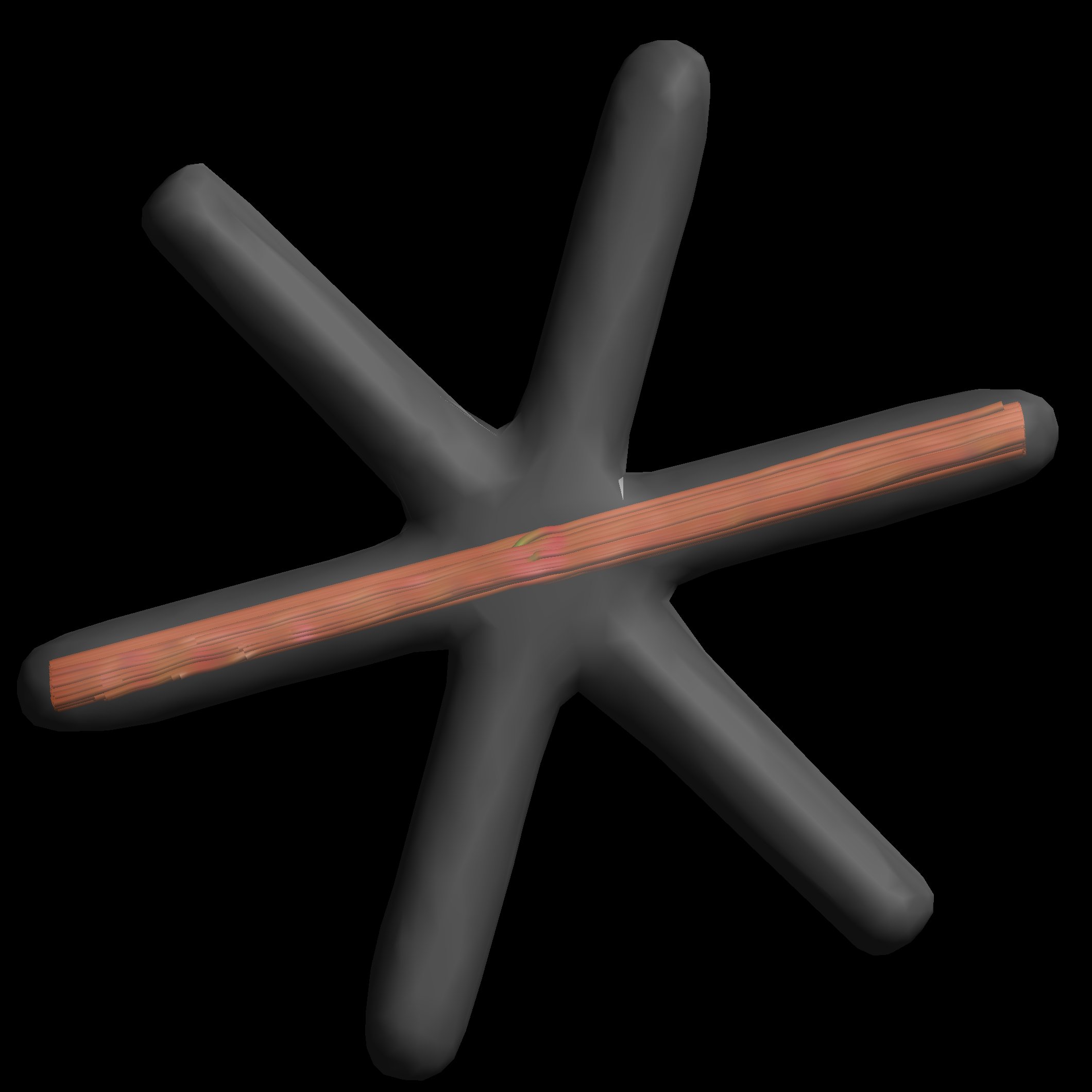}}
            \annotatedFigureBox{0.08,0.9}{A}
        \end{annotatedFigure}
        \hfill
        \begin{annotatedFigure}{\includegraphics[width=0.32\linewidth]{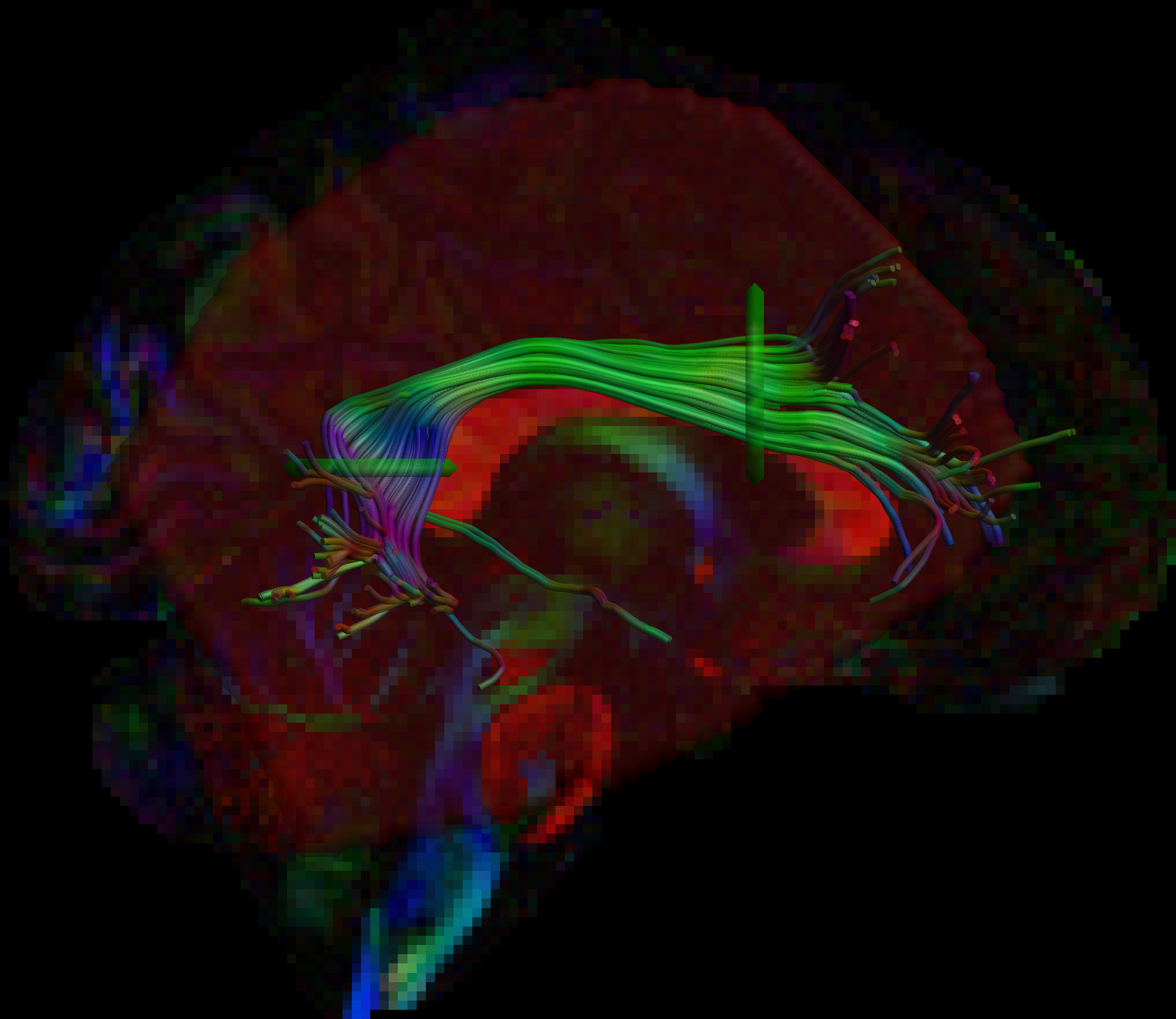}}
            \annotatedFigureBox{0.1,0.9}{B}
        \end{annotatedFigure}
        \hfill
        \begin{annotatedFigure}{\includegraphics[width=0.32\linewidth]{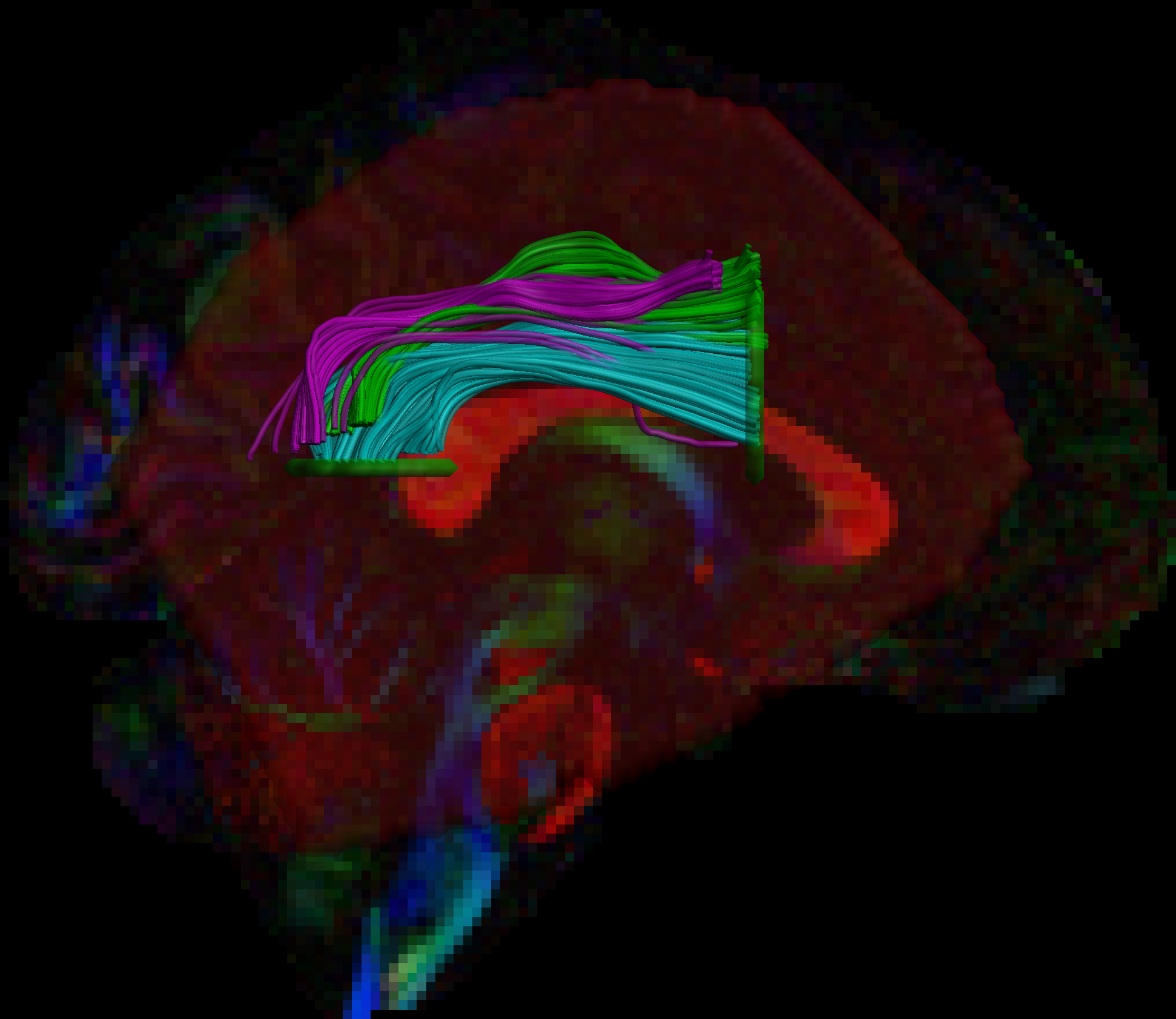}}
            \annotatedFigureBox{0.9,0.9}{C}
        \end{annotatedFigure}
        \begin{annotatedFigure}{\includegraphics[width=0.49\linewidth]{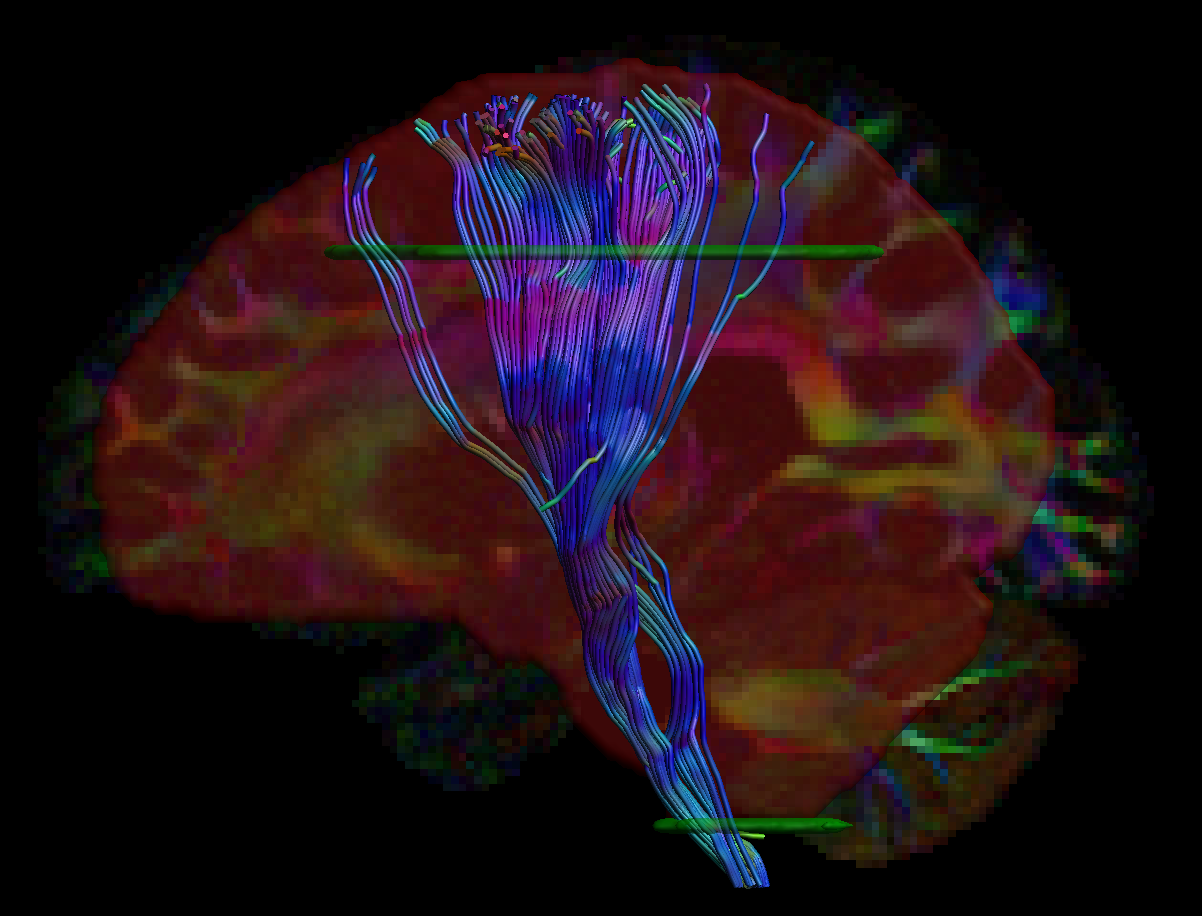}}
            \annotatedFigureBox{0.1,0.9}{D}
        \end{annotatedFigure}
        \hfill
        \begin{annotatedFigure}{\includegraphics[width=0.49\linewidth]{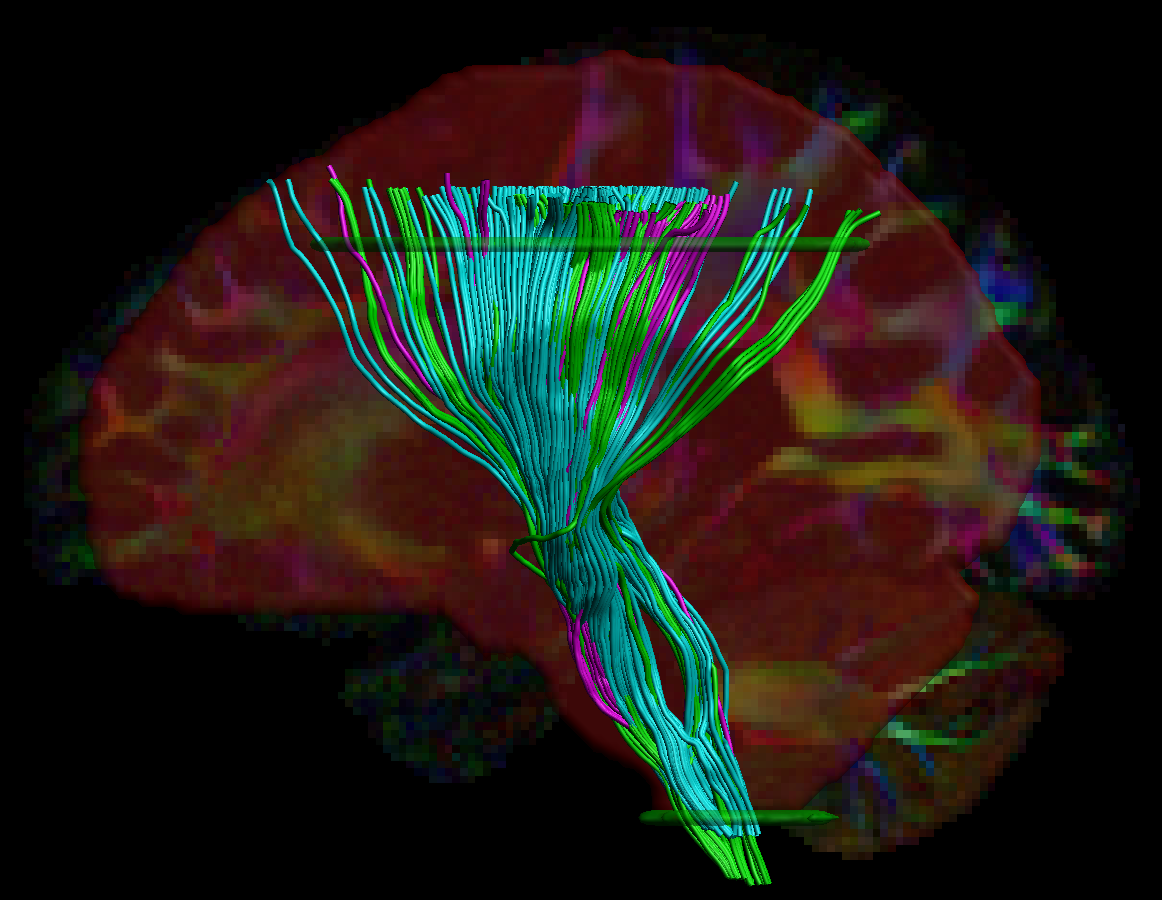}}
            \annotatedFigureBox{0.9,0.95}{E}
        \end{annotatedFigure}
    \end{mdframed}
\caption{The synthetic bundles dataset and the locations of the ROIs used to segment some of the in vivo bundles on the exemplar subject
with their automatically extracted counterpart for three subjects.
In the top row, \textbf{A)} streamlines in a straight bundle of the synthetic datasets.
Note that the streamlines are not truncated at the end points, but rather cover the full length of the red bundle
so that they cross exactly at the center.
The two inclusions (in green) and one exclusion (in red) ROIs segmenting
\textbf{B)} the right AF on the exemplar subject and
\textbf{C)} three automatically extracted right AF drawn in the exemplar subject native space (shown in green, cyan and magenta).
On the bottom row, \textbf{D)} the left CST on the exemplar subject and
\textbf{E)} three automatically extracted left CST bundles (shown in green, cyan and magenta) drawn in the exemplar subject native space.
Note that each subject's bundle would correspond roughly to the same anatomical location in its own native space.
}
\label{fig:tractography_rois}
\end{figure}

Tractography was performed with an angle threshold of 30 degrees and a seeding grid resolution of 0.5 mm on each axis to ensure a dense coverage of each bundle.
Only the streamlines with a length of at least 10 mm and up to 150 mm were kept to prevent the presence of spurious streamlines.
Two ROIs were manually drawn on one bundle to select only straight streamlines belonging to this bundle as shown in \cref{fig:tractography_rois}.
The streamlines were kept at their full extent, including some small variations near the end points due to partial voluming,
which ensures that the intersection of the three bundles is approximately at the center.
To mimic similar representative streamlines extracted from various subjects,
150 streamlines were randomly selected and cut randomly from 1\% up to 10\% of their total length at both end points.
Two sets of representative streamlines were created using classical resampling to the same number of points
and our novel two-step resampling strategy, which is detailed in \cref{sec:resampling}.
In the first case, all streamlines were resampled to 50 points, which is approximately one unit point size per voxel.
As each synthetic representative streamline had a different length after truncation,
resampling to the same number of points allows a direct comparison between each coordinate,
even if they do not match the same \enquote{anatomical} location by design of the experiment.
No resampling was needed to simulate our proposed resampling strategy as the distance between each point is already equal
for this particular synthetic example.

\paragraph{Tractography parameters for the HCP datasets}

Whole-brain tractography was performed with an angle threshold of 45 degrees and a seeding grid resolution of 2 mm on each axis.
Only the streamlines with a length of at least 10 mm and up to 300 mm were kept to limit the presence of spurious streamlines.
ROIs were manually drawn to segment the left and right arcuate fasciculus (AF) and the left and right corticospinal tract (CST) on an exemplar subject \citep{Wakana2007}
as shown in \cref{fig:tractography_rois}.
This exemplar subject FA map was used as a template and subsequently non linearly registered
to each other subject respective FA map using Elastix \citep{Klein2010}.
The obtained transformation was then applied on each ROIs drawn on the exemplar subject defining the four bundles,
therefore warping the original ROIs unto each subject's respective diffusion space as in \citet{Lebel2008a}.
Only the segments between the ROIs were kept \review{to only retain the straight sections and to remove spurious end points \eg before the fanning in the CST}.
An alternative approach could be to extract the bundles automatically using a parcellation of the white matter obtained from each subject's T1-weighted MR image \citep{Wassermann2016,Cousineau2017a}.
\rereview{This would capture the full extent of the bundle instead of only keeping the sections between ROIs as done in the present work, but at the expense of possibly increasing variability.
Such an approach may be useful if important anatomical information is contained in these end regions.}

\paragraph{Extracting representative streamlines for the HCP datasets}

To extract the representative streamline of each subject,
all streamlines forming a given bundle were linearly resampled to the same number of points,
chosen as the number of points of the top 5\% longest streamlines to reduce the effect of possible outliers.
This choice is robust to possible outliers which might be longer (or much smaller) than the rest of the streamlines due to spurious results from tractography
while also keeping a high sampling resolution, a desirable property for \cref{eq:crosscorr}.

\begin{figure}
    \begin{annotatedFigure}{\includegraphics[width=\linewidth]{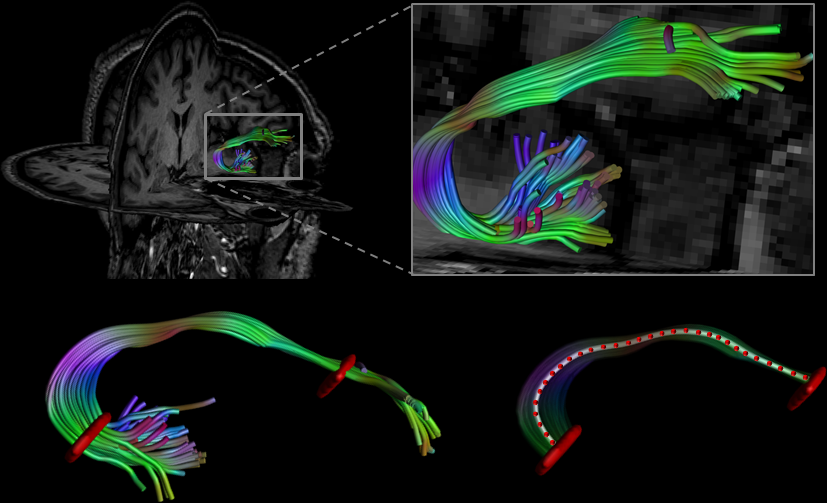}}
        \annotatedFigureBox{0.05,0.9}{A}
        \annotatedFigureBox{0.32,0.2}{B}
        \annotatedFigureBox{0.8,0.2}{C}
    \end{annotatedFigure}
    \caption{An example of along-tract analysis.
    \textbf{A)} The uncinate fasciculus is first segmented from a whole-brain tractography on an exemplar subject.
    \textbf{B)} The two ROIs (shown in red) that were defined to segment the uncinate fasciculus.
        Warping these ROIs to each subject provides an automatic dissection of the bundle.
    \textbf{C)} Only the portion of the mean streamline (shown in white) between the two ROIs is discretized (shown by the red dots),
    which allows mapping scalar metrics along the bundle itself.
    }
    \label{fig:along_bundle_example}
\end{figure}

In the present work, the mean streamline per bundle was extracted and finally resampled in two different ways:
1) using a fixed number of points for all subjects and 2) ensuring an equal distance between each point.
For the classical resampling strategy, we resampled all subjects to 70 points for the arcuate fasciculi and 105 points for the corticospinal tracts.
The second resampling strategy ensured that the distance $\delta_{\min}$ (in mm) between each point is the same for all subjects.
This also means that the representative streamlines of each subject do \emph{not}
have the same number of points and can not be compared directly at this stage when using this resampling strategy.
\cref{fig:along_bundle_example} shows an example of selecting a representative segment between two ROIs as would be done for the uncinate fasciculus.

\subsection{Extracting diffusion metrics for along-tract analysis}
\label{sec:metric_extraction}

Once every representative streamline has been obtained, it can be used to collect diffusion derived metrics along the 3D pathway indexing a volume of interest.
We collected the values of MD, FA and AFD for each subject along the streamline trajectory as in \citet{Colby2012}.
The resulting 1D segment is a vector of values varying along the length of the representative streamline.
This single representative pathway can now be realigned in a pointwise fashion to ensure correspondence between subjects before moving on to statistical analysis.

\subsection{Applying the diffusion profile realignment on representative streamlines}
\paragraph{Realignment of uniformly resampled and variable length streamlines}

To evaluate the reduction in variability brought by our proposed DPR algorithm,
we estimated the coefficient of variation (CV) at each coordinate along the streamlines
before and after realignment using both resampling strategies.
The CV, defined as $CV = \sigma / \mu$ with $\sigma$ the standard deviation and $\mu$ the mean of each metric,
is a unitless standardized measure of dispersion where a lower CV indicates a lower standard deviation around the mean value.
For all experiments, we used a maximum displacement threshold of 15\% to find the subject serving as a template during realignment.
We computed the CV before and after realignment of the representative streamlines using both resampling strategies.
To compare the variability due to truncation of the end points, only the segments where
1\% (at least one streamline present), 50\%, 75\% and 100\% (all streamlines are fully overlapping) of the realigned streamlines were kept for computing the CV.
In the synthetic datasets experiments, we weighted the CV by the number of points at each coordinate to account for the different number of points of the unaligned bundle.
For experiments with the HCP datasets, we instead did a final resampling to the same number of points (if appropriate)
after the realignment as previously used for the classical resampling strategy in order to ensure a fair comparison between both approaches.

\paragraph{Simulating abnormal values of diffusion metrics in HCP subjects}

An example application of the along-tract analysis framework could be to study neurological changes in a given population.
These changes would presumably affect some specific white matter bundles and their underlying scalar values extracted from dMRI.
Both the location and magnitude of these changes could reveal an effect of interest
that might be hidden at first due to potential misalignment between subjects.
To simulate a change in scalar metrics, 50 subjects were chosen randomly and
had their representative streamlines profile modified while the other 50 subjects were left untouched.
These 50 modified subjects are now classified as the \enquote{altered} subjects and
the other untouched 50 subjects as the \enquote{controls} subjects in the subsequent experiments.
For each altered subject, a location \review{covering two times the affected length on both sides was chosen at random starting from the middle} and the metrics were modified at this location.
Two separate set of experiments were performed where the changes in metric was at first $+10\%$ and then $-10\%$ of its original value \review{over 15\% of the length}.
\review{An additional set of experiments simulating highly focused damage of $\pm 25\%$ and $\pm 50\%$ of the metrics over 5\% or 1\% of the bundle length was performed.
For the three cases, the randomly chosen location was at a position from 20\% to 80\%, 40\% to 60\% and 48\% to 52\% of the bundle length.}
This process is repeated for each metric and each bundle, creating a different set of randomly modified subjects every time.
The representative streamlines were finally realigned separately per group.
\review{As the control and altered subjects likely have different 1D profiles, realigning them separately makes it possible to select the best template for each group by itself.
This strategy implicitly assumes that the neurological changes induce a similar increase or decrease in the diffusion metrics of each subject and that after realignment,
each anatomical location is in correspondence between both groups.
Correspondence between groups is also implied in classical along-tract analysis when resampling to the same number of points for comparison.
Limiting the maximum displacement allowed also ensures that information carried by the diffusion metrics stays locally around the same position.
The correspondence after separate realignment is assumed by resampling to the same number of points as the final step before analysis.}
In a clinical study setting, this could reflect neurological changes as induced by, \eg a neurodegenerative disease or aging.
The idea is to induce some changes in the extracted scalar values \emph{only},
without modifying the underlying raw data or performing tractography and representative streamlines extraction once again.
This choice of working in the extracted metric space only is to assess the changes on the metrics and realignment,
in opposition to changes affecting the raw data itself.
As the tractography process and extracted streamlines
would most likely be slightly different due to the inherent challenges in reproducing tractography \citep{Maier-Hein2017},
the subsequent interpretation of the results could be confounded if tractography would be done anew.

\paragraph{Statistical tests between HCP subjects}

We conducted a Student's t-test for \rereview{independent} samples between the controls and altered HCP subjects with
a correction for the false discovery rate (FDR) of $\alpha = 0.05$ \citep{Benjamini1995} for one metric on each bundle.
The t-test was realized on the datasets before and after realignment of the representative streamlines metrics.
However, the FDR correction only ensures an upper bound on the occurrence of false effects and do not indicate their location nor how many are present.

%% file: tex/results.tex
\section{Results}
\label{sec:results}
\subsection{Simulations with the synthetic datasets}
\label{sec:simulations}

\begin{figure}
    \begin{annotatedFigure}{\includegraphics[width=\linewidth]{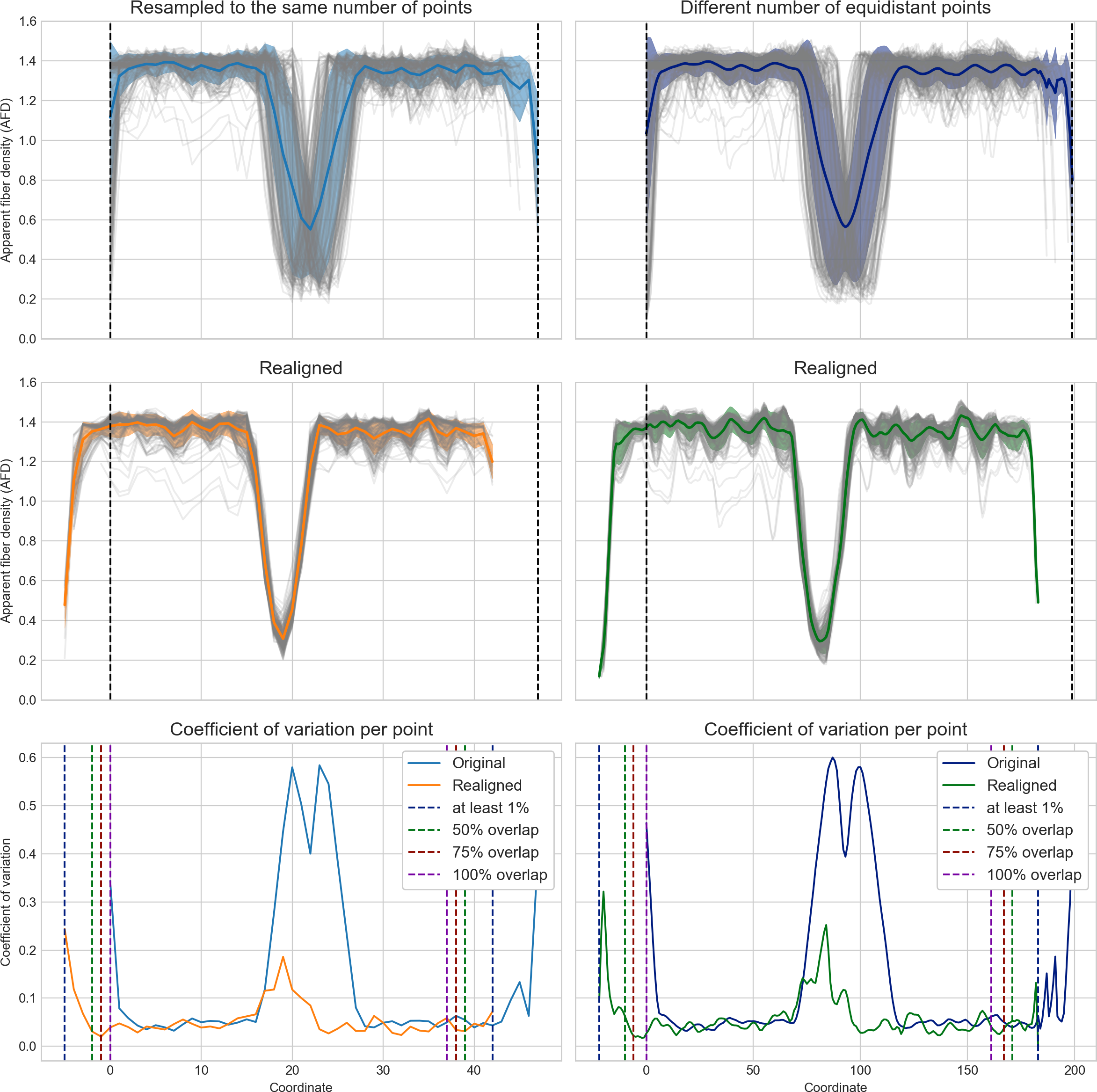}}
        \annotatedFigureBox{0.5,0.96}{A}
        \annotatedFigureBox{0.5,0.64}{C}
        \annotatedFigureBox{0.5,0.32}{E}
        \annotatedFigureBox{0.98,0.96}{B}
        \annotatedFigureBox{0.98,0.64}{D}
        \annotatedFigureBox{0.98,0.32}{F}
    \end{annotatedFigure}
    \caption{Realignment of representative streamlines resampled to 50 points (left column) and with an equal distance $\delta_{\min}$ (right column)
    for the AFD case at SNR 20 and \bval{3000}.
    Each individual streamline is plotted in light gray, with the mean value in color and the standard deviation as the shaded area.
    The black vertical bars indicate the location of the original, non realigned streamlines.
    The colored vertical bars indicate the number of overlapping streamlines,
    ranging from at least 1 (all subjects, purple lines)
    to all of them (100\%, red lines).
    Panels \textbf{A)} and \textbf{B)} show the streamlines before realignment.
    Note how individual streamlines are rather dispersed around the mean.
    Panels \textbf{C)} and \textbf{D)} show the streamlines after realignment, with the mean value being closer to all of the subjects
    and a smaller standard deviation than in panels \textbf{A)} and \textbf{B)}.
    However, due to the realignment, the end points have less subjects contributing to the mean value and
    should be truncated according to the number of overlapping subjects.
    Panels \textbf{E)} and \textbf{F)} show the coefficient of variation (CV, where lower is better) for each point,
    which is in general lower or equal than the non realigned version in both cases.
    Note how the largest reduction in CV is in the crossing region, where the standard deviation is approximately
    three times smaller in the realigned case than for the unaligned case.
    }
    \label{fig:phantomas_realign}
\end{figure}

\begin{figure}
    \includegraphics[width=\linewidth]{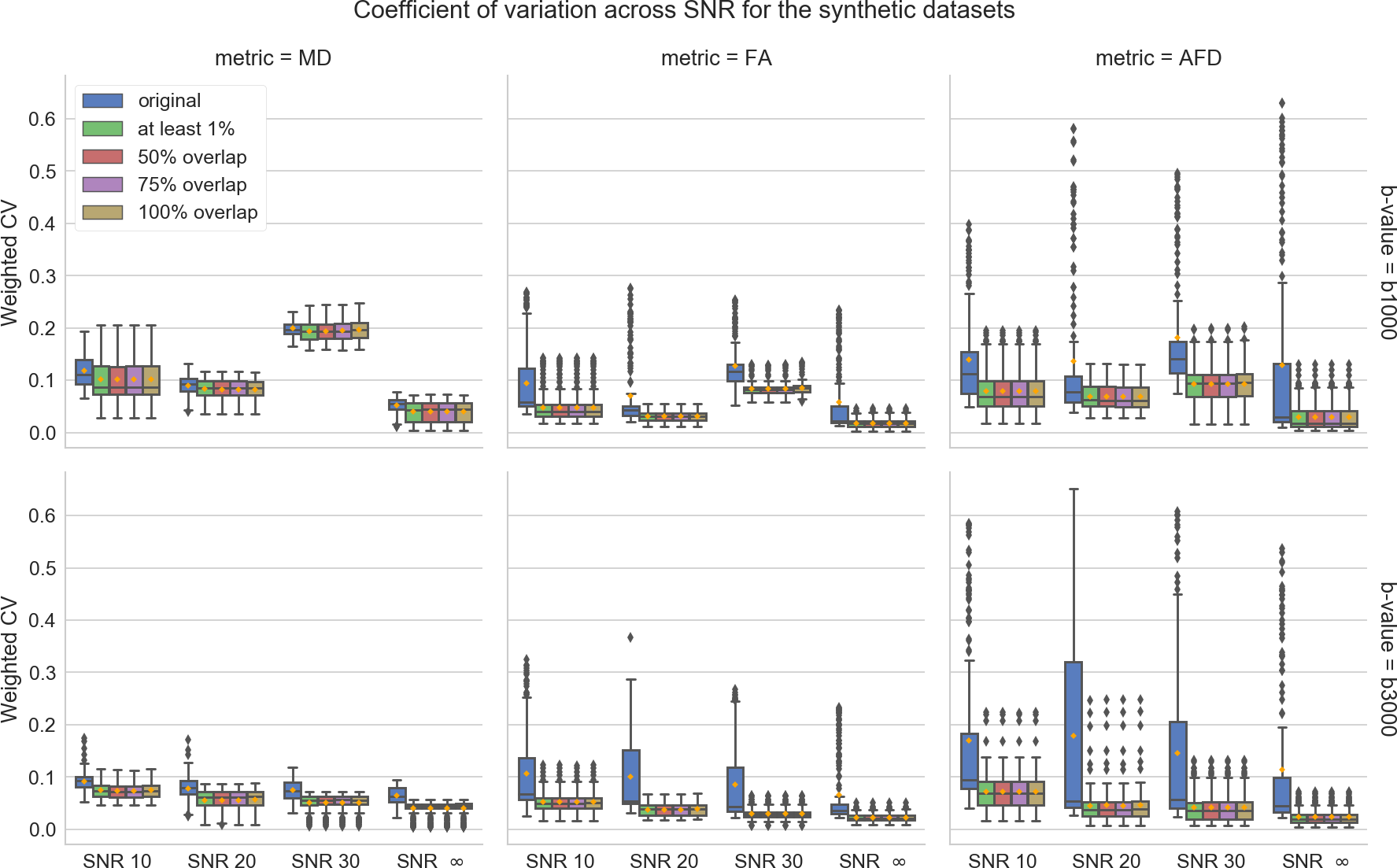}
    \caption{Boxplots of the CV for each point weighted by the number of overlapping subjects,
    for the MD (left), FA (center) and AFD (right) metrics and their respective mean value (in orange).
    The top row shows results for \bval{1000} on the synthetics datasets at SNR 10, 20, 30
    and in the noiseless case while the bottom row shows results for \bval{3000}.
    In all cases, the realigned metrics (for any truncation percentage)
    have a lower or equal CV on average than the non realigned metrics (in blue).
    The FA and AFD metrics have a CV in the realigned case which is on average approximately
    two times smaller than the non realigned case across all SNRs and both b-values.
    This gain is smaller for the MD, which might be due to the relative homogeneity of the MD values.
    }
    \label{fig:phantomas_trends}
\end{figure}

We now present numerical simulations involving the synthetic datasets presented in \cref{sec:experiments},
comparing the two resampling strategies from \cref{sec:resampling} before applying the DPR algorithm.
\cref{fig:phantomas_realign} shows the reduced CV for the realignment of the AFD metric
on the SNR 20 dataset at \bval{3000} when compared to their non realigned counterpart.
After realignment, the standard deviation at each coordinate is now generally lower,
especially in the center portion where the three bundles are crossing.
\sout{However, the end points have a higher CV when all streamlines are resampled to 50 points and less than 50\% of the streamlines are overlapping.}
In the case of resampling to an equal distance $\delta_{\min}$, a few streamlines are overlapping at the end points,
which might reduce statistical power for these regions during subsequent analyses.
As previously mentioned in \cref{sec:algorithm}, portions where only a few streamlines are overlapping should be truncated accordingly
to prevent these degenerate cases.
\cref{fig:phantomas_trends} shows summary boxplots of the CV in addition to the mean CV across all coordinates
for the synthetic datasets for the MD, FA and AFD.
In all cases, realignment provides a lower CV than the non realigned synthetic streamlines.

\clearpage

\subsection{Realignment of the in vivo HCP datasets}
\label{sec:hcp_datasets}
\paragraph{Realignment of the arcuate fasciculi and corticospinal tracts}

\begin{figure}
    \includegraphics[width=\linewidth]{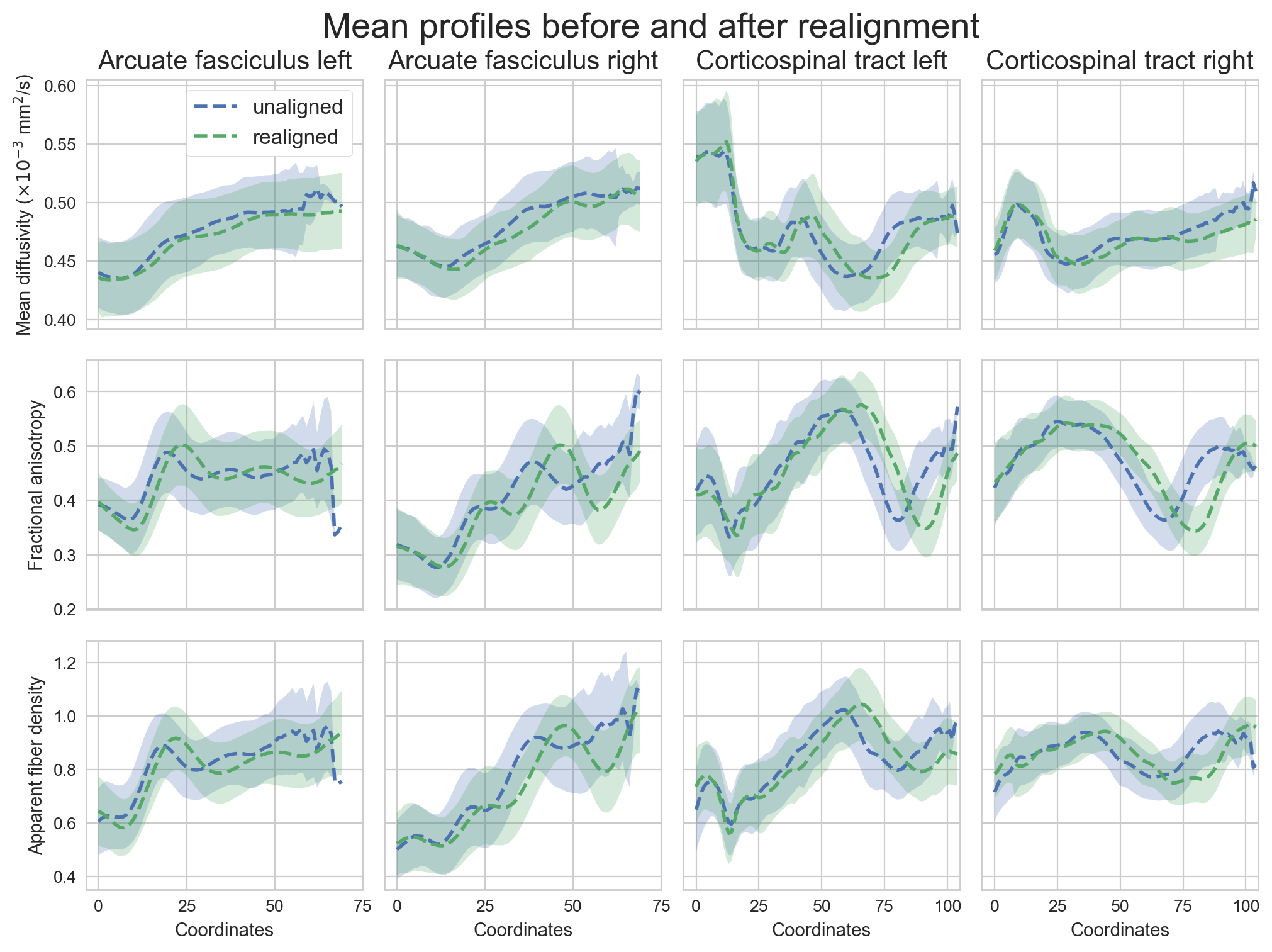}
    \caption{Along-tract averaged profiles (and standard deviation as the shaded area) of the unaligned (blue) and realigned (green) HCP subjects truncated to 75\% of overlap
    with a final resampling to the same number of points.
    Each row shows the profile for one diffusion metric (MD, FA and AFD)
    while each column shows one of the studied bundles (AF left/right \review{from anterior (coordinate 0) to posterior} and CST left/right \review{from inferior (coordinate 0) to superior}).
    After realignment and truncation, the profiles are slightly different from their unaligned version at the end points while the center profile is similar.
    This is likely due to the misalignment mostly affecting the initial end points which are defined by the original truncation
    from the ROIs.
    }
    \label{fig:hcp_realignment}
\end{figure}

\begin{figure}
    \includegraphics[width=\linewidth]{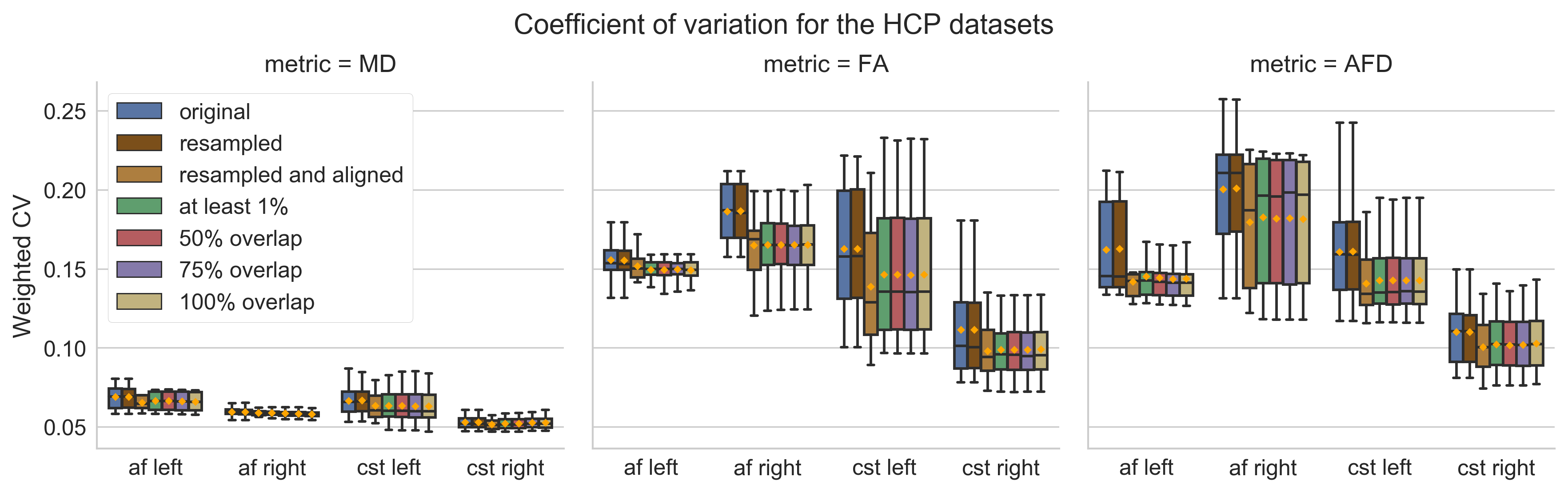}
    \caption{Boxplots of the CV for each point weighted by the number of overlapping subjects,
    for the MD (left), FA (center) and AFD (right) metrics and their respective mean value (in orange) for the four studied bundles.
    Similar to the synthetic datasets experiments, the in vivo datasets have a lower CV
    after realignment (green, red, purple and yellow boxplots)
    than when they are unaligned (brown boxplots).
    Even if the representative streamlines are truncated to the shortest number of points (yellow boxplot)
    or are resampled to the same length (light brown boxplot), the CV is smaller in the realigned cases than in the unaligned cases (brown and blue boxplots respectively).
    The gain in CV is once again smaller for the MD but larger for the FA and AFD in favor of the realigned cases,
    which is in line with the synthetic experiments.
    }
    \label{fig:hcp_coeff_variation}
\end{figure}

To quantify the improvements brought by the DPR algorithm for the in vivo datasets,
we realigned the representative streamlines extracted from the 100 HCP datasets.
\cref{fig:hcp_realignment} shows the final outcome with the two previously discussed pipelines for producing along-tract averaged profiles:
resampling to the same number of points as is conventionally done and after realignment with the DPR algorithm.
For the realigned case, we kept only the segments where at least 75\% of the subjects are overlapping
and finally resampled all subjects to the same number of points.
This last resampling step could be considered optional and is used to allow an easier visual comparison between the unaligned and realigned group profiles.
While the overall shape of each profile\sout{s} is similar between the unaligned and realigned version,
the end points \review{and location of salient features} are slightly different due to the realignment and the truncation threshold we used.
As the maximum displacement threshold dictates which subject is used as a template for the realignment, average group profiles using
a maximum displacement threshold of 5, 10 and 20\% are shown in the supplementary materials \cref{sec:supp_realignment}.
To assess the effect of truncation on variance near the end points,
we computed the CV for each metric at various truncation thresholds and for the unaligned metrics.
\cref{fig:hcp_coeff_variation} shows the CV for the HCP datasets when the bundles are first resampled to the same number of points
and after realignment (in brown). In all cases, the CV is approximately equal or lower after realignment with the DPR algorithm
than when the representative streamlines are unaligned and resampled to the same number of points.
We also show the CV in the unaligned case where all streamlines have an equal distance $\delta_{\min}$ between points
and for four truncation thresholds after applying the DPR algorithm (no truncation, 50\%, 75\% and 100\% of overlap).
In this particular case, the resampled and realigned bundles (light brown) and the realigned bundles with no truncation (green)
are mostly equivalent as they are resampled to the same final number of points after realignment for comparison purposes.
The main tendency shows a lower mean CV after realignment when compared to the non realigned cases.
The CV values are also generally lower with increasing truncation thresholds as the number of overlapping points per coordinates is also increasing,
contributing to a lower standard deviation of each metric.

\paragraph{Robustness of the shapes of averaged profiles towards different metrics}

\begin{figure}
    \includegraphics[width=\linewidth]{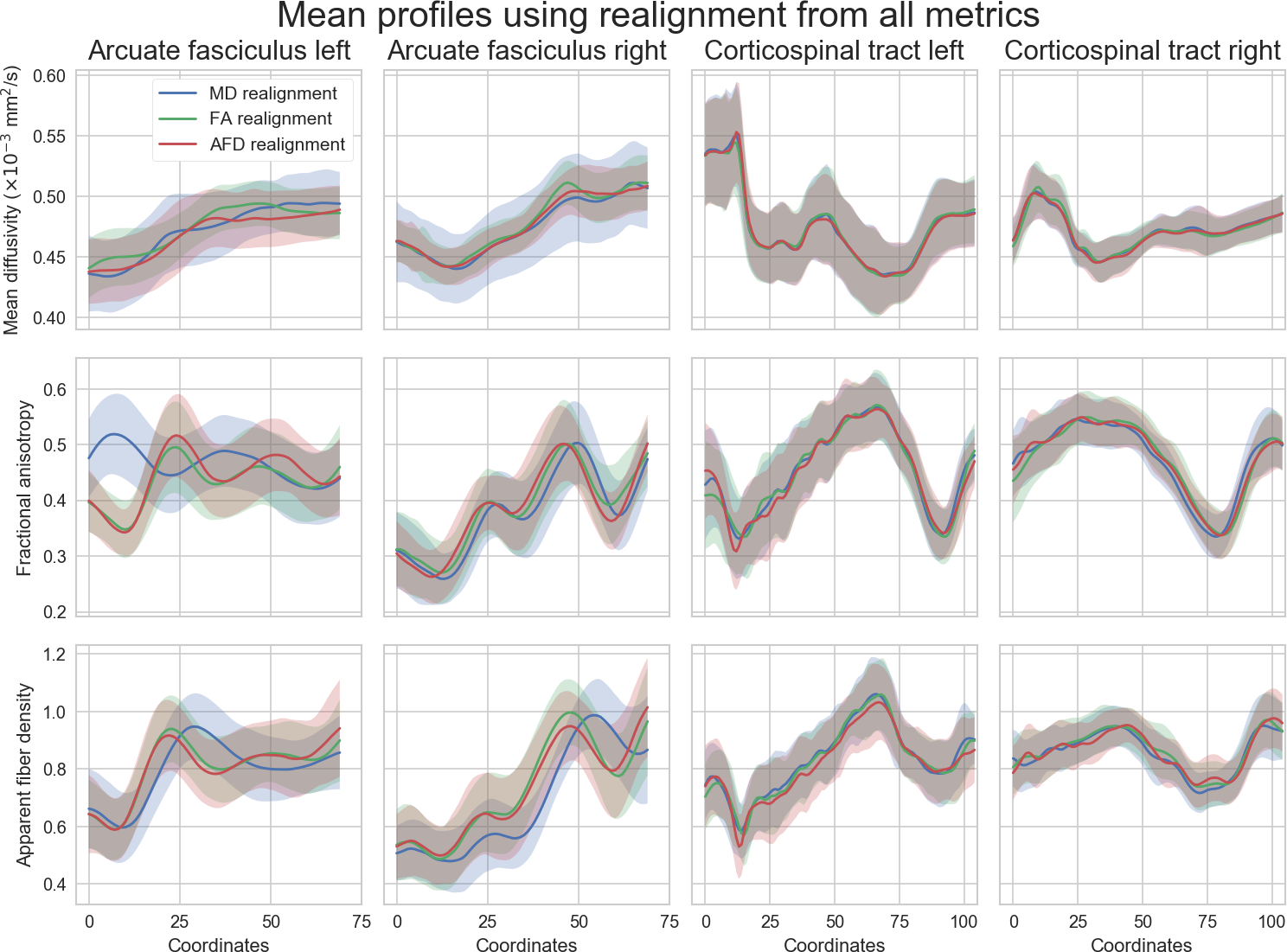}
    \caption{Along-tract averaged profiles (and standard deviation as the shaded area) of the white matter fiber bundles (columns) from the HCP datasets
    after realignment for each studied metric (rows). The metrics were truncated to 75\% of overlap after realignment
    with a final resampling to the same number of points.
    On each row, the along-tract profile after realignment is shown for a given metric (MD on the first row, FA on the second row and AFD on the third row)
    using the displacement computed by the MD (blue), FA (green) and AFD (red).
    \review{The AF are displayed from anterior (coordinate 0) to posterior and the CST from inferior (coordinate 0) to superior.}
    }
    \label{fig:hcp_displacement}
\end{figure}

When performing an along-tract analysis, tractography plays a key role as a spatial indexation method for \emph{extracting}
the 1D metric profiles along the streamline.
Given a particular subject representative streamline, the various scalar metrics that can be extracted each have their own distinct 1D profile along the streamline.
In order to assess the robustness of our proposed DPR algorithm,
we investigated whether for a given metric and template the resulting average group profile would be similar using the displacement computed from the other metrics.
As the displacement depends on the spectrum of each 1D profile,
each metric may use a different template and apply a different displacement for each subject.
This may ultimately lead to a different group average profile due to our algorithm automatically choosing the template amongst the subjects.
However, the \emph{relative} displacement due to a change of template (and hence the resulting group average 1D profile) may be unaffected by this choice,
leading to a similar group average profile.
\cref{fig:hcp_displacement} shows the resulting average group profiles for each metric
when using the original realignment and the realignment that would be applied from the two other remaining metrics with a maximum displacement threshold of 15\%.
As the AF is slowly varying in terms of diffusion metrics along its extracted path,
the realignment of the MD metric is similar even when using the displacement computed from the FA or AFD metric.
On the other hand, applying the realignment given by the MD to the FA and AFD profiles
leads to different optimal realignments and a change in their overall profile.
For the CST, as the representative streamline crosses other anatomical bundles along its path,
the 1D profiles have more variation along coordinates than in the AF case.
This is mostly notable in the MD metric profile which is now similarly realigned when using either the FA or AFD.
Due to these anatomical \enquote{landmarks}, the displacement given by the MD also yields similar profiles when applied to the FA and AFD metrics.
Results for maximum displacement thresholds of 5, 10 and 20\% produced similar trends which are shown in the supplementary materials \cref{sec:supp_displacement}.

\paragraph{Realignment with simulated diffusion abnormalities in HCP datasets}

\begin{figure}
    \begin{annotatedFigure}{\includegraphics[width=0.49\linewidth]{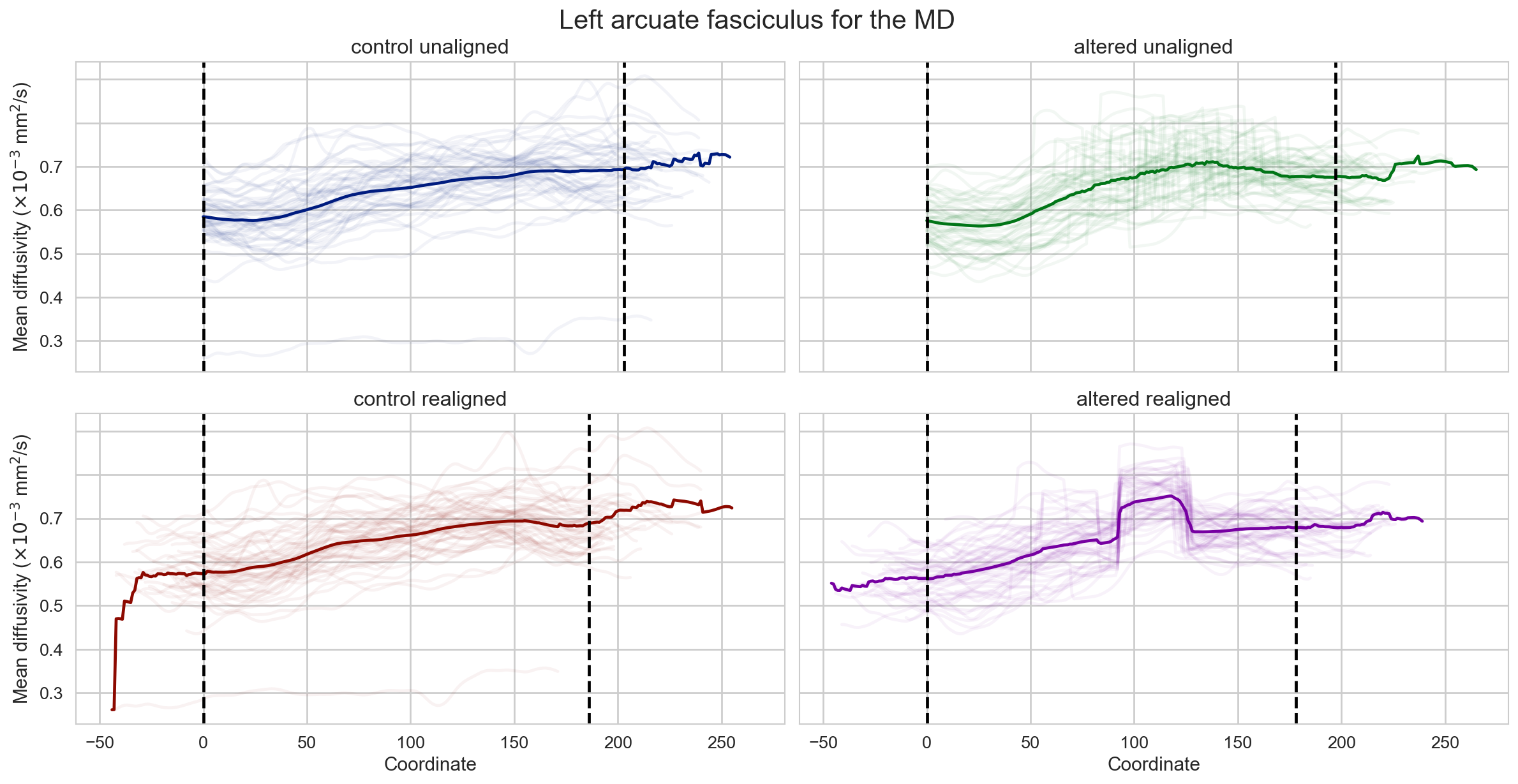}}
        \annotatedFigureBox{0.1,0.94}{A}
        \draw[color=red,thick] (0.3, 0.25) rectangle (0.4, 0.4);
        \draw[color=red,thick] (0.3, 0.7) rectangle (0.4, 0.85);
        \draw[color=red,thick] (0.71, 0.25) rectangle (0.81, 0.4);
        \draw[color=red,thick] (0.71, 0.7) rectangle (0.81, 0.85);
    \end{annotatedFigure}
    \hfill
    \begin{annotatedFigure}{\includegraphics[width=0.49\linewidth]{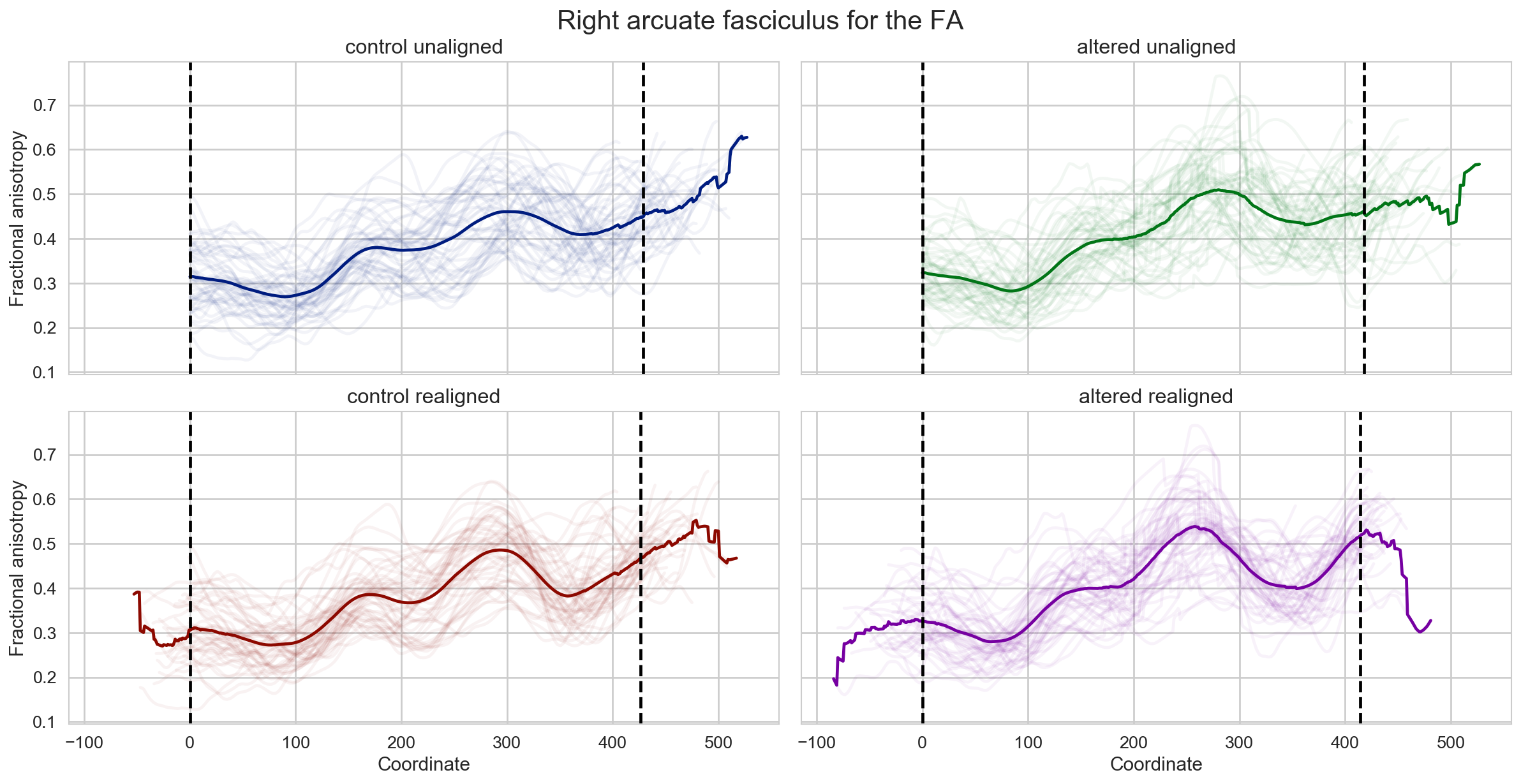}}
        \annotatedFigureBox{0.1,0.94}{B}
        \draw[color=red,thick] (0.3, 0.65) rectangle (0.45, 0.8);
        \draw[color=red,thick] (0.75, 0.65) rectangle (0.9, 0.8);
        \draw[color=red,thick] (0.3, 0.2) rectangle (0.45, 0.35);
        \draw[color=red,thick] (0.75, 0.2) rectangle (0.9, 0.35);
    \end{annotatedFigure}

    \begin{annotatedFigure}{\includegraphics[width=0.49\linewidth]{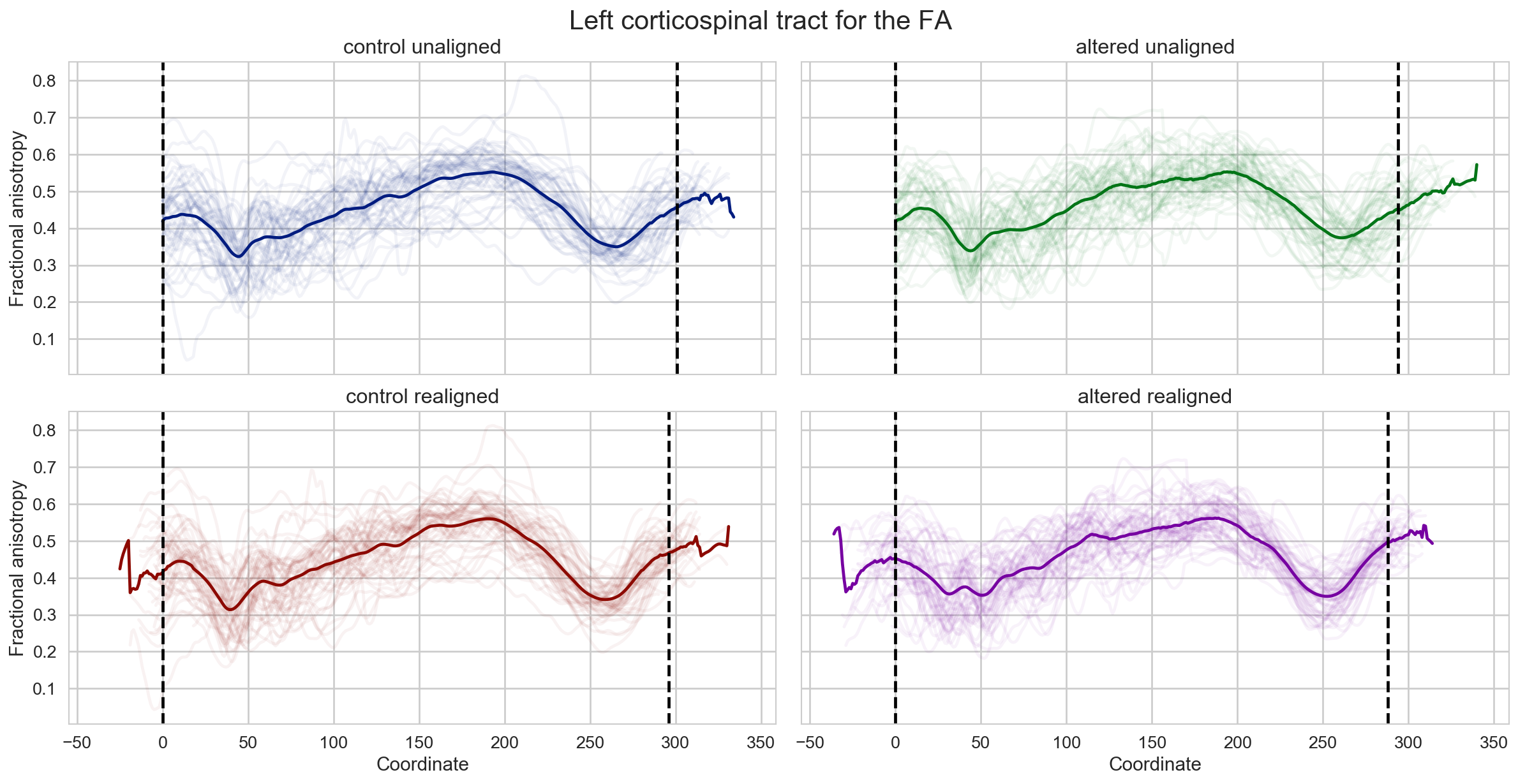}}
        \annotatedFigureBox{0.1,0.94}{C}
        \draw[color=red,thick] (0.11, 0.65) rectangle (0.23, 0.75);
        \draw[color=red,thick] (0.58, 0.65) rectangle (0.7, 0.75);
        \draw[color=red,thick] (0.11, 0.22) rectangle (0.23, 0.32);
        \draw[color=red,thick] (0.58, 0.22) rectangle (0.7, 0.32);
    \end{annotatedFigure}
    \hfill
    \begin{annotatedFigure}{\includegraphics[width=0.49\linewidth]{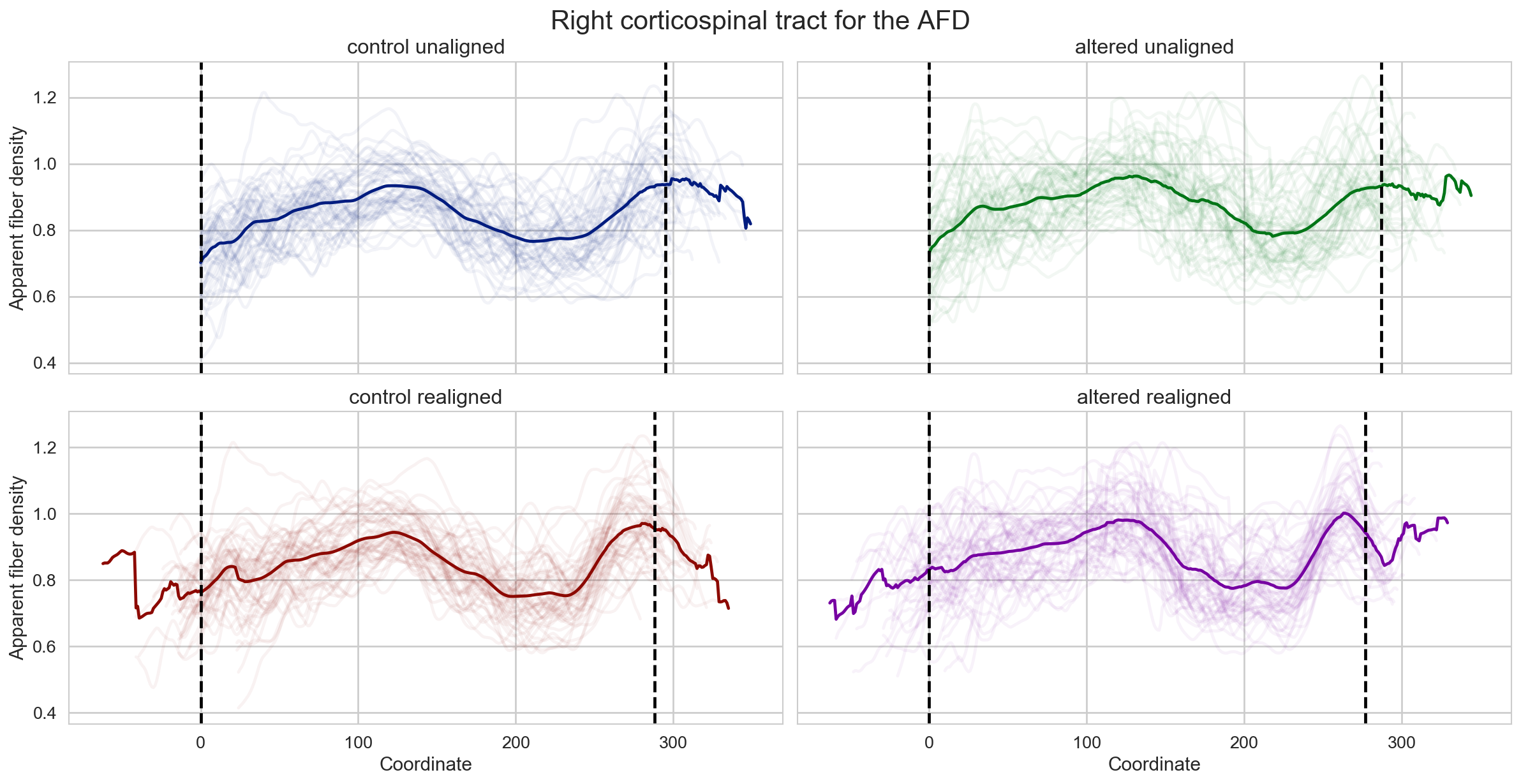}}
        \annotatedFigureBox{0.1,0.94}{D}
        \draw[color=red,thick] (0.33, 0.65) rectangle (0.45, 0.82);
        \draw[color=red,thick] (0.8, 0.65) rectangle (0.92, 0.82);
        \draw[color=red,thick] (0.33, 0.23) rectangle (0.45, 0.4);
        \draw[color=red,thick] (0.8, 0.23) rectangle (0.92, 0.4);
    \end{annotatedFigure}
    \caption{Comparisons between the unaligned and realigned profiles for the HCP datasets without (control column) and with (altered column) simulated diffusion alterations in the white matter fiber bundles.
    A different bundle for a specific metric is shown in each subfigure:
    \textbf{A)} AF left for the MD, \textbf{B)} AF right for the FA, \textbf{C)} CST left for the FA and \textbf{D)} CST right for the AFD.
    \review{The AF are displayed from anterior (coordinate 0) to posterior and the CST from inferior (coordinate 0) to superior.}
    Each subject representative streamline is rendered transparently and the group average representative streamline is represented by the solid line.
    The black bars indicate where at least 75\% of the subjects are overlapping.
    Some key visual differences (red boxes) are hidden by misalignment between the control and altered subject data when they are unaligned,
    while realignment helps to uncover those hidden degeneracies.
    Note that the red boxes in the subgraphs have the same size and are aligned for easier visual comparison.
    The most striking example is in \textbf{A)} where the change in MD is easier to see after realignment
    as the control subjects are keeping their original shape while the altered datasets exhibit a drop in their scalar value around the same region.
    The unaligned group average streamline however makes this difference harder to uncover.
    }
    \label{fig:hcp_disease}
\end{figure}

We first focus on the new strategy of resampling the representative streamlines,
while ensuring that the distance between each point $\delta_{\min}$ is the same.
As one can always resample to a common number of points after realignment,
this prevents a reduced sampling resolution when using \cref{eq:crosscorr}.
Automatically selecting a template from the subjects themselves allows the DPR algorithm to be as flexible as possible.
The changes in scalar metrics (\eg introduced by local alteration of tissue microstructure following disease) might not be obviously identified on the group average for the unaligned streamlines case,
but the variations in shape of the realigned group average may be uncovered by selecting a new template.
\cref{fig:hcp_disease} shows four examples of the unaligned and realigned profiles of the scalar metrics
for the datasets with and without simulated diffusion abnormalities for each white matter fiber bundle.
Note how the original and altered unaligned streamlines have a similar profile for both metrics at first,
but the realigned altered streamlines now have a different profile which was uncovered by realignment with the DPR algorithm (see the red boxes in \cref{fig:hcp_disease}).
This is especially prevalent in the case of the MD metric where the unaligned profiles are similar for the control and altered subject data,
while realignment uncovers the higher MD values that were originally simulated.

\paragraph{Statistical hypothesis testing}

\begin{figure}
    \begin{subfigure}{0.47\linewidth}
        \begin{annotatedFigure}{\includegraphics[width=\linewidth,align=c,valign=t]{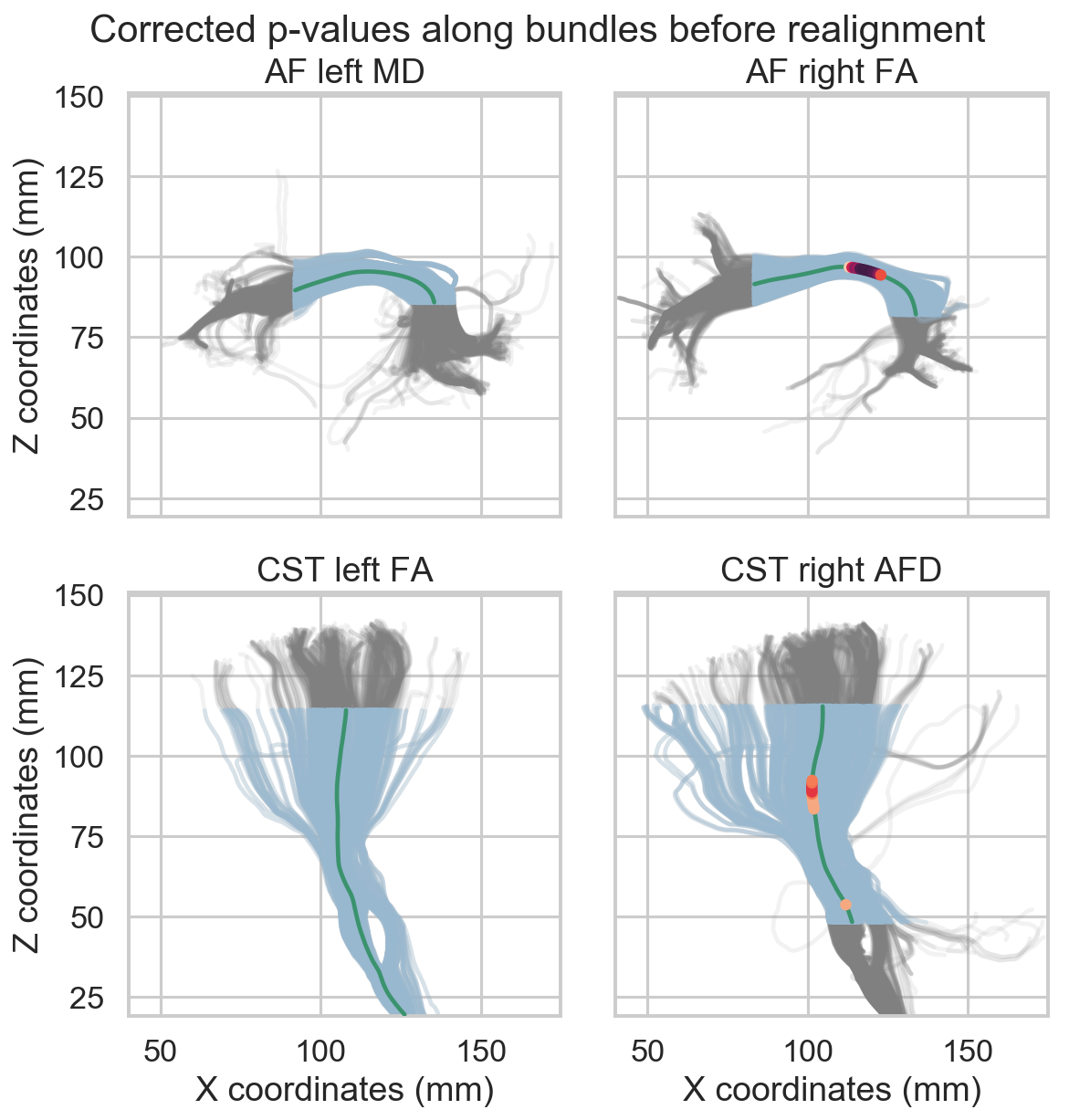}}
            \annotatedFigureBox{0.15,0.9}{A}
        \end{annotatedFigure}
    \end{subfigure}
    \hfill
    \begin{subfigure}{0.04\linewidth}
        \includegraphics[width=\linewidth,align=c,valign=t]{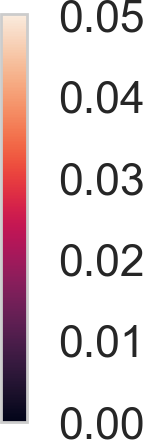}
    \end{subfigure}
    \hfill
    \begin{subfigure}{0.47\linewidth}
        \begin{annotatedFigure}{\includegraphics[width=\linewidth,align=c,valign=t]{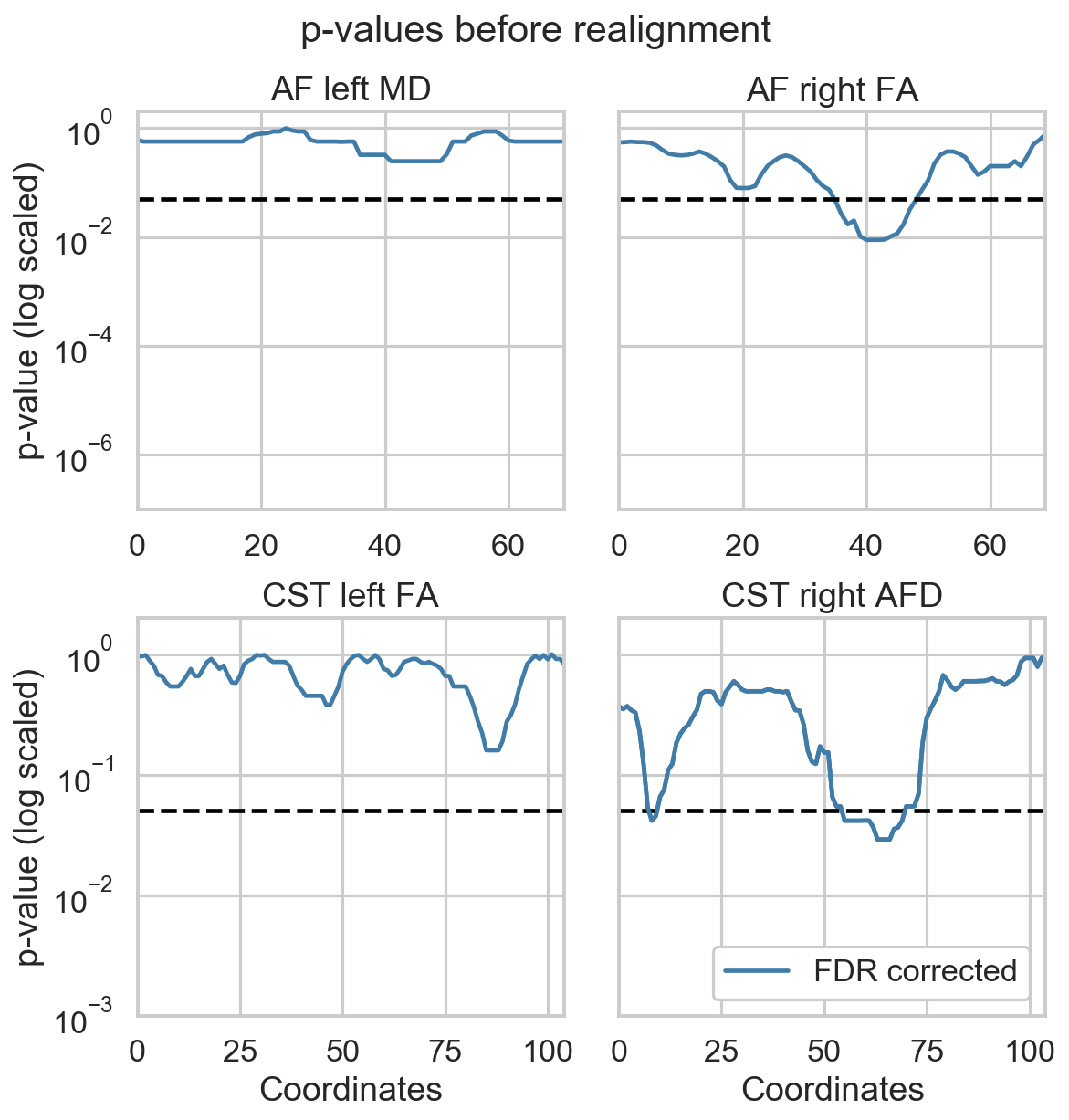}}
            \annotatedFigureBox{0.15,0.9}{B}
        \end{annotatedFigure}
    \end{subfigure}

    \rule{0.75\linewidth}{1pt}

    \begin{subfigure}{0.47\linewidth}
        \begin{annotatedFigure}{\includegraphics[width=\linewidth,align=c,valign=t]{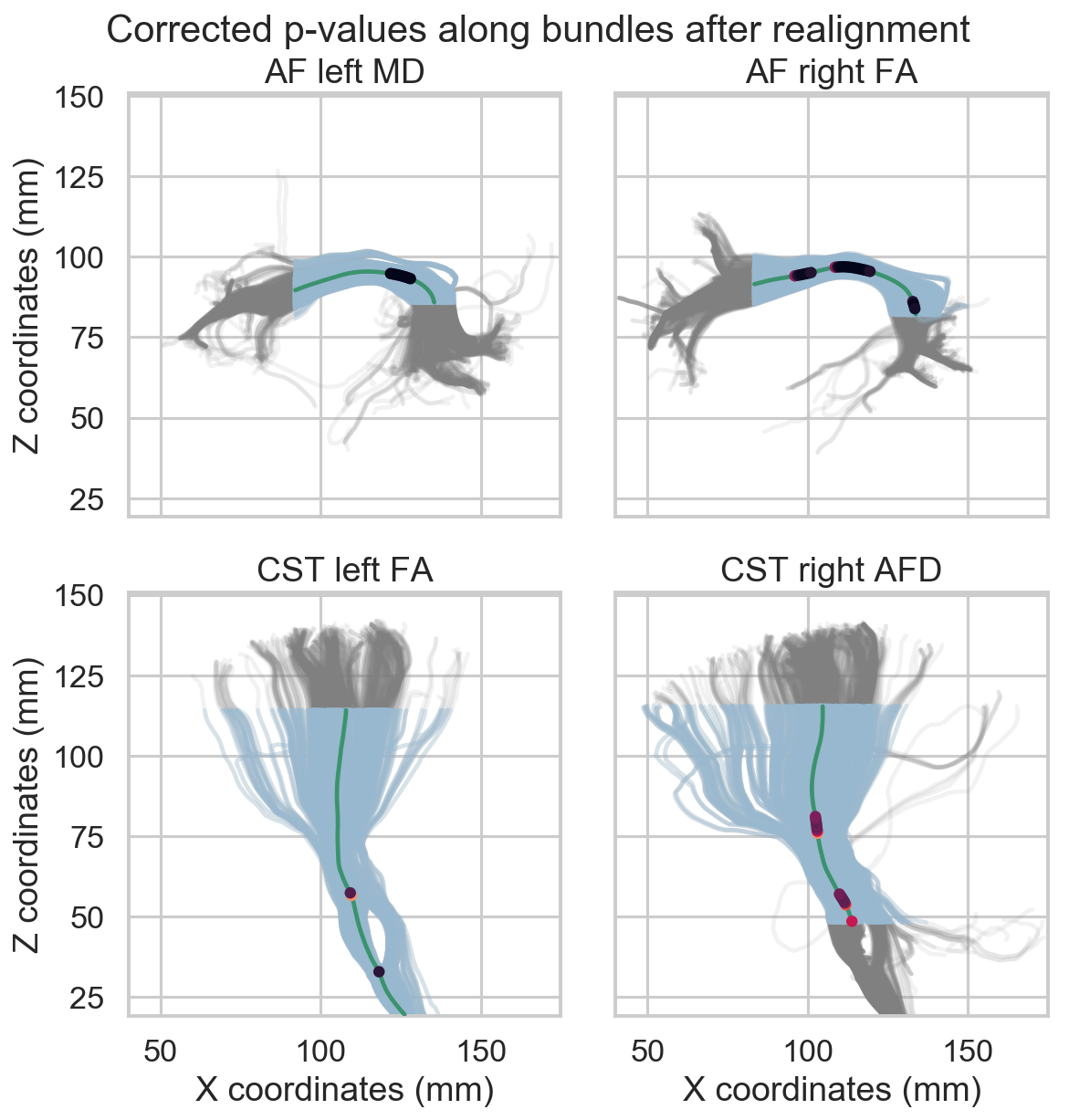}}
            \annotatedFigureBox{0.15,0.9}{C}
        \end{annotatedFigure}
    \end{subfigure}
    \hfill
    \begin{subfigure}{0.04\linewidth}
        \includegraphics[width=\linewidth,align=c,valign=t]{results/hcp_bundle_pval_cbar}
    \end{subfigure}
    \hfill
    \begin{subfigure}{0.47\linewidth}
        \begin{annotatedFigure}{\includegraphics[width=\linewidth,align=c,valign=t]{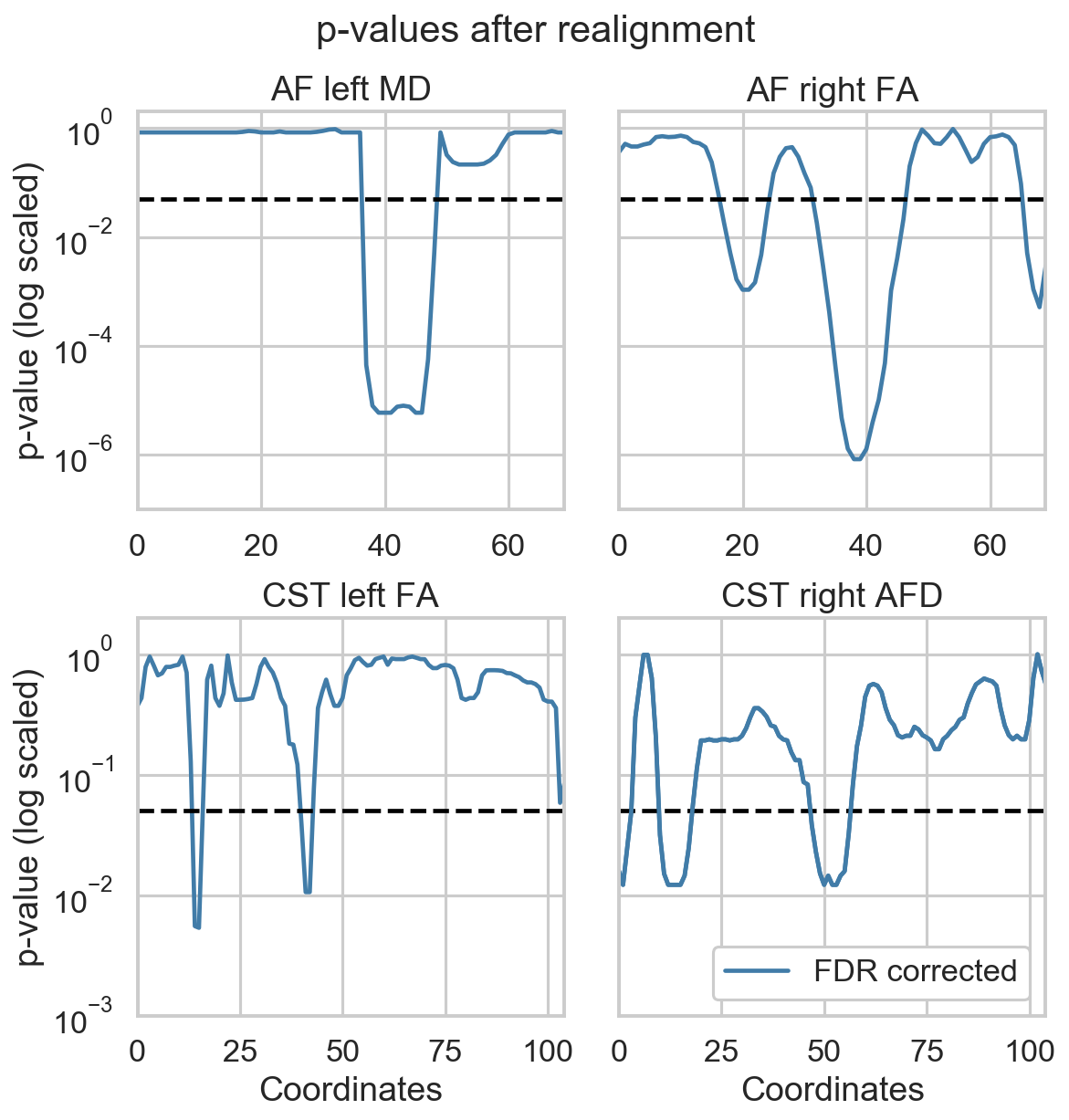}}
            \annotatedFigureBox{0.15,0.9}{D}
        \end{annotatedFigure}
    \end{subfigure}
    \caption{\rereview{Unpaired} t-test corrected for false discovery rate (FDR) at $\alpha = 0.05$ overlaid on the exemplar subject bundle for the same cases as in \cref{fig:hcp_disease}.
    On the left, fiber trajectories of the exemplar subject (in gray) and truncated portions of these pathways between the ROIs (in blue) expressed in world coordinates
    \textbf{A}) before realignment and \textbf{C}) after realignment with the DPR algorithm.
    The p-values at locations deemed statistically significant in the present work $(p < 0.05)$ are overlaid on the average streamline (in green).
    On the right, the p-values on a log scale after FDR correction along the average streamlines \textbf{B}) before realignment
    and \textbf{D}) after realignment with the DPR algorithm, but expressed as along-tract 1D point coordinates.
    The horizontal black bar is located at p-value = 0.05.
    In the realigned data case, the p-values are lower in the significant regions (corticospinal tract right) or even show affected regions which are not detected
    when the data is unaligned (arcuate fasciculi \review{and corticospinal tract left}).
    The most prominent case is for the left arcuate fasciculus, where the affected portion is not identified in the unaligned case (for our chosen significance threshold of 0.05),
    but has a corrected p-value of approximately $10^{-5}$ after realignment.
    }
    \label{fig:hcp_ttest}
\end{figure}

\begin{figure}
    \figuretitle{Focused alterations before and after realignment}
    \begin{annotatedFigure}{\includegraphics[width=0.495\linewidth]{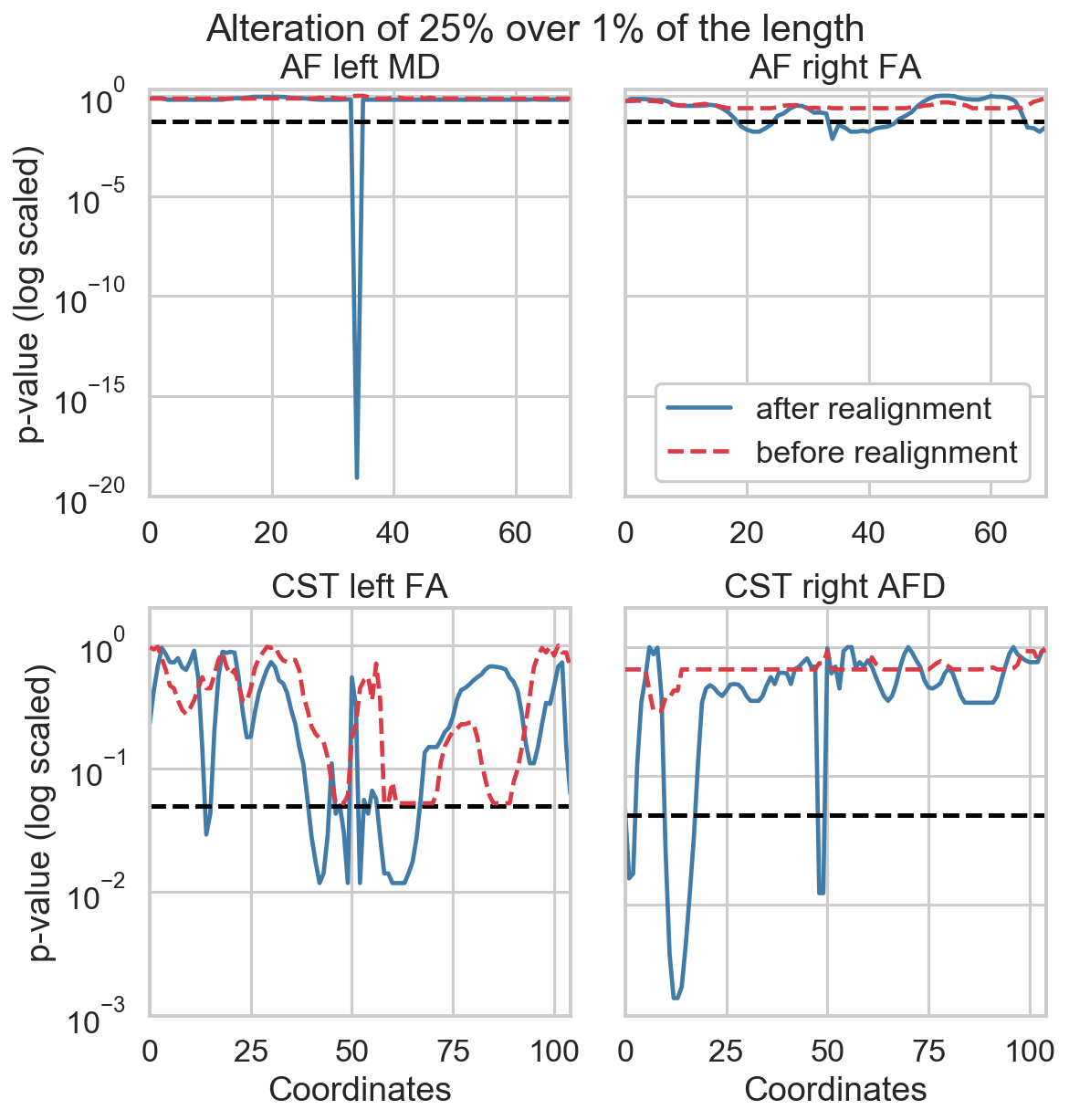}}
        \annotatedFigureBox{0.1,0.97}{A}
    \end{annotatedFigure}
    \hfill
    \begin{annotatedFigure}{\includegraphics[width=0.495\linewidth]{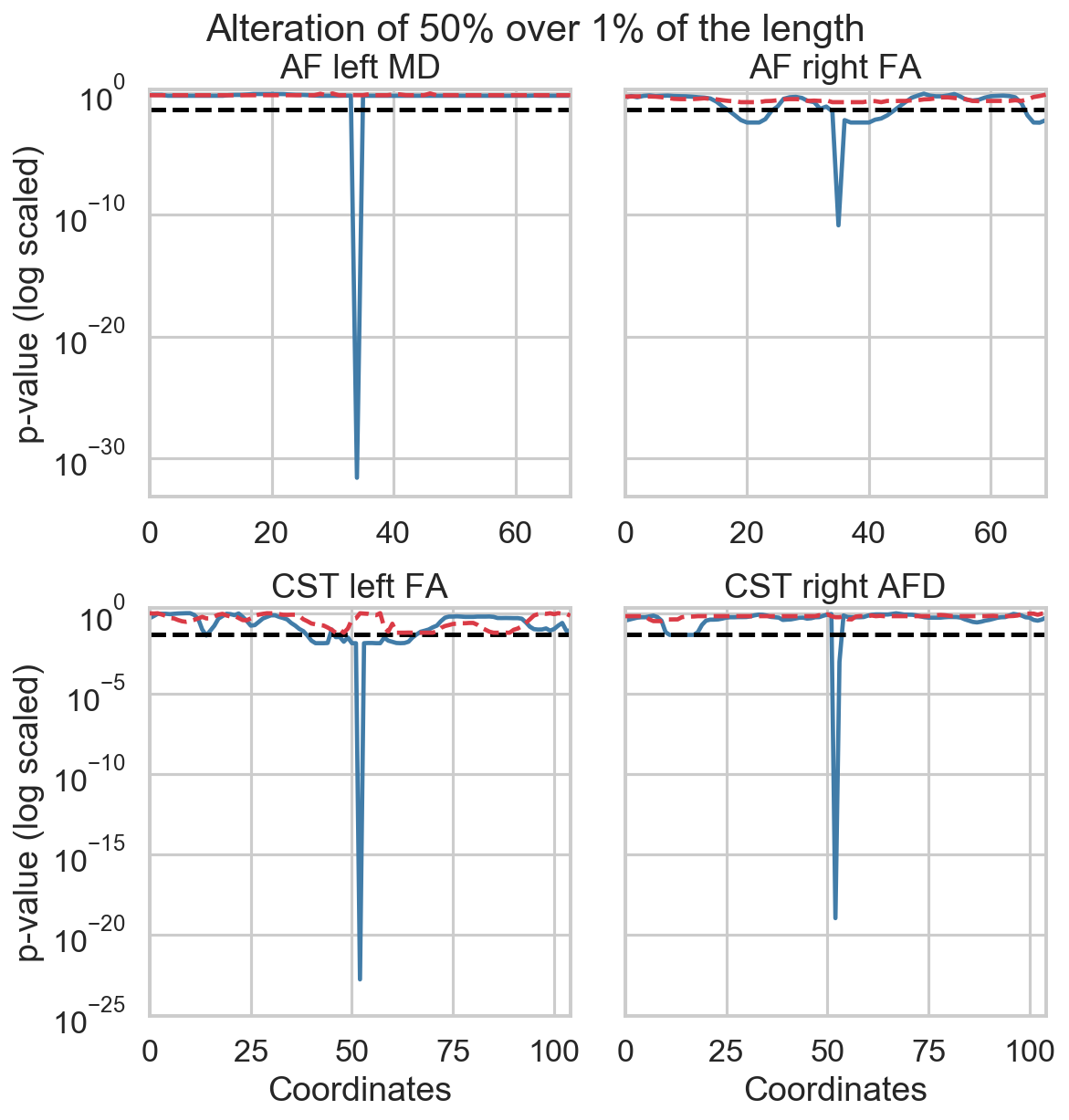}}
        \annotatedFigureBox{0.1,0.97}{B}
    \end{annotatedFigure}
    \begin{annotatedFigure}{\includegraphics[width=0.495\linewidth]{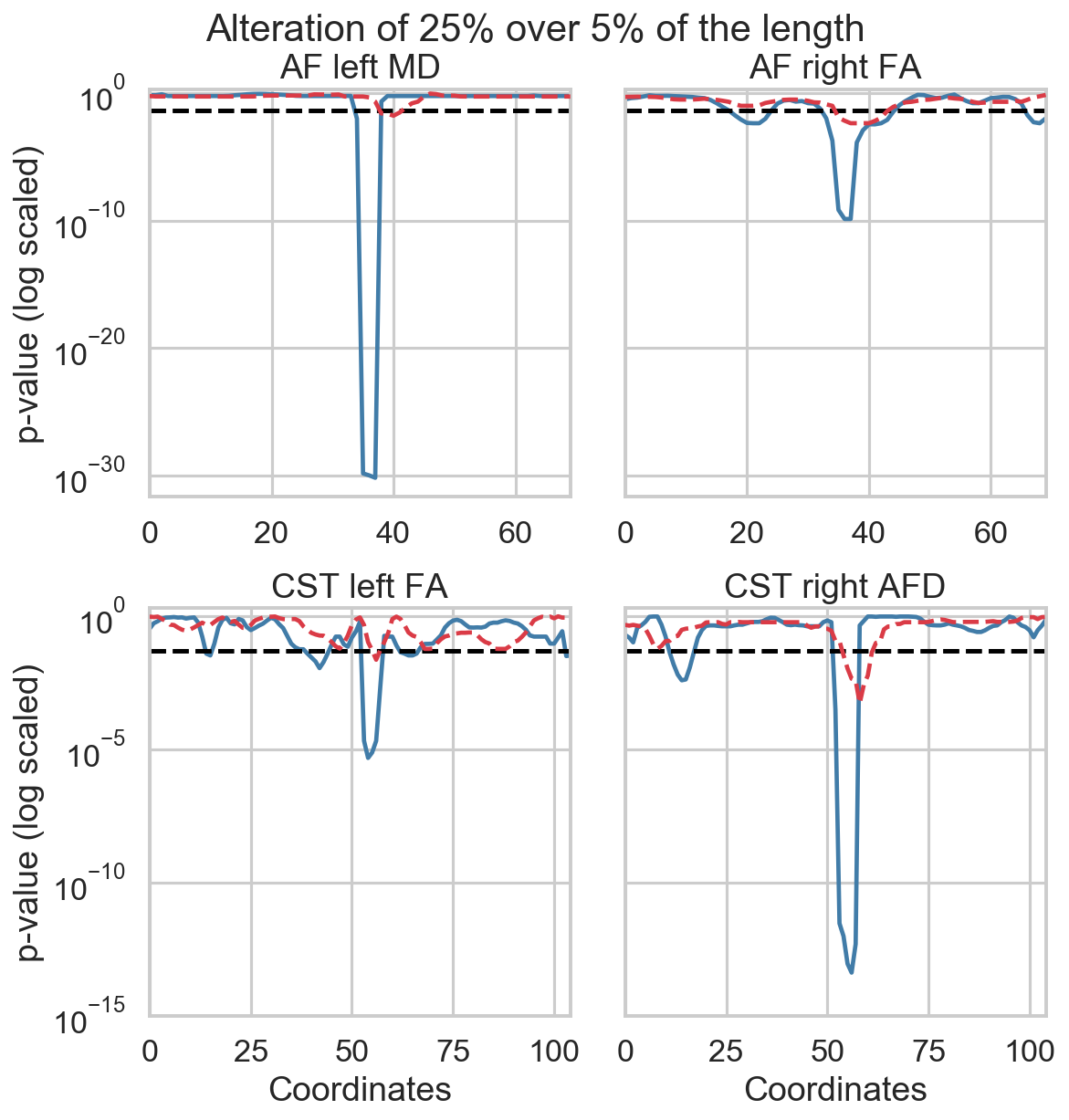}}
        \annotatedFigureBox{0.1,0.97}{C}
    \end{annotatedFigure}
    \hfill
    \begin{annotatedFigure}{\includegraphics[width=0.495\linewidth]{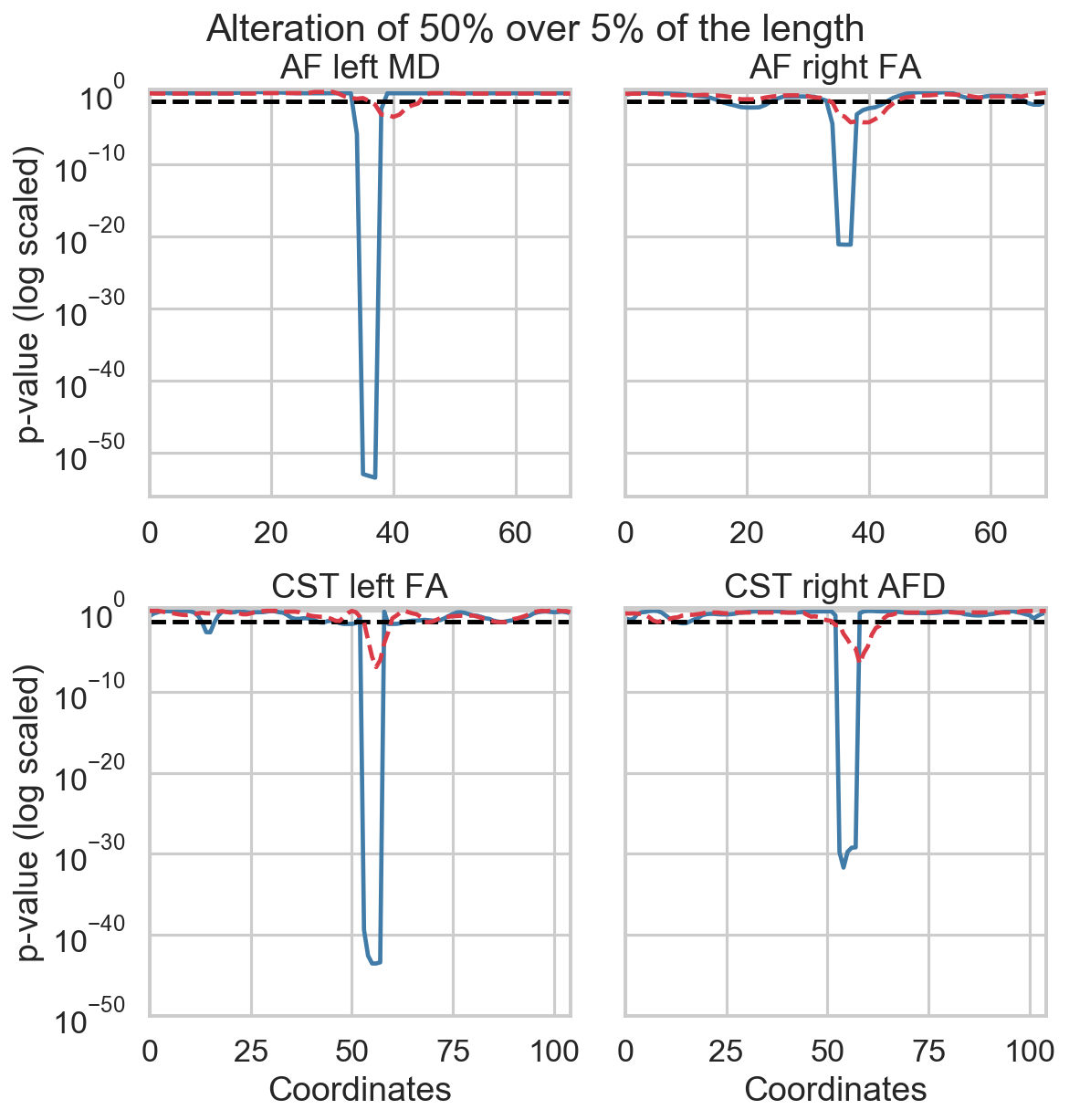}}
        \annotatedFigureBox{0.1,0.97}{D}
    \end{annotatedFigure}
    \caption{\review{\rereview{Unpaired} t-test (FDR corrected at $\alpha = 0.05$) with focused alterations of the metrics for each bundle of
    \textbf{A)} 25\% over 1\% of the length, \textbf{B)} 50\% over 1\% of the length, \textbf{C)} 25\% over 5\% of the length and \textbf{D)} 50\% over 5\% of the length.
    The AF left/right are represented from anterior (coordinate 0) to posterior and the CST left/right from inferior (coordinate 0) to superior.
    The p-values are on a log scale along the average streamline before realignment (dashed red lines) and after realignment (solid blue lines) with the DPR algorithm.
    The horizontal dashed black lines indicate p-value = 0.05.
    When alterations cover 1\% of the length, the affected profiles are identified only after realignment.
    At 5\% of the length, the uncovered regions after realignment are concentrated around smaller sections than their counterpart before realignment.}}
    \label{fig:hcp_focused}
\end{figure}

We now look at uncovering groupwise differences between the control and altered HCP subjects over the affected regions.
\cref{fig:hcp_ttest} shows the results of the \rereview{unpaired} t-test for the HCP datasets before and after realignment
for the \textbf{A)} AF left with the MD metric, \textbf{B)} AF right with the FA metric, \textbf{C)} CST left with the FA metric
and \textbf{D)} CST right with the AFD metric, as previously shown in \cref{fig:hcp_disease}.
All of the regions uncovered before using realignment are also identified as statistically significant at the level of \review{p-value <} 0.05 after realignment.
This indicates that findings for the unaligned case are preserved when using our proposed algorithm,
with the addition of new affected regions which might have been averaged out due to misalignment in the first place.
For example, the left AF \review{and left CST} \sout{, for the particular case of the MD metric,} showcase\sout{s} an affected portion which is statistically significant only after realignment.
\sout{No affected region is present for the left CST when studying the FA metric for both the unaligned and realigned case.}
However, using a lower statistical threshold or a higher level $\alpha$ for the FDR might reveal more affected regions at the cost of introducing potential false positives.
\review{\cref{fig:hcp_focused} shows a second set of experiments on the four bundles realized with large alterations of the metrics which are spatially focused \eg in the presence of tumors.
Specifically, alterations in the metrics of 25\% or 50\% were induced over 1\% or 5\% of each bundle length and each group subsequently realigned with DPR.
\rereview{Unpaired} t-test before and after realignment are conducted between the two groups at each location as in \cref{fig:hcp_ttest}.
Almost all affected regions are identified before and after realignment when the affected length is of 5\%.
For the CST left, the affected region is only identified after realignment when the alteration is of 25\%.
When only 1\% of the bundle length is affected, no changes are identified before realignment,
but are \rereview{uncovered after realignment with the DPR algorithm in all cases.}
Results obtained with maximum displacement thresholds of 5\%, 10\% and 100\% are shown in the supplementary materials \cref{sec:supp_focused}.}

%% file: tex/discussion.tex
\section{Discussion}
\label{sec:discuss}
\subsection{Reducing variability along bundles}

Using simulations, we have shown how residual misalignment may hide the expected average profile of an along-tract analysis.
\cref{fig:phantomas_realign} shows this effect directly as the group mean profile from a set of streamlines only roughly corresponds to their individual, but in truth identical, shape
as their spatial location differs due to small differences in their length.
Realignment not only restores the expected group profile, but also reduces the pointwise variability of the metrics as the unequal streamlines are now aligned as reflected by the lower overall CV.
Each individual subject is therefore participating to the group average instead of being spread out and biasing the estimated mean scalar value of the overall bundle in the crossing region.
This is also true if the streamlines are first resampled to the same number of points.
In this case the variance at the end points is larger, possibly due to a loss in spectral resolution caused by resampling to a lower number of points than originally present.
Resampling early in the along-tract analysis pipeline may not only inadvertently hide information for the realignment,
but also hamper statistical testing by reducing the spatial specificity of the data \citep{ODonnell2009a}.

For the realignment of the in vivo HCP datasets, \cref{fig:hcp_realignment} shows that realignment alters the group profile
at the end points while preserving the overall shape and the central portion of the bundle.
This leads to a reduction of the CV, likely due to the reduction in variance at the end points while the overall mean profile is preserved as shown in \cref{fig:hcp_coeff_variation}.
As the realigned end points will also have less data from different subjects present at each coordinate, subsequent truncation further reduces the CV once again.
The change of shape after realignment is possibly due to the difference in length between each subject and the subsequent mapping to their 1D metric profile.
This 1D space hides the spatial 3D coordinates misalignment which may be present between subjects. %
However, this misalignment can still be mitigated afterwards.
Even if the representatives streamlines are shifted as a whole with the realignment, preservation of the overall shape and center portion
might indicate that only the end points were dissimilar.
The lower end point variance effect is also present when using the classical resampling strategy and subsequently realigning the representative streamlines.
The misalignment at the end points between subjects is due in part to the truncation effect of the ROIs
and to the nature of tractography itself and its many user defined settings \citep{Chamberland2014}.
The use of termination criteria (\eg FA threshold, white matter mask, maximum curvature) or seeding strategy (\eg white matter versus cortex seeding) \citep{Girard2014}
may prematurely terminate tractography in the middle of a white matter bundle,
contributing in producing shorter streamlines which end before fully reaching the gray matter \citep{Maier-Hein2017}.
New algorithms and seeding strategies are developed to enhance tractography end points near the cortex \citep{St-Onge2018a} and could help to reduce this truncation effect.

\subsection{Effect of exchanging metrics for realignment}

We have shown in \cref{fig:hcp_displacement} the effect of applying the realignment computed from different metrics on the mean group tract profile.
From these results, we can observe the different displacement values obtained
from the dMRI metrics, even though the representative streamlines arise from the same anatomical location.
This is due to the fact that our framework is fully driven by the 1D profiles of the studied metric which all have different shapes and features,
leading to slightly different realignment outcome depending on the bundle and the metric that is used.
As the FA and AFD profiles are similar in the four studied bundles, exchanging their value still leads to the same overall profile in most cases.
For the MD, results showed that the CST is also stable.
This is most likely due to the complex 1D profile of the CST for the three metrics, as it defines unique landmarks that are picked by our algorithm for accurate realignment.
Regarding the AF, exchanging the displacement from the FA or AFD yields similar profiles, an observation which does not hold for the MD metric.
As the MD metric for the AF has a rather flat profile, the algorithm might pick up a spurious displacement due to the lack of well defined features to exploit.
\citet{Avants2011} also reached a similar conclusion in the context of 3D volume registration when using different metrics
such as the mean square difference, cross correlation or mutual information; using different metrics, type of registration or registering subject A onto subject B
(and vice versa) leads to slightly different outcomes.
We have fixed the maximally allowed displacement to 15\% of the length of the bundle, but similar conclusions also applied for 10\% and 20\%
of maximum displacement as shown in \cref{sec:supp_displacement}.
When the maximum displacement is only 5\%, the AF show similar mean profiles
for the three metrics, whereas the opposite is seen in the CST.
This indicates that the maximally allowed displacement should be chosen per bundle and is data dependent.
Short, straight and simpler bundles, such as the AF, might only need small realignment, whereas more complex structures
with fanning, intersecting bundles and possibly large anatomical variations between subjects, such as the CST,
likely benefit from larger maximum displacement thresholds to find their full overlap between subjects.

\subsection{Identifying brain regions affected by abnormalities along-tract}

One of the end goal of along-tract analysis is to uncover alteration of the white matter due to, \eg disease at their specific locations.
This is at the cost of trading the sensitivity of ROI averaging based analysis for additional specificity along the bundle,
which also depends on the discretization of the points forming the streamlines \citep{ODonnell2009a}.
Using simulated changes in scalar metrics from the HCP subjects, we have shown in \cref{fig:hcp_disease} how misalignment can artificially reduce
the specificity of along-tract analysis.
As the affected portion of the bundle is usually unknown \textit{a priori}, morphological differences between subjects
might map the affected area to different points in their 1D profile during the representative streamlines extraction.
The unaligned metrics might exhibit similar mean profiles between the control and altered subjects in this case, as the affected portions
would be originating from an adjacent anatomical location in each subject's original space, but would not be aligned in the 1D space.
The mean representative streamline at the group level could therefore average out each subject's individual difference due to residual misalignment,
hiding the effect of interest in the process.
As we have previously mentioned in \cref{sec:introduction}, this effect of averaging out important information has also been theorized by \citet{ODonnell2009a}.
However, the same effects can also become easier to detect after realignment since the control subjects mean profile will potentially be different
from the altered subjects mean profile.
This is thanks to the particular features of their 1D profile now being realigned instead of averaged out.
\review{In a similar way, if changes in the diffusion metrics are potentially present across the whole length of a white matter bundle,
the maximum displacement threshold should be increased.
This may reduce the number of subjects identified as outliers by using a smaller maximum displacement, which
would not have been realigned in the first place.
The tradeoff in allowing a larger maximum displacement is a potential reduction in statistical power or false discoveries
as less subjects may be present at each along-tract location for statistical testing.}

In our simulations, changes on the left AF \review{and left CST} \sout{with the MD metric} are identifiable only after realignment whereas
the original control and altered average profiles are mostly similar since each individual contribution is lost in the unaligned group averaging.
After realignment, the altered region can be identified as each realigned subject now contributes to the group average at the same location.
This effect is similar to what we observed in our simulations in \cref{sec:simulations},
where the CV is lower in the crossing-bundles region after realignment and how the mean group profile is also lower after realignment.
It is also noticeable on the right AF bundle with the FA metric or on the CST bundles,
but to a lesser extent, as the overall morphology of the CST bundles stays relatively similar even after altering the scalar metrics.
Interestingly, the altered group profile seems to be subject to larger morphometric changes after realignment than the control group counterpart.
This might indicate that sharp profile changes in each subject's shape due to disease are automatically picked up by our algorithm, providing realignment based on this change.

We also conducted \rereview{unpaired} Student's t-tests to statistically identify the altered regions on the same bundles and metrics as shown in \cref{fig:hcp_ttest}.
While we used an FDR correction of $\alpha = 0.05$, different results could be obtained by choosing a different value of $\alpha$.
However, the main conclusion should still be valid; statistical testing performed on the realigned datasets uncovered affected regions
which were not identified in the unaligned case as shown from the global p-values plot.
This difference could be partly due to the residual misalignment between subjects inadvertently canceling out the effect of interest as coordinates are not overlapping.
In this study, we considered statistical testing at the spatial resolution in the order of magnitude of one voxel size (1.25 mm in our case),
but studying larger bundle segments could be used as a compromise between averaging data over the whole bundle
in order to uncover effects of interest at the expense of spatial specificity \citep{ODonnell2009a}.

\subsection{Mapping to 1D space versus registration methods}

In the present work, we concentrated on reducing the effect of residual misalignment between representative streamlines.
As tractography is a mandatory step before using our approach, registration methods for raw dMRI datasets would likely not reduce
the misalignment resulting from streamlines extraction.
Some registration methods specifically work directly on the streamlines or bundles space (\eg \citet{Leemans2006}),
but the same transformation should be applied to the underlying 3D volume containing the metric of interest.
This is because we work on metrics extracted from representative streamlines, and not directly in the streamlines space itself,
see \eg \citet{ODonnell2017a,Glozman2018} and references therein for a review of registration methods in dMRI.

\citet{ODonnell2009a} state that \enquote{because within a bundle fibers have varying lengths and their point correspondence is not known a priori,
it is not possible to directly average fiber coordinates to calculate a mean fiber};
care must be taken during the representative streamline extraction step that is at the core of the along-tract analysis framework.
As such, the required step dictating this possible misalignment is the mapping strategy
used to extract the representative streamline and how its end points are defined.
Various schemes have been proposed such as assignment to perpendicular planes \citep{Corouge2006},
variants reducing the effect of outliers by additionally considering the spatial distance between streamlines \citep{ODonnell2009a},
extracting representative core streamlines with splines \citep{Chamberland2018}
or resampling to a common number of points \citep{Colby2012}.
All these choices inevitably lead to differences and a mismatch across subjects after metric extraction,
even if the original underlying anatomy would be perfectly aligned as we have shown in our synthetic experiments in \cref{fig:phantomas_realign}.
Assignment and truncation strategies between the common points of bundles have been explored in \citet{Colby2012}
with the authors noting that all compared methods are generally successful in extracting a meaningful
(but slightly different) representation as they use different strategies and parameters.
Close similarities in the extracted metrics using the representative streamline could explain why 1D misalignment,
while still present, had not been thoroughly investigated previously.
Reliably extracting the information from fanning regions (\eg CST towards the motor cortex) or from a splitting configuration (\eg anterior pillars of the fornix)
in a single representative streamline still remains an open problem \citep{Chamberland2018}.

\subsection{Assumptions of the DPR algorithm and limitations of this study}

In the present work, we exchanged the classical assumption of 1D \emph{spatial correspondence} between points for the assumption of an equal 1D \emph{spatial distance} between points.
This latter requirement is usually fulfilled with the use of a fixed step size during tractography, but might be void by the representative streamline extraction.
Without loss of generality, we chose to resample each subjects' representative streamline a second time to ensure an equal distance $\delta_{\min}$ between each point.
We advocate resampling to a larger number of points than initially present to reduce possible complications due to aliasing or using windowing functions for filtering \citep{Stoica2004}.
While this theoretically increases the computational complexity of the DPR algorithm, it also preserves the full spectra when applying \cref{eq:crosscorr}.
This is not a problem in practice owing to the existence of efficient FFT implementations;
our algorithm can realign the 100 HCP subjects in less than 3 seconds on a standard desktop with a 3.5 GHz Intel processor.
The resulting realigned metrics can then be resampled back to approximately one point per unit voxel size to minimize the effect of multiple comparisons during statistical testing.
With the development of new methods that go beyond fixed step size tractography, such as the use of compressed streamlines \citep{Rheault2017a},
it might be beneficial to avoid this resampling step for computational reasons after sampling metrics along non regularly spaced streamlines.
Another approach to remove the need of resampling could be to use an FFT implementation
dealing with non equal sampling of \review{the data \citep{Dutt1993,Scargle1989}}, %
but such implementations may not be as widely (and easily) available as the classical equispaced version of the FFT algorithm.

Due to the difficulty in reproducing tractography \citep{Maier-Hein2017},
our simulations on the in vivo datasets were designed around altered versions of already extracted scalar values.
One would however expect true neurodegenerative changes to additionally influence the steps prior to tractography
such as the main orientations extracted from tensors or fODF.
The results we obtained should translate as long as a representative streamline for each bundle of interest can be reliably delineated for all subjects.
Similar recommendations apply if the white matter bundle of interest is largely affected by disease or
altered when compared to the expected overall shape from a healthy subject.
Specific care should also be taken during the prior step of extracting the representative streamlines in these cases
to ensure that relevant portions of the bundles of interest are present in all subjects \citep{Parker2016}.

Although not considered in the present work, any quantitative diffusion metric such as
the diffusion kurtosis metrics (\eg mean kurtosis (MK)) \citep{Jensen2010},
the axon diameter \citep{Assaf2008},
or metrics provided by NODDI \citep{Zhang2012d}
could be studied using our proposed framework.
\review{In cases of physical alterations of the white matter (\eg tumors, lesions), the diffusion metrics themselves may not provide accurate landmarks for realignment
due to differences in tractography when extracting the representative streamline of each subject.
The use of shape descriptors, such as torsion or curvature of the bundles themselves \citep{Leemans2006},
could also be employed with DPR instead of diffusion metrics as done in the present work.
These descriptors may also be useful in cases where using a large maximum displacement threshold may yield false positives detections if the effects are small,
see the supplementary materials \cref{sec:supp_focused} for examples.}
In a similar fashion, any other volume (\eg T1 or T2 relaxometry values \citep{Deoni2008a}) providing anatomical information of interest
can be used once co-registered to each subject's native diffusion space.
\rereview{Combining the realignment information from multiple or complementary metrics (\eg computing their average displacement) may improve the robustness of the DPR framework.
When white matter alterations are affecting the diffusion metrics to an unacceptable extent, the average displacement
from these independent anatomical features (which are presumably less affected by these effects) could be used to circumvent this issue.}

We did not investigate realignment of lateralized bundles
(\eg realignment of the left and right AF together instead of separately)
which can be useful for studying intra-hemispheric differences between subjects \citep{Catani2007}.
Variations between left and right anatomical locations also implicitly assumes that each coordinate in the 1D space is matched against its
inter hemispheric counterpart.
To facilitate this mapping between hemispheres, \citet{ODonnell2009a} proposed to mirror all streamlines
from one hemisphere to the other, allowing a direct correspondence between the subsequently extracted representative streamlines as they would be effectively identical.
However, the 3D volume used to extract the scalar metrics of interest would possibly be different in each hemisphere.
In this context, the realignment could be done separately for each side, providing different profiles reflecting lateralization.

%% file: tex/conclusion.tex
\section{Conclusion}
\label{sec:conclusion}

In this paper, we developed a new correction strategy, the diffusion profile realignment (DPR),
which is designed to address residual misalignments between subjects in along-tract analysis.
Through simulations on synthetic and in vivo datasets, we have shown how realignment based on our novel approach can
reduce variability at the group level between subjects.
Furthermore, realignment of the in vivo datasets provided new insights and improved sensitivity
about the location of the induced changes, which could not be completely identified at first when misalignment was present.
The DPR algorithm can be integrated in preexisting along-tract analysis pipelines
as it comes just before conducting statistical analysis.
It can be used to reveal effects of interest which may be hidden by misalignment
and has the potential to improve the specificity in longitudinal population studies beyond the traditional ROI based analysis
and along-tract analysis workflows.

%% file: tex/appendix.tex
\appendix
\gdef\thesection{\Alph{section}} %
\makeatletter
\renewcommand\@seccntformat[1]{Appendix \csname the#1\endcsname.\hspace{0.5em}}
\makeatother

\section{The diffusion profile realignment algorithm}
\label{sec:appendix}

This appendix outlines the diffusion profile realignment (DPR) algorithm.
Our implementation is also freely available at \url{https://github.com/samuelstjean/dpr} \citep[{[Code]}][]{st-jean2019c}
and will be a part of \textit{ExploreDTI} \citep{Leemans2009a}.
The synthetic datasets and metrics extracted along the representative streamlines of the HCP datasets
used in this manuscript are also available \citep[{[Dataset]}][]{St-jean2018b}.

\review{To complement \cref{eq:crosscorr},
the shift needed to maximize the overlap between the vector $x$ and $y$ is the maximum of the CCF, given by
\begin{equation}
    \text{shift}(x, y) = \argmax(\text{CCF}(x, y)).
    \label{eq:displacement}
\end{equation}
In practice, $x$ and $y$ are discrete and must be both zero-padded sufficiently, that is, zeros are appended to each vector and make them artificially longer
to prevent border effects when computing the linear cross-correlation \citep{Stoica2004}.}

\begin{algorithm}
    \SetAlgoLined
    \KwData{Metrics extracted from streamlines discretized (with an equal distance $\delta_{\min}$ and stationary metrics),
        displacement threshold $\mathbf{t}$, percentage of overlap $\mathbf{p}\%$}
    \KwResult{Realigned metrics}
    \vspace{0.2cm}
    \textbf{Step 1} : \emph{Finding a common template}\;
        \ForEach{streamline}{
            Compute the displacement $\mathbf{d}$ with each other streamline using \cref{eq:displacement}\;
        }
        \vspace{0.2cm}
        Define the template as the subject which realigns the most streamlines below the threshold $\mathbf{t}$\;
        \vspace{0.2cm}
        \ForEach{streamline}{
        \eIf{$|\mathbf{d}| \le \mathbf{t}$}{
            Realign the streamline unto the candidate template by its displacement $\mathbf{d}$\;
        }{
            Do not touch the streamline and flag it as an outlier\;}
        }
    \vspace{0.2cm}
    \textbf{Step 2} : \emph{Realigning outliers}\;
    \ForEach{outlier}{
        Compute the new displacement $\mathbf{nd}$ between the template, the outlier and each other non outlier\;
        \eIf{$\min (|\mathbf{d + nd}|) < \mathbf{t}$}{
            Realign the streamline unto the template using the new displacement $\mathbf{d + nd}$ (see \cref{fig:spectrum_of_retemplater})\;
            Add the streamline to the pool of non outliers candidates such that it can now be used\;
        }{
            Do not touch the streamline and flag it as an outlier\;
        }
    }
    \vspace{0.2cm}
    \textbf{Step 3} : \emph{Truncating to overlapping coordinates}\;
    Truncate the realigned metrics to have at least $\mathbf{p}\%$ of overlapping streamlines\;
    If outliers are still present from \textbf{Step 2},
    (optionally) exclude them from further analysis as they can\\ not be realigned inside the chosen displacement threshold $\mathbf{t}$\;
    \caption{The proposed diffusion profile realignment (DPR) algorithm.}
    \label{alg:main}
\end{algorithm}

%% file: tex/supplementary.tex
\clearpage

\section{Supplementary materials}
\label{sec:supp}
\subsection{Realignment of the HCP datasets}
\label{sec:supp_realignment}

\Cref{sec:supp_realignment} presents counterpart results to \cref{fig:hcp_realignment}, comparing along-tract averaged profiles
before and after realignment, but instead using a maximally allowed displacement of 5\%, 10\% or 20\%.
\review{Coordinates for the AF are from anterior (coordinate 0) to posterior and the CST are drawn from inferior (coordinate 0) to superior.}
In general, the overall mean profile is similar for every value of the maximally allowed displacement that were tested.

\begin{figure}
    \includegraphics[width=\linewidth]{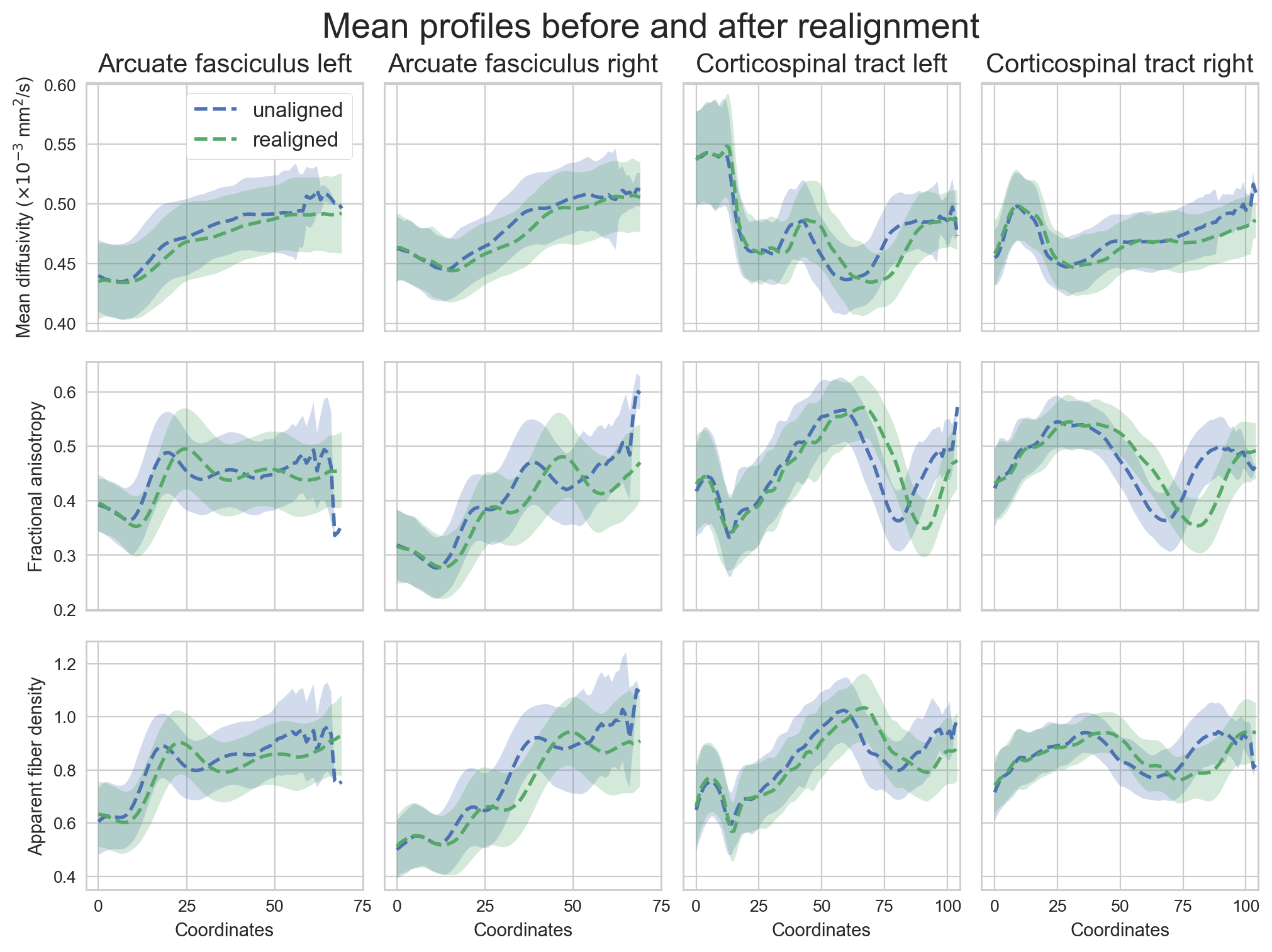}
    \caption{Along-tract averaged profiles (and standard deviation as the shaded area) of the unaligned (blue) and
    realigned (green) HCP subjects truncated to 75\% of overlap with a final resampling to the same number of points.
    These results are obtained by using a maximally allowed displacement of 5\%.}
    \label{fig:hcp_realignment5}
\end{figure}

\begin{figure}
    \includegraphics[width=\linewidth]{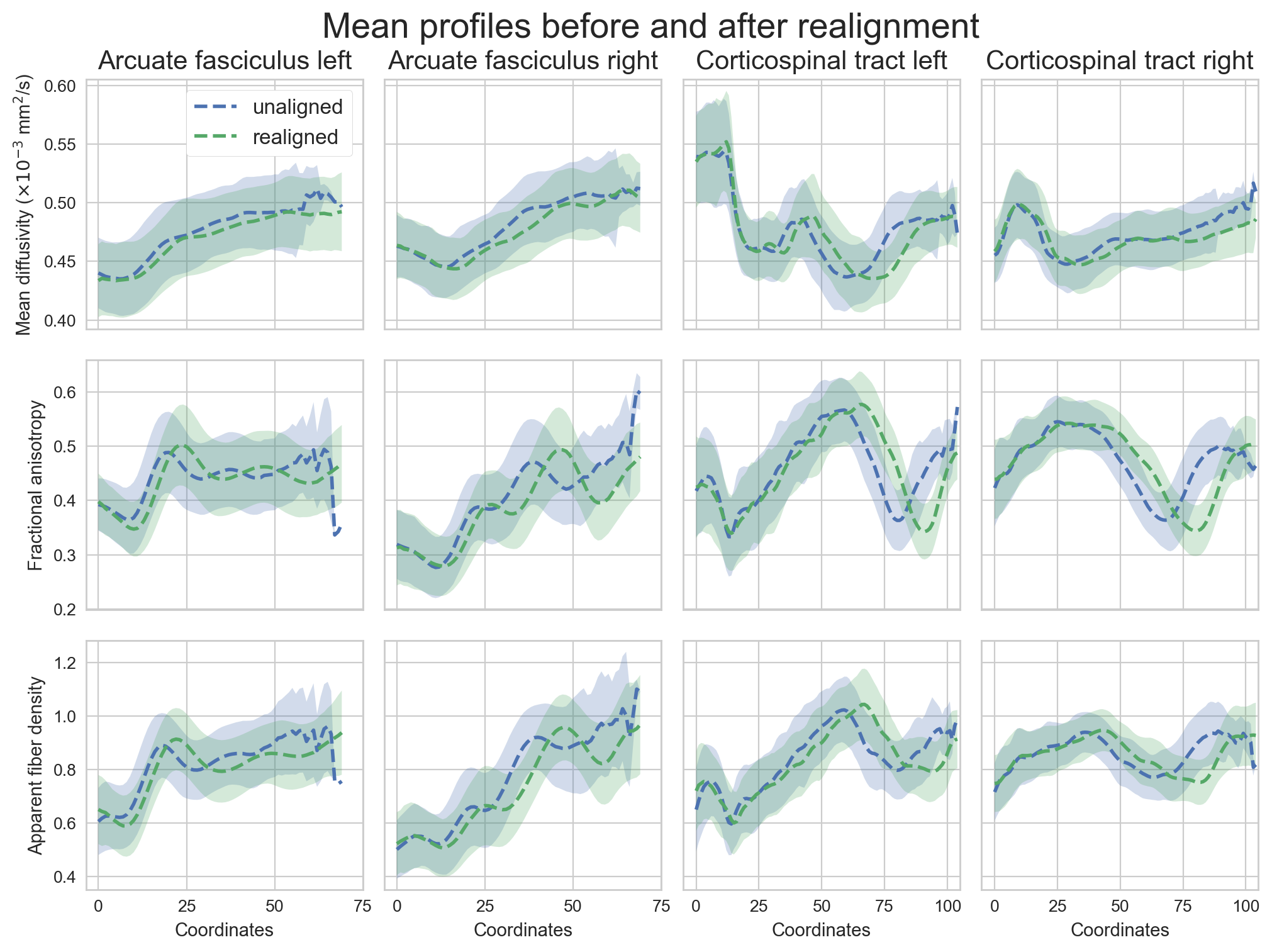}
    \caption{Along-tract averaged profiles (and standard deviation as the shaded area) of the unaligned (blue) and
    realigned (green) HCP subjects truncated to 75\% of overlap with a final resampling to the same number of points.
    These results are obtained by using a maximally allowed displacement of 10\%.}
    \label{fig:hcp_realignment10}
\end{figure}

\begin{figure}
    \includegraphics[width=\linewidth]{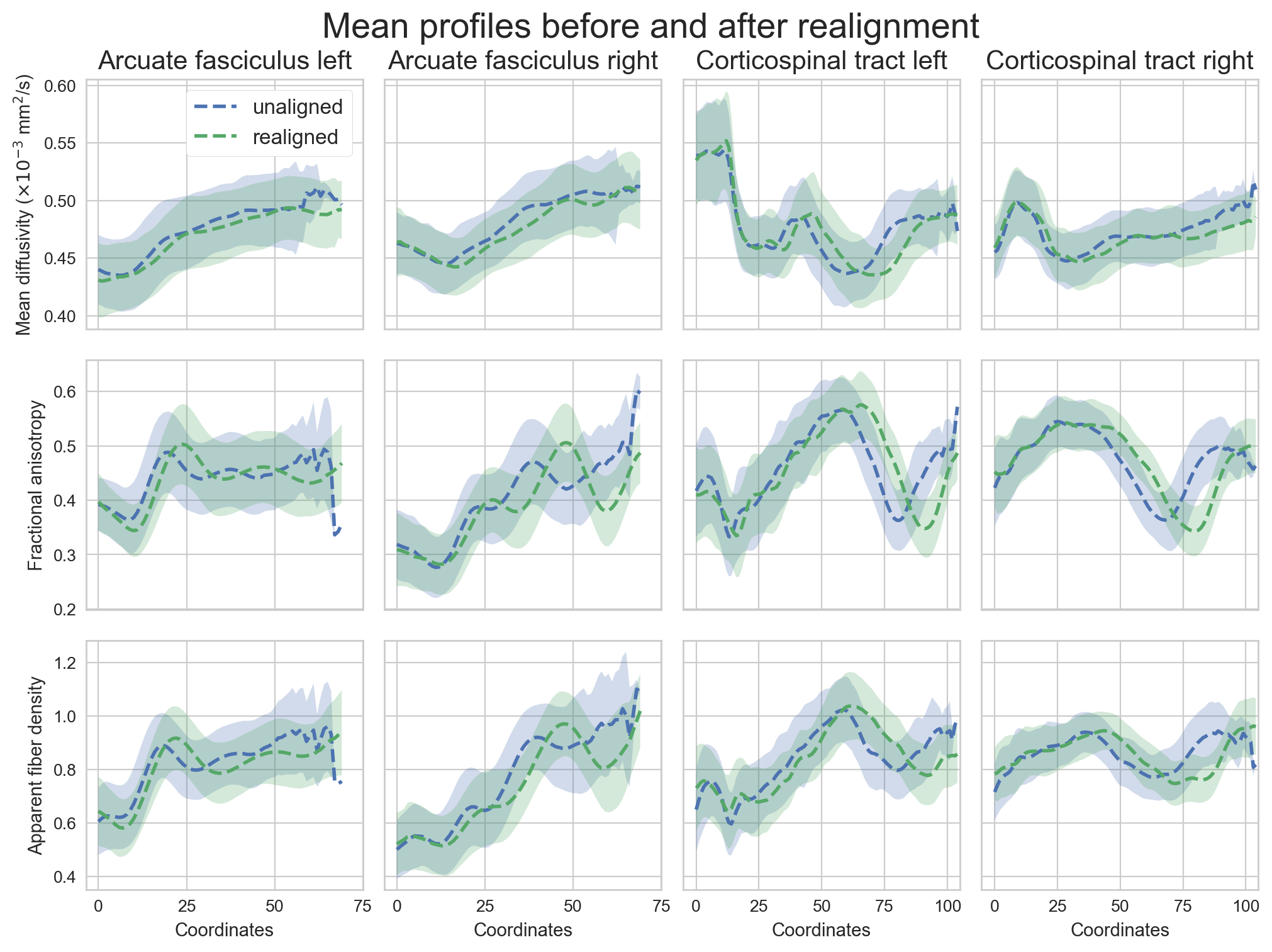}
    \caption{Along-tract averaged profiles (and standard deviation as the shaded area) of the unaligned (blue) and
    realigned (green) HCP subjects truncated to 75\% of overlap with a final resampling to the same number of points.
    These results are obtained by using a maximally allowed displacement of 20\%.}
    \label{fig:hcp_realignment20}
\end{figure}

\FloatBarrier

\subsection{Displacement of the HCP datasets}
\label{sec:supp_displacement}

\Cref{sec:supp_displacement} presents counterpart results to \cref{fig:hcp_displacement} using realignment from other metrics,
but instead using a maximally allowed displacement of 5\%, 10\% or 20\%.
\review{Coordinates for the AF are from anterior (coordinate 0) to posterior and the CST are drawn from inferior (coordinate 0) to superior.}
Most of the trends observed previously when the displacement was of 15\% are still valid.

\begin{figure}
    \includegraphics[width=\linewidth]{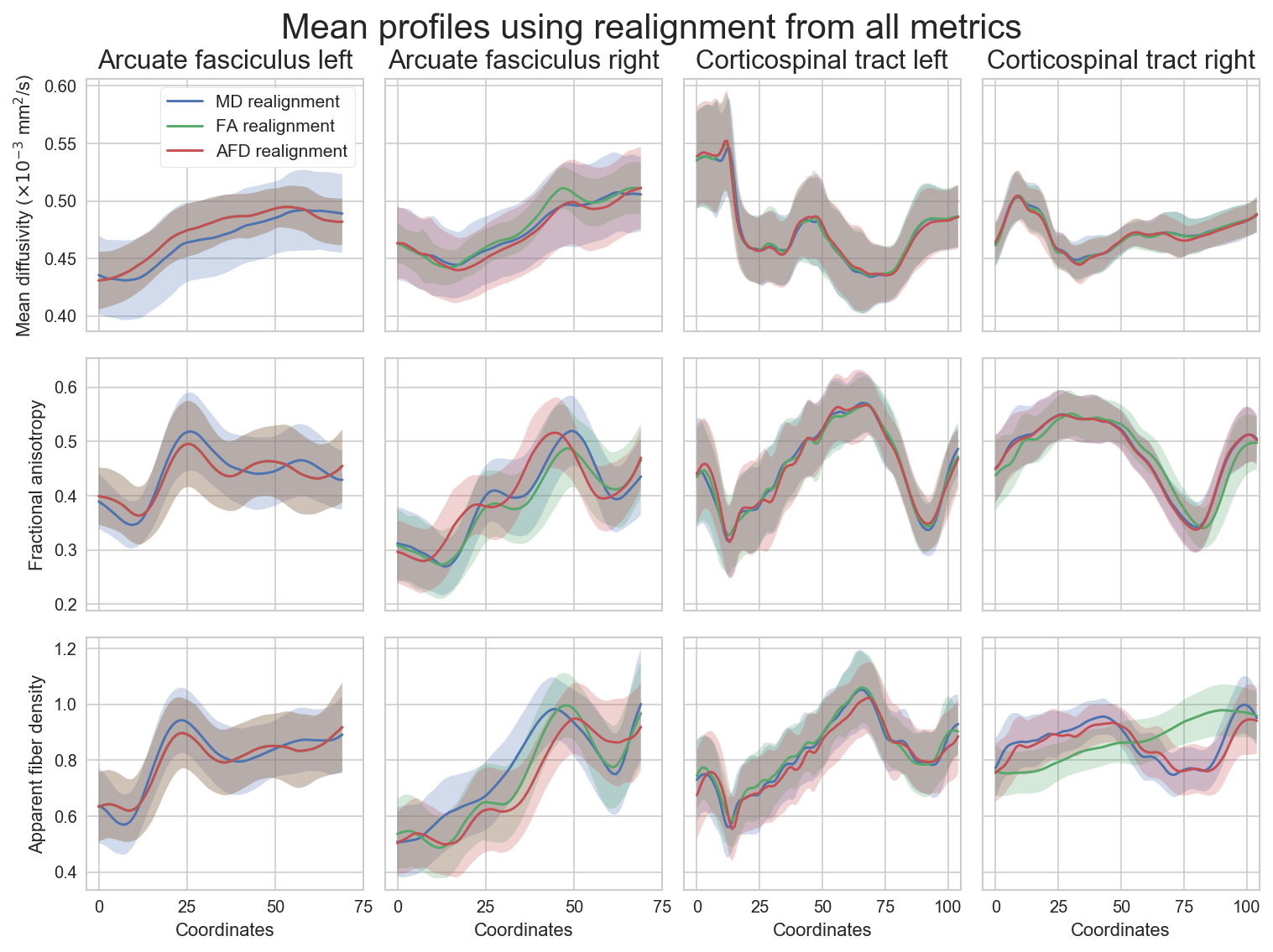}
    \caption{Along-tract averaged profiles (and standard deviation as the shaded area) of the white matter fiber bundles
    (columns) from the HCP datasets after realignment for each studied metric (rows).
    These results are obtained by using a maximally allowed displacement of 5\%.
    }
    \label{fig:hcp_displacement5}
\end{figure}

\begin{figure}
    \includegraphics[width=\linewidth]{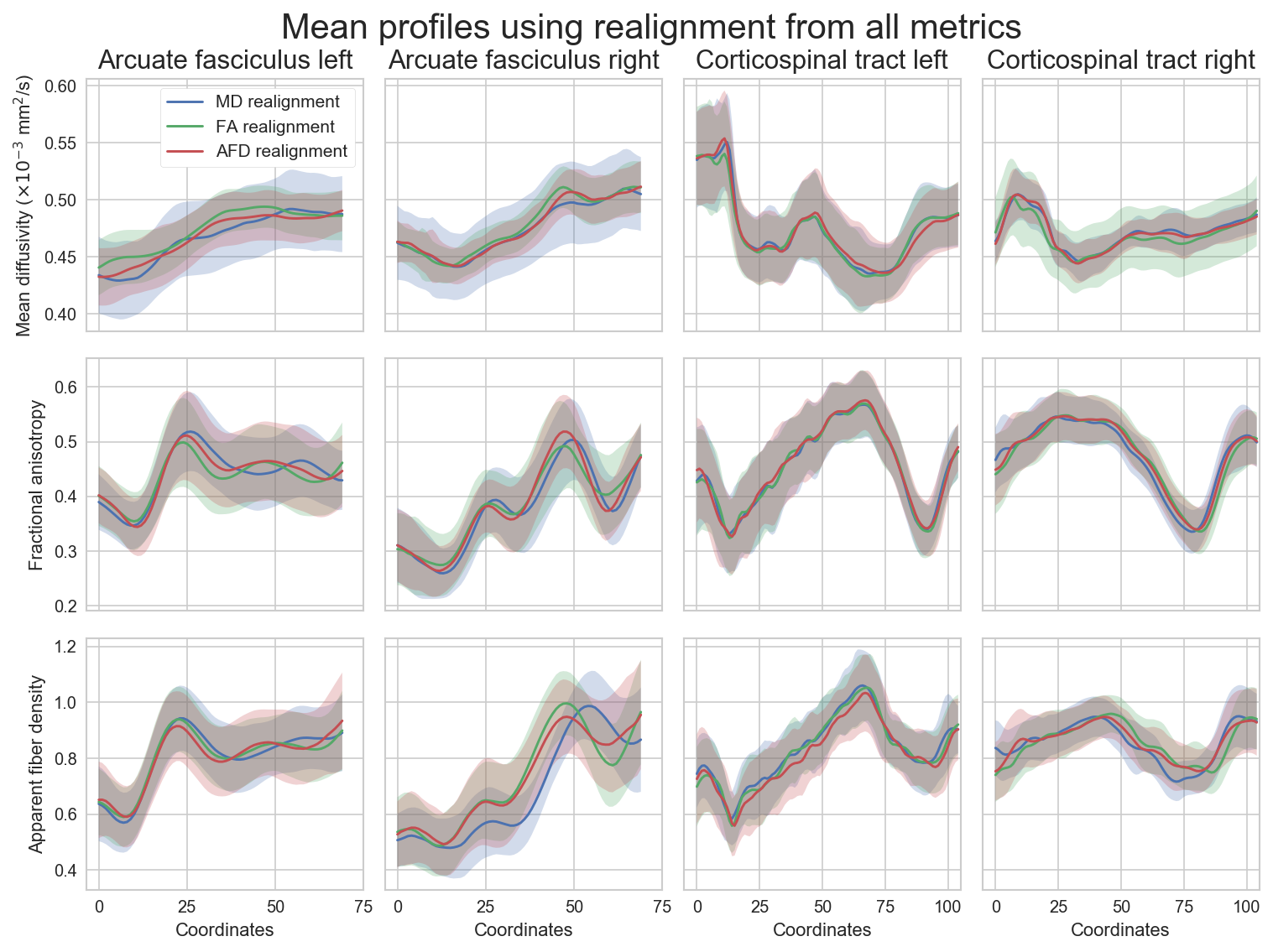}
    \caption{Along-tract averaged profiles (and standard deviation as the shaded area) of the white matter fiber bundles
    (columns) from the HCP datasets after realignment for each studied metric (rows).
    These results are obtained by using a maximally allowed displacement of 10\%.
    }
    \label{fig:hcp_displacement10}
\end{figure}

\begin{figure}
    \includegraphics[width=\linewidth]{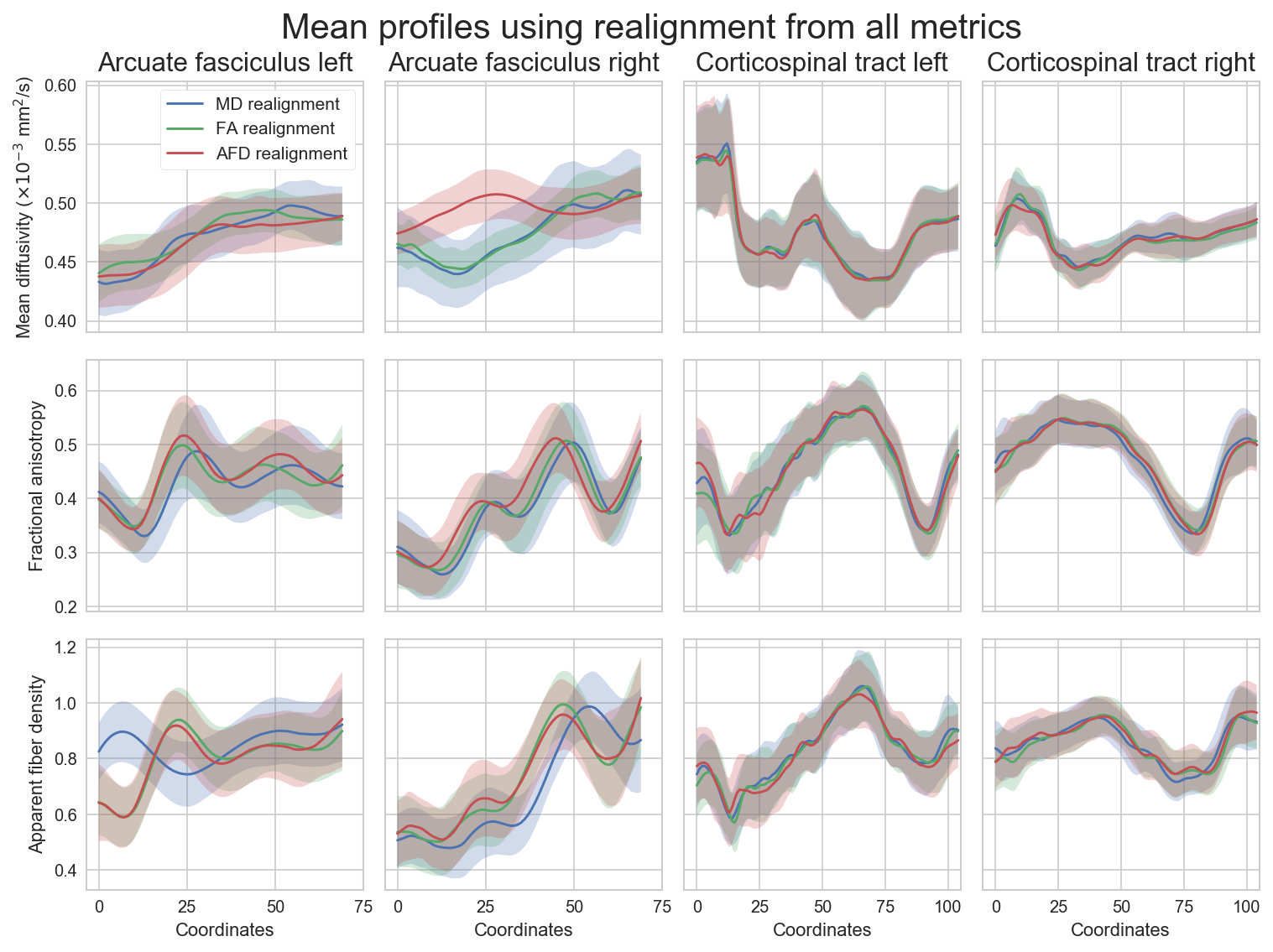}
    \caption{Along-tract averaged profiles (and standard deviation as the shaded area) of the white matter fiber bundles
    (columns) from the HCP datasets after realignment for each studied metric (rows).
    These results are obtained by using a maximally allowed displacement of 20\%.
    }
    \label{fig:hcp_displacement20}
\end{figure}

\FloatBarrier

\subsection{Localized alterations of the HCP datasets}
\label{sec:supp_focused}

\review{\Cref{sec:supp_focused} presents counterpart results to \cref{fig:hcp_focused},
but instead using a maximally allowed displacement of 5\%, 10\% or 100\% (no limit).
\rereview{Unpaired} t-test (FDR corrected at $\alpha = 0.05$) with focused alterations of the metrics for each bundle of
\textbf{A)} 25\% over 1\% of the length, \textbf{B)} 50\% over 1\% of the length, \textbf{C)} 25\% over 5\% of the length and \textbf{D)} 50\% over 5\% of the length are shown.
The AF left/right are represented from anterior (coordinate 0) to posterior and the CST left/right from inferior (coordinate 0) to superior.
The p-values are on a log scale along the average streamline before realignment (dashed red lines) and after realignment (solid blue lines) with the DPR algorithm.
The horizontal dashed black lines indicate p-value = 0.05.
In general, alterations \rereview{occurring over} 5\% of the \rereview{length of the} bundle can be detected,
whereas small \rereview{local} changes \rereview{over} 1\% \rereview{of the length are detected only after realignment with the DPR algorithm.}}

\begin{figure}
\figuretitle{Focused alterations with a maximum displacement of 5\%}
    \begin{annotatedFigure}{\includegraphics[width=0.495\linewidth]{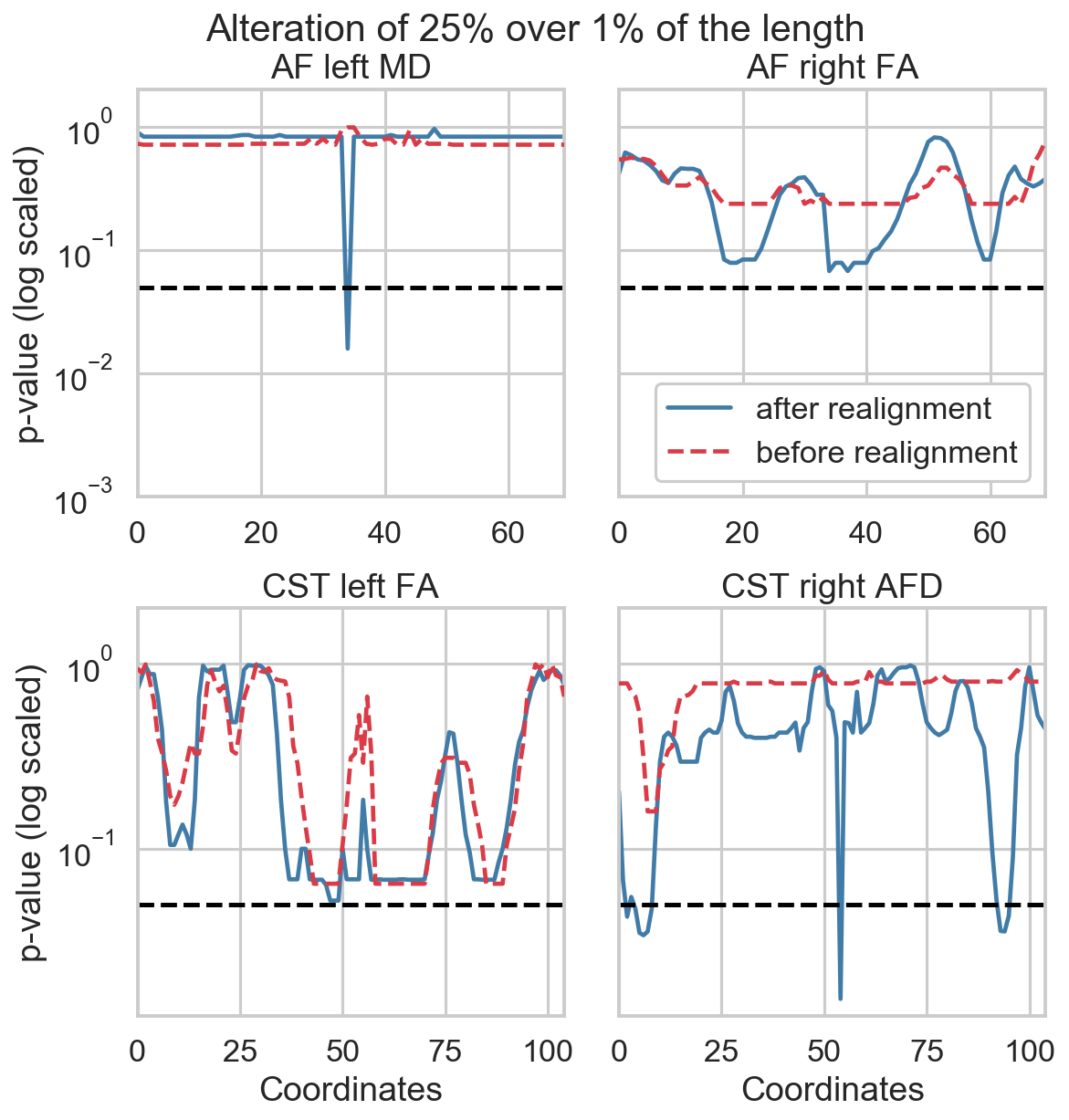}}
        \annotatedFigureBox{0.1,0.97}{A}
    \end{annotatedFigure}
    \hfill
    \begin{annotatedFigure}{\includegraphics[width=0.495\linewidth]{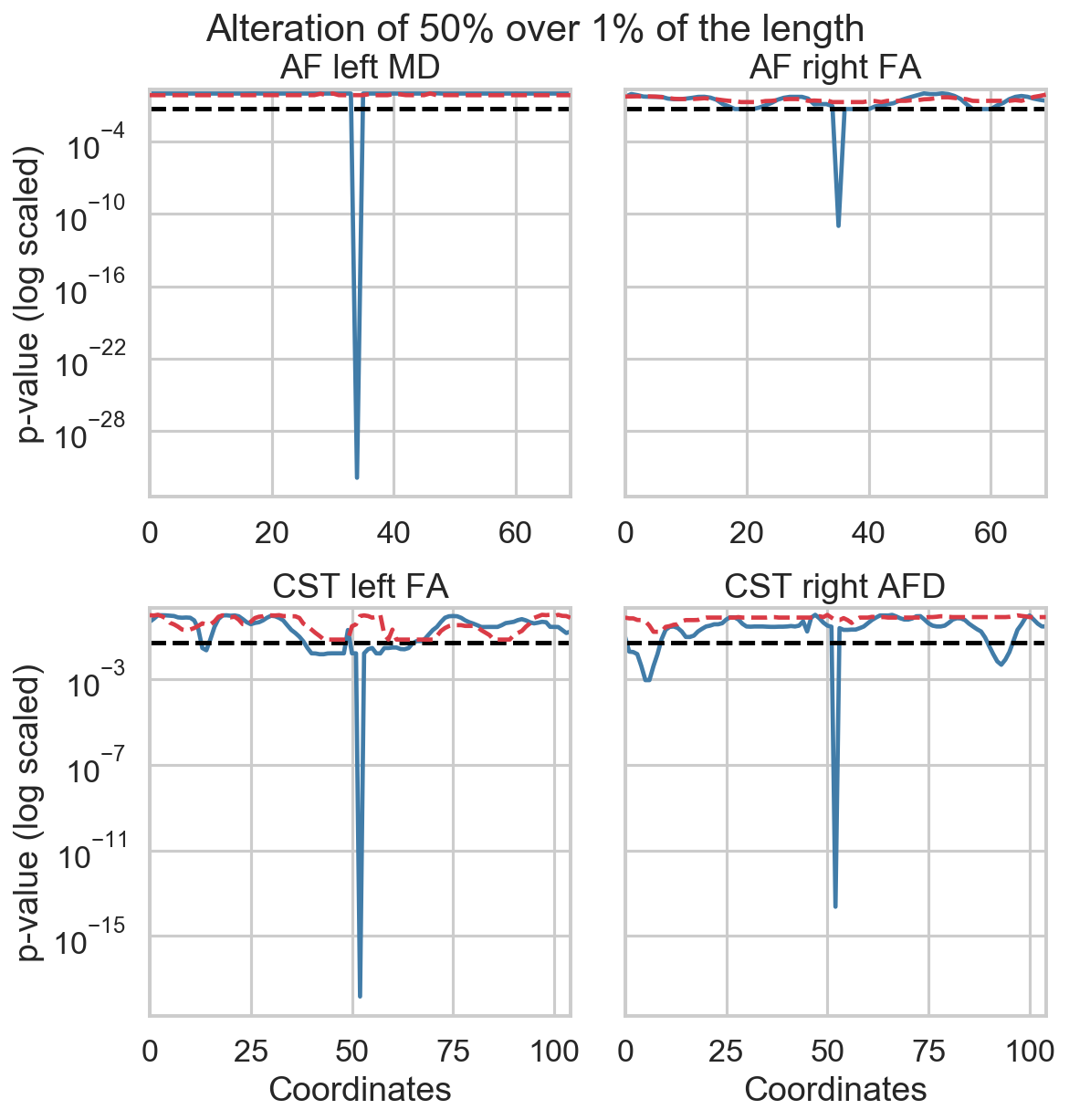}}
        \annotatedFigureBox{0.1,0.97}{B}
    \end{annotatedFigure}

    \begin{annotatedFigure}{\includegraphics[width=0.495\linewidth]{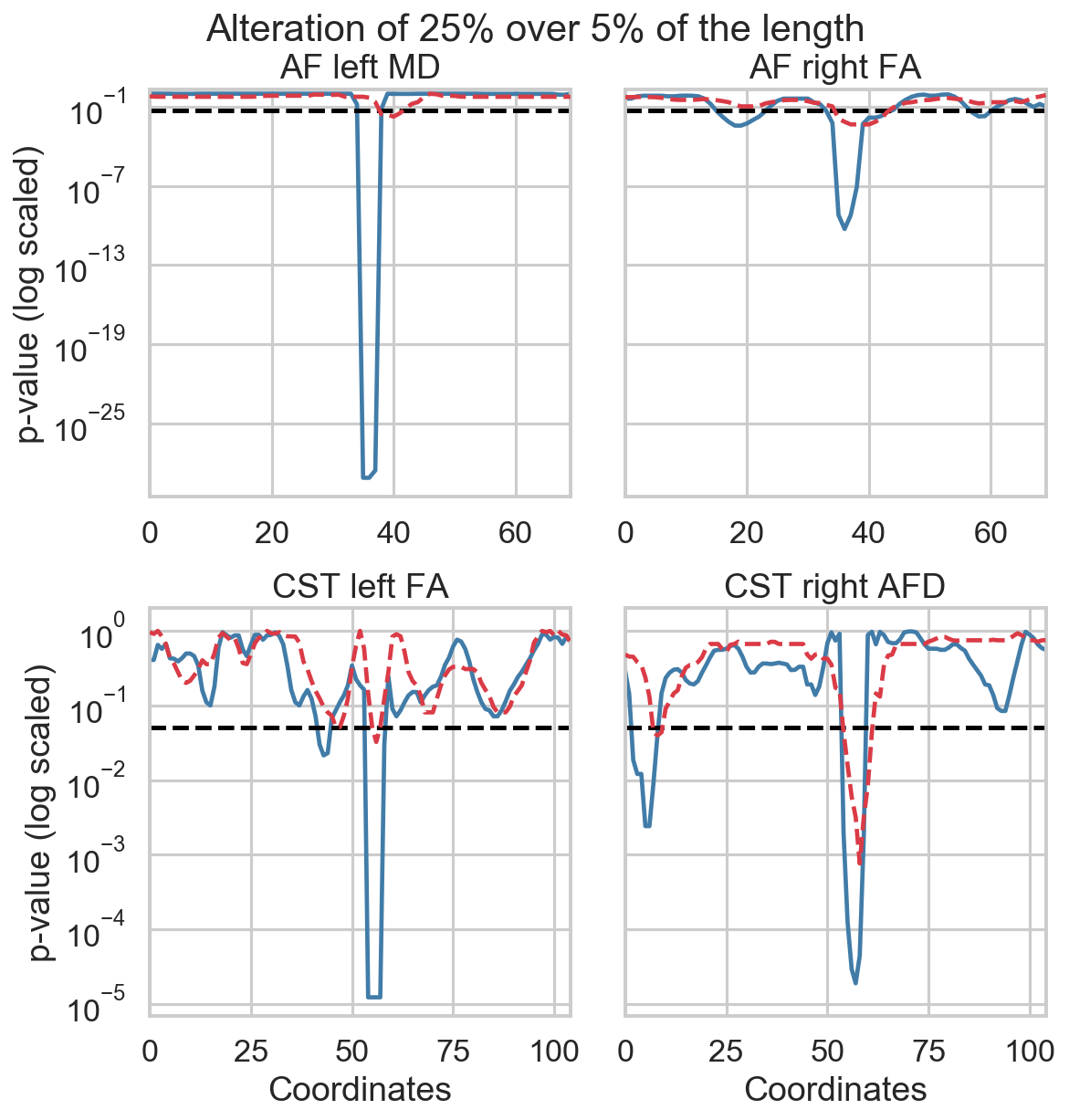}}
        \annotatedFigureBox{0.1,0.97}{C}
    \end{annotatedFigure}
    \hfill
    \begin{annotatedFigure}{\includegraphics[width=0.495\linewidth]{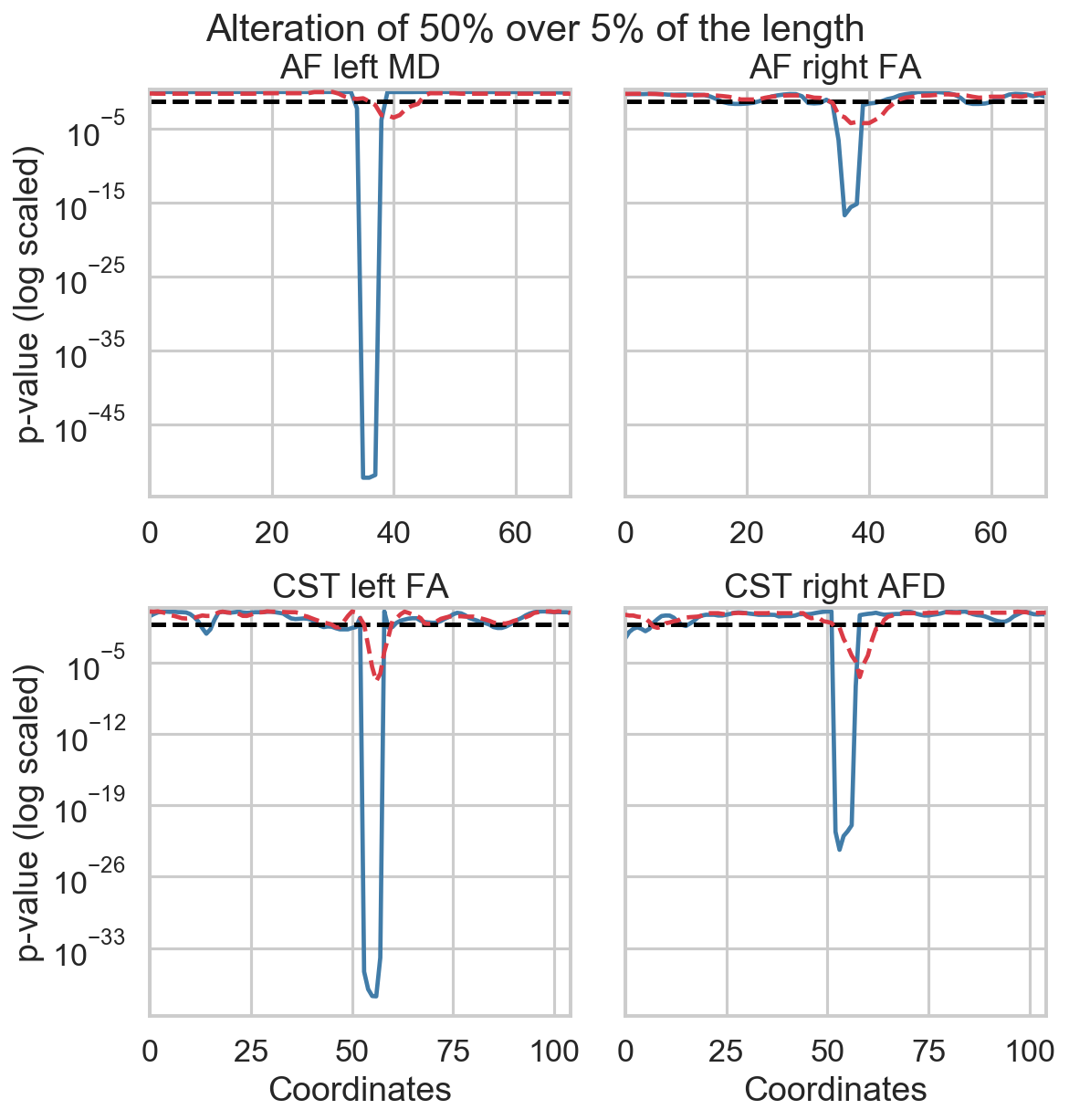}}
        \annotatedFigureBox{0.1,0.97}{D}
    \end{annotatedFigure}
    \caption{\review{\rereview{Unpaired} t-test before and after realignment for the four bundles.
    These results are obtained by using a maximally allowed displacement of 5\%.}}
    \label{fig:hcp_focused5}
\end{figure}

\begin{figure}
    \figuretitle{Focused alterations with a maximum displacement of 10\%}
    \begin{annotatedFigure}{\includegraphics[width=0.495\linewidth]{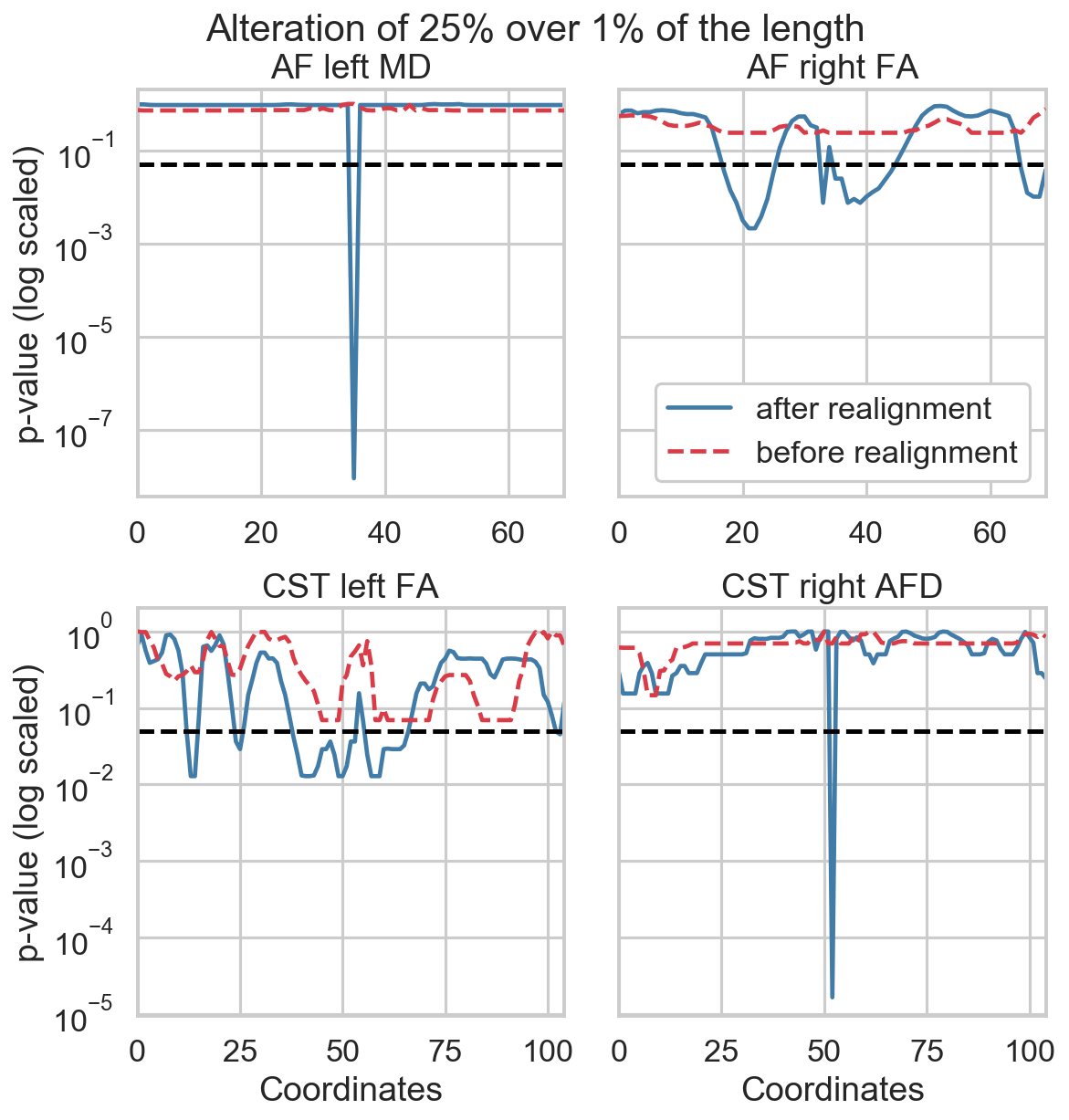}}
        \annotatedFigureBox{0.1,0.97}{A}
    \end{annotatedFigure}
    \hfill
    \begin{annotatedFigure}{\includegraphics[width=0.495\linewidth]{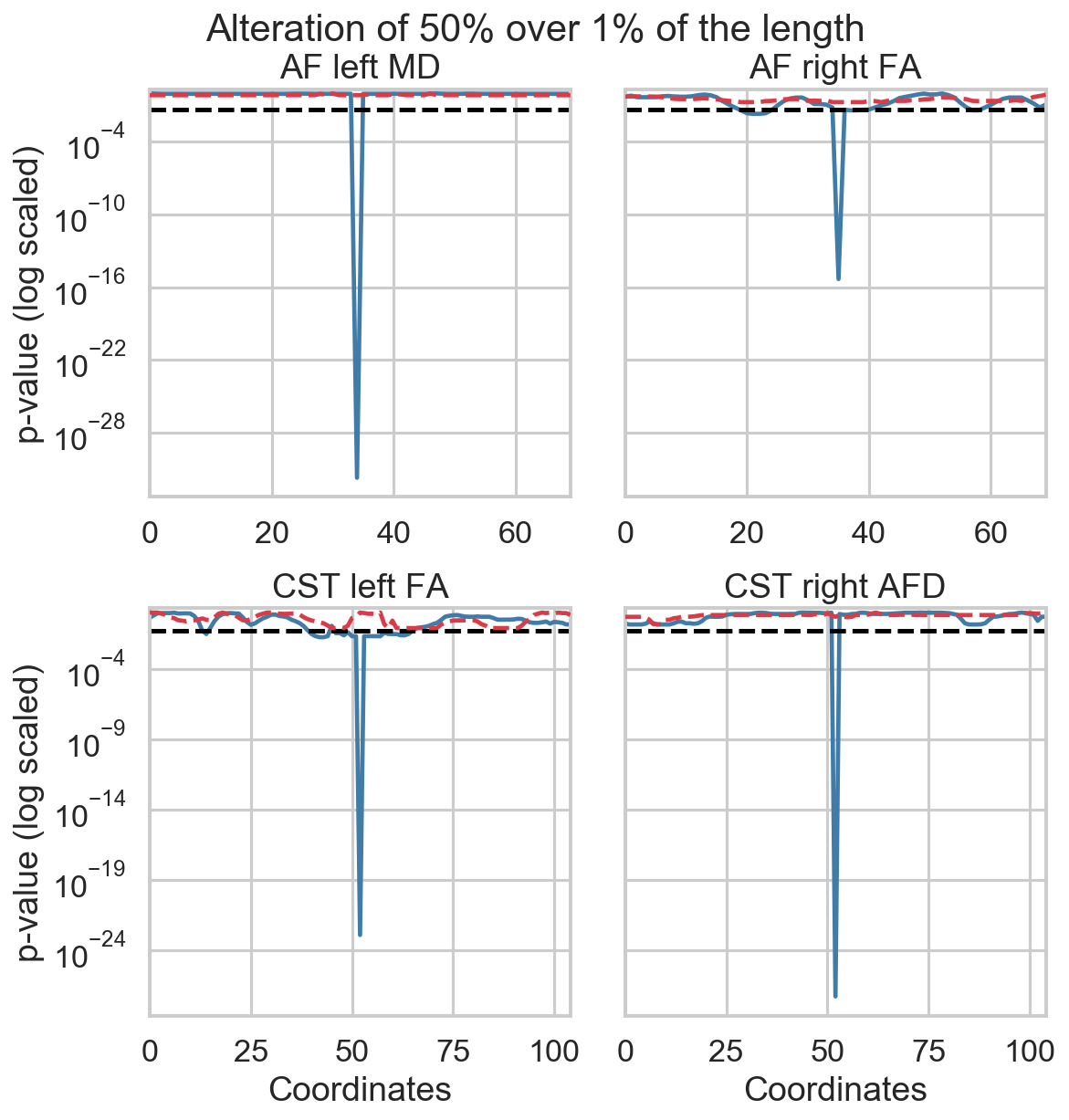}}
        \annotatedFigureBox{0.1,0.97}{B}
    \end{annotatedFigure}

    \begin{annotatedFigure}{\includegraphics[width=0.495\linewidth]{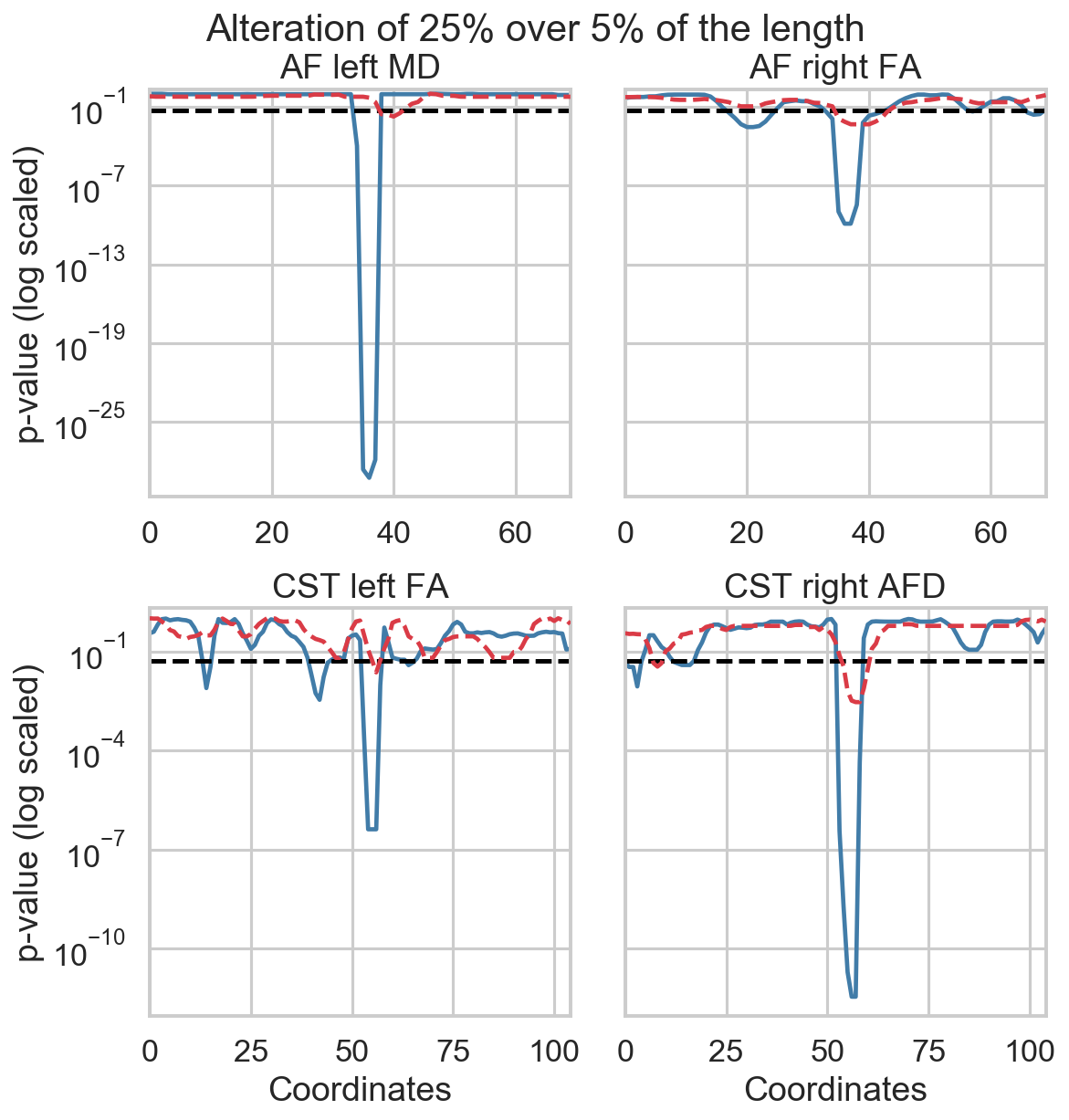}}
        \annotatedFigureBox{0.1,0.97}{C}
    \end{annotatedFigure}
    \hfill
    \begin{annotatedFigure}{\includegraphics[width=0.495\linewidth]{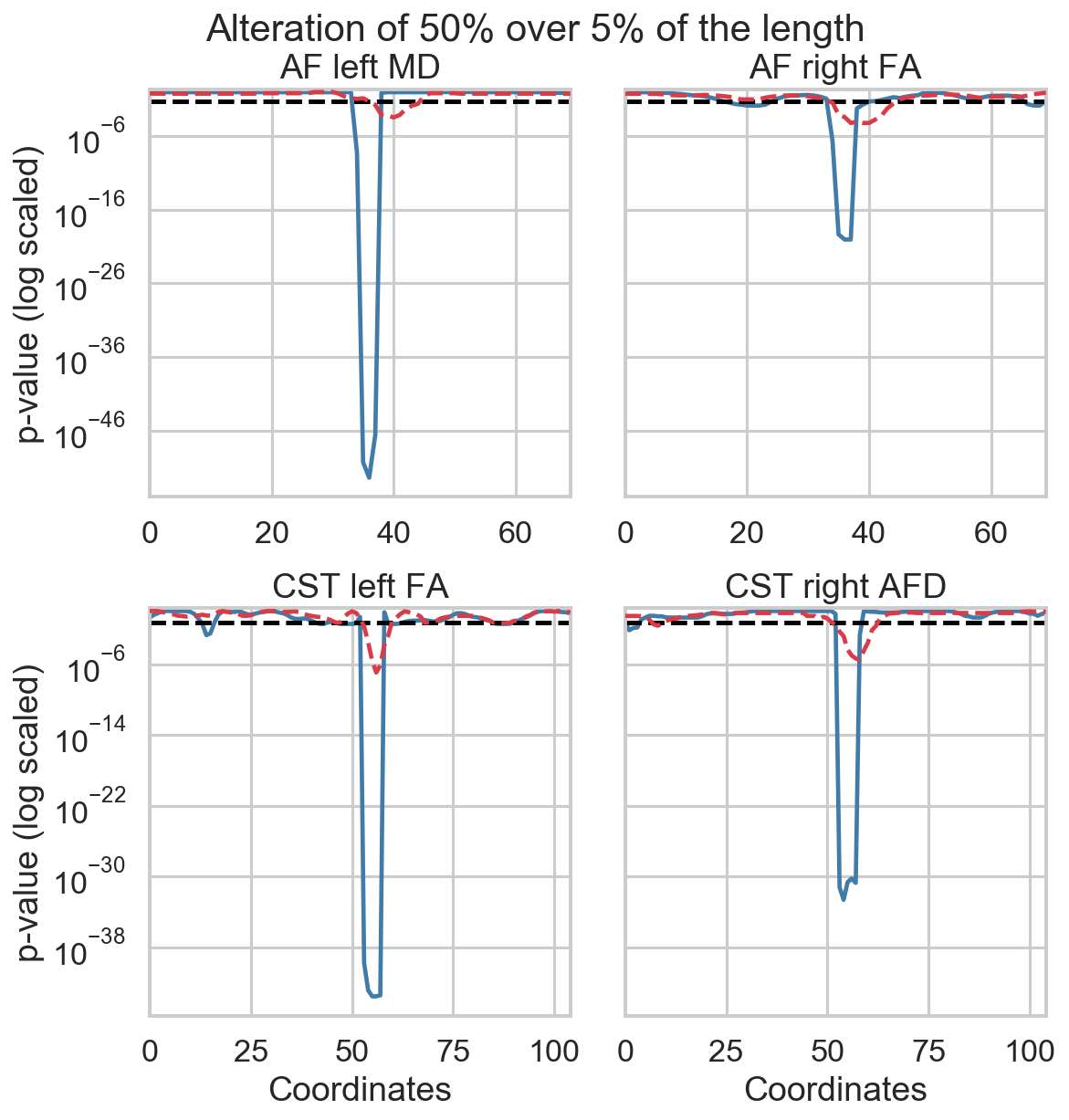}}
        \annotatedFigureBox{0.1,0.97}{D}
    \end{annotatedFigure}
    \caption{\review{\rereview{Unpaired} t-test before and after realignment for the four bundles.
    These results are obtained by using a maximally allowed displacement of 10\%.}}
    \label{fig:hcp_focused10}
\end{figure}

\begin{figure}
    \figuretitle{Focused alterations without limiting the maximum displacement}
    \begin{annotatedFigure}{\includegraphics[width=0.495\linewidth]{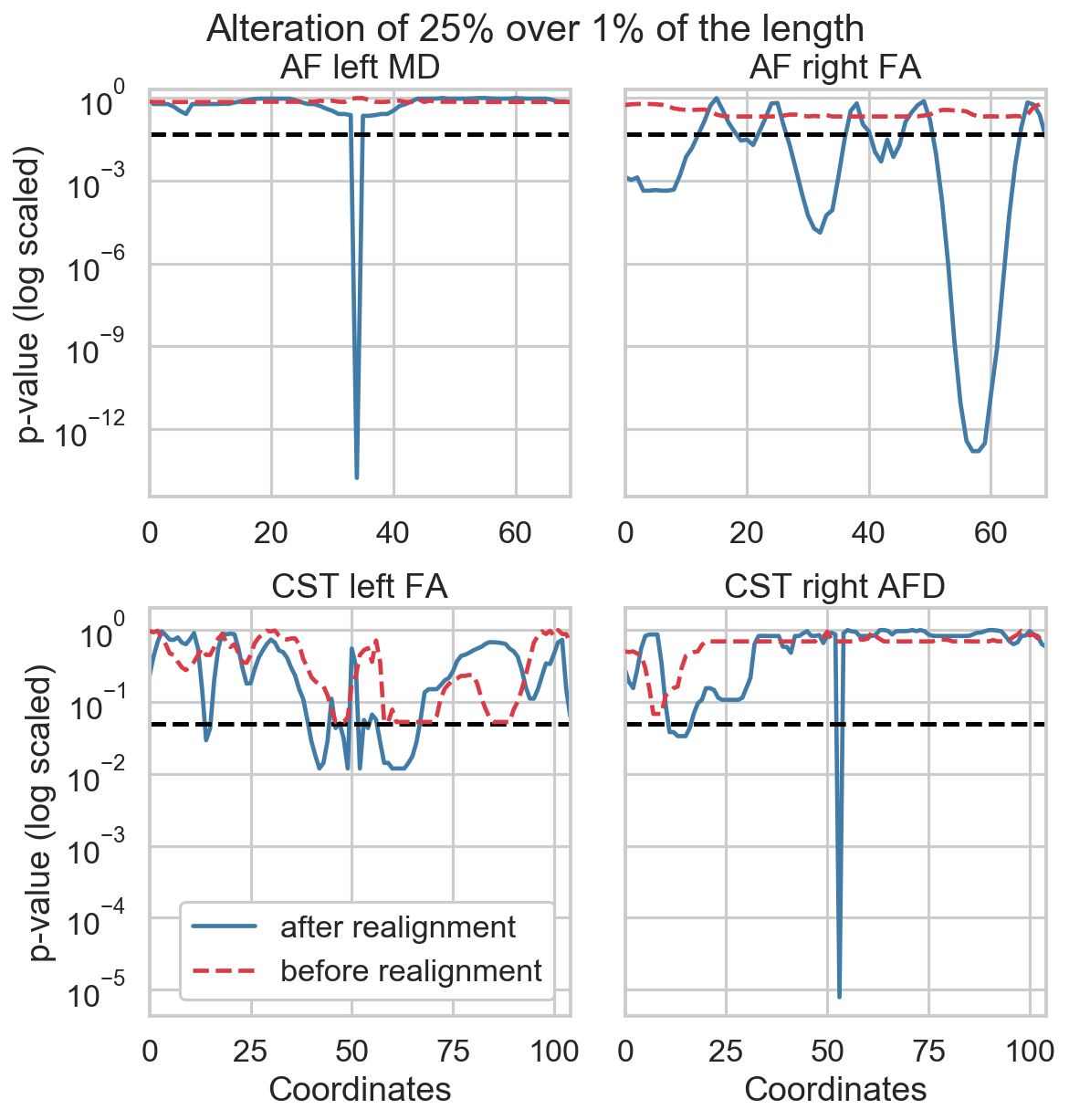}}
        \annotatedFigureBox{0.1,0.97}{A}
    \end{annotatedFigure}
    \hfill
    \begin{annotatedFigure}{\includegraphics[width=0.495\linewidth]{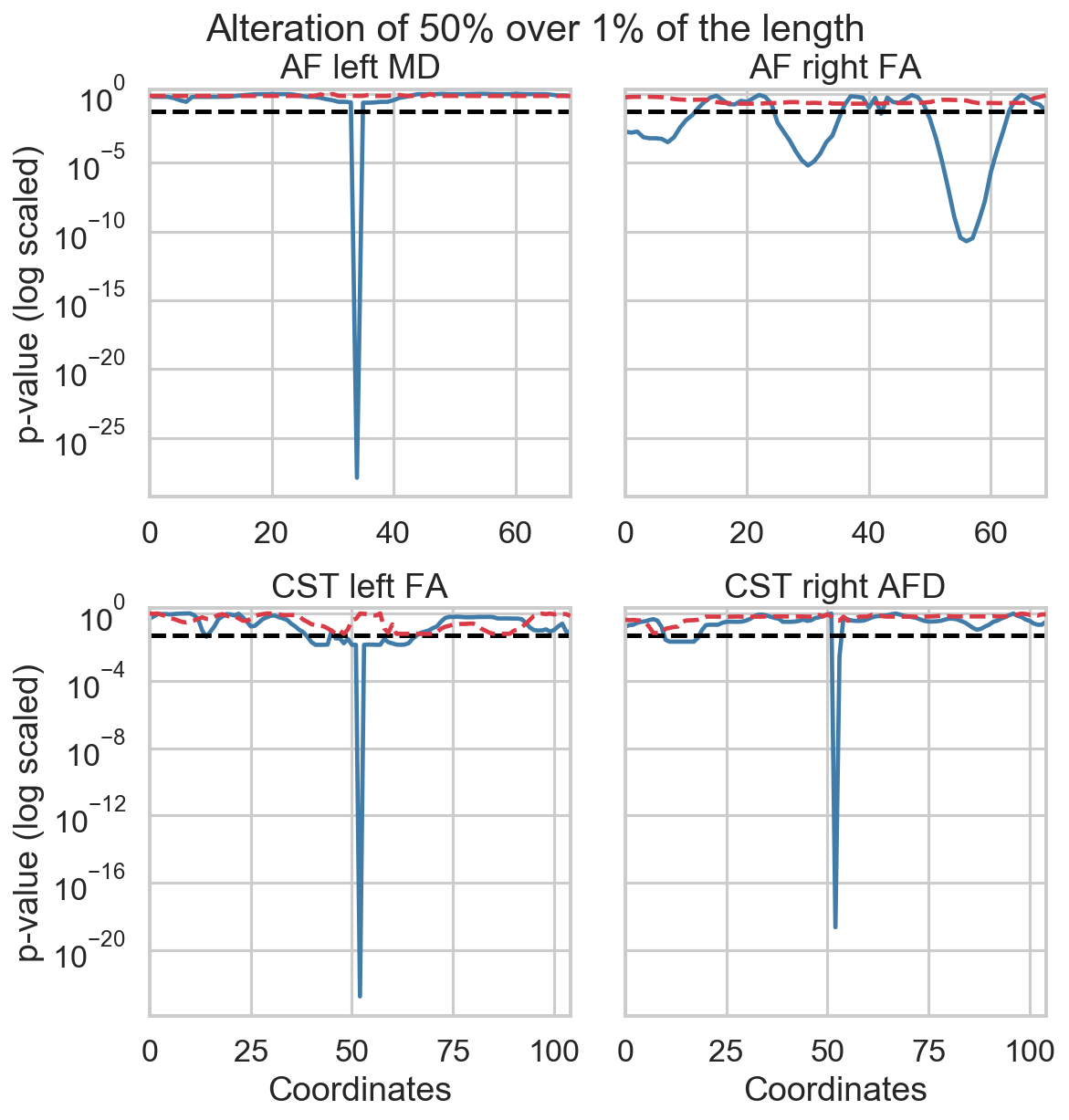}}
        \annotatedFigureBox{0.1,0.97}{B}
    \end{annotatedFigure}

    \begin{annotatedFigure}{\includegraphics[width=0.495\linewidth]{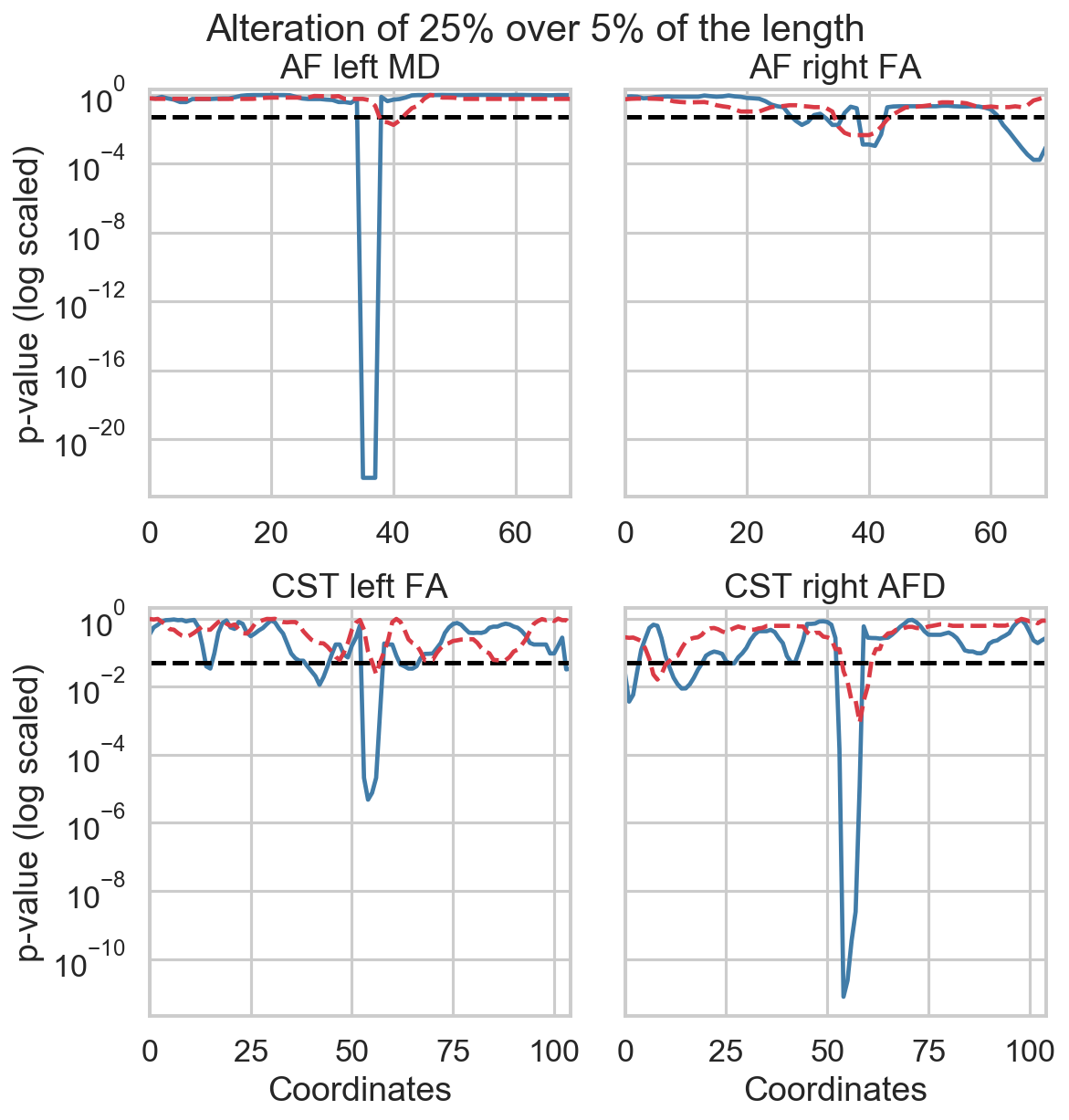}}
        \annotatedFigureBox{0.1,0.97}{C}
    \end{annotatedFigure}
    \hfill
    \begin{annotatedFigure}{\includegraphics[width=0.495\linewidth]{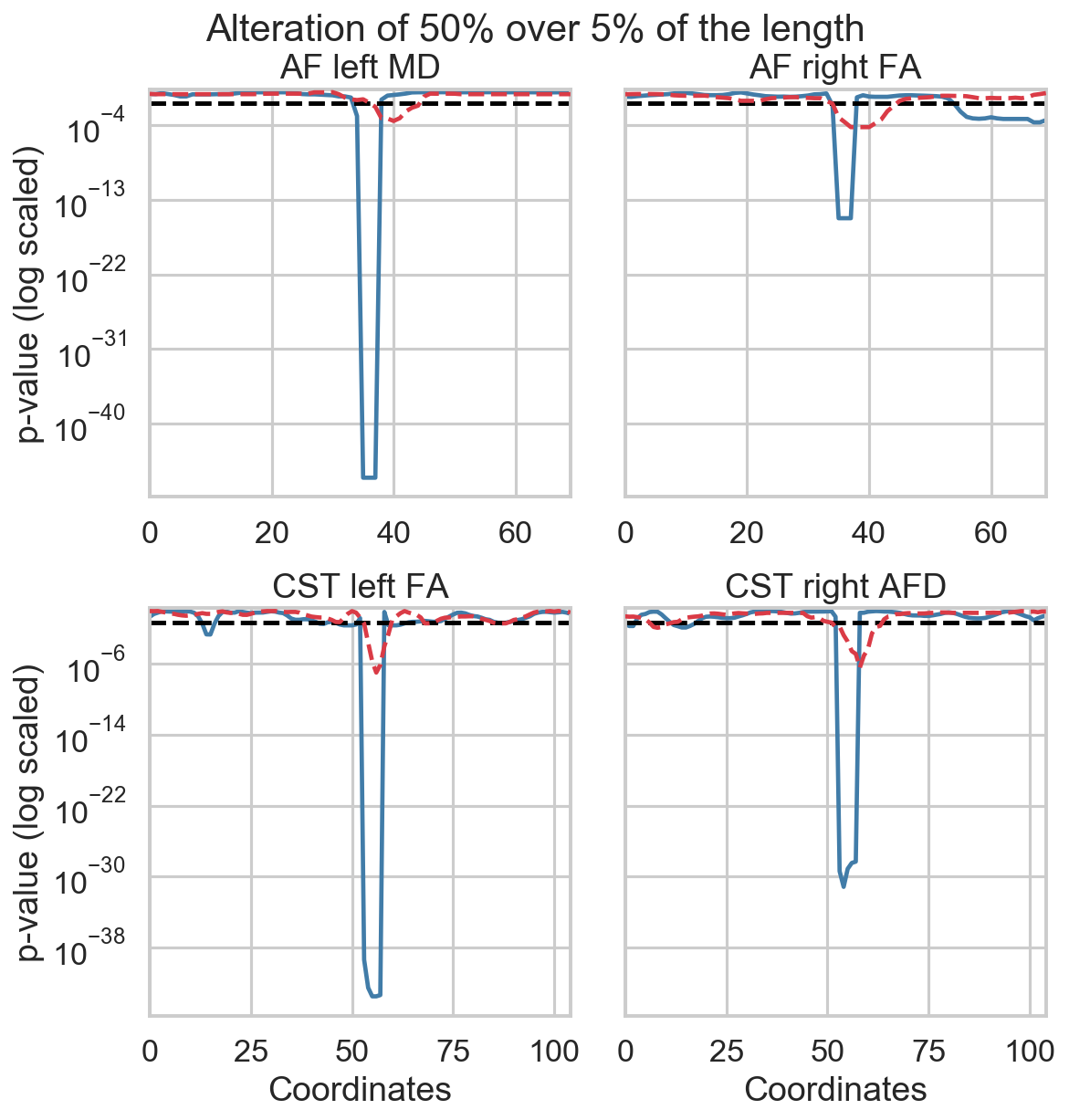}}
        \annotatedFigureBox{0.1,0.97}{D}
    \end{annotatedFigure}
    \caption{\review{\rereview{Unpaired} t-test before and after realignment for the four bundles.
    These results are obtained without limiting the allowed maximum displacement.
    This leads to false effects for the AF right bundles, presumably because structural differences,
    rather than local alterations, are driving the realignment process.}}
    \label{fig:hcp_focused100}
\end{figure}